\documentclass[aps,pra,% journal choice determines style
			   reprint,% or preprint for single column 1.5 line spacing
			   a4paper,% for UK printers
			   superscriptaddress,% for PRA style author list
			   longbibliography, % includes article titles, useful for drafting and review
			   floatfix, % produces extra stuck float errors, but might help fix them too
		       nofootinbib %footnotes on bottom of page, they weren't showing up otherwise
			  ]{revtex4-1}	% this may depend on your latex installation
\usepackage[colorlinks=true,linkcolor=blue,urlcolor=blue,citecolor=blue]{hyperref}
\usepackage{graphicx}	% to include eps figures
\graphicspath{ {QWAQCfig_files/} } %% tidy subdir for figure files

\RequirePackage{amsmath}
\RequirePackage{mathtools} % fixes broken stuff in amsmath
\usepackage{amsfonts}
\usepackage{amssymb}
\usepackage{amsbsy}
\usepackage{dsfont}   % for a nice identity symbol \openone
\usepackage{latexsym}
\newcommand{\bra}[1]{\langle#1 |}
\newcommand{\ket}[1]{|#1 \rangle}
\newcommand{\braket}[2]{\left \langle #1 | #2 \right \rangle}
\newcommand{\ketbra}[2]{\vert #1 \rangle \! \langle #2 \vert}
\newcommand{\average}[1]{\langle #1 \rangle}
\newcommand{\sandwich}[3]{\left \langle #1 \left\lvert #2 \right\rvert #3 \right\rangle}
\newcommand{\List}{\text{\textsc{List}}}
\newcommand{\bin}[2]{\left(\begin{array}{c}#1\\#2\\ \end{array}\right)}
\begin{document}  % note this comes earlier than standard latex article class
%%%%%%%%%%%%%%%%%%%%%%%%%%%%%%%%%%%%%%%%%%%%%%%%%%%%%%%%%%%%%%%%%%%%%%%%%%%%%%%
%
\title{Quantum search with hybrid adiabatic--quantum walk algorithms and realistic noise}
\date{Friday 10th August, 2018} 
%
% author list:
\author{James G.~Morley}
\email{james.morley.15@ucl.ac.uk}	% corresponding author emails only
\affiliation{Department of Physics, UCL, 
Gower Street, London, UK}
\author{Nicholas Chancellor}
\affiliation{Department of Physics, Durham University, 
South Road, Durham, UK}
\author{Sougato Bose}
\affiliation{Department of Physics, UCL, 
Gower Street, London, UK}
\author{Viv Kendon}
\email{viv.kendon@durham.ac.uk}
\affiliation{Department of Physics, Durham University, 
South Road, Durham, UK}

%%%%%%%%%%%%%%%%%%%%%%%%%%%%%%%%%%%%%%%%%%%%%%%%%%%%%%%%%%%%%%%%%%%%%%%%%%%%%%%
\begin{abstract}
Computing using a continuous-time evolution, based on the natural interaction 
Hamiltonian of the quantum computer hardware, is a promising route to building 
useful quantum computers in the near-term.  Adiabatic quantum computing, quantum
annealing, computation by continuous-time quantum walk, and special purpose 
quantum simulators all use this strategy.  In this work, we carry out a detailed
examination of adiabatic and quantum walk implementation of the quantum search 
algorithm, using the more physically realistic hypercube connectivity, rather 
than the complete graph, for our base Hamiltonian.  We calculate optimal 
adiabatic schedules both analytically and numerically for the hypercube, and 
then interpolate between adiabatic and
quantum walk searching, obtaining a family of hybrid algorithms.  We show that 
all of these hybrid algorithms provide the quadratic quantum speed up when run 
with optimal parameter settings, which we determine and discuss in detail.  
We incorporate the effects of multiple runs of the same algorithm, noise applied
to the qubits, and two types of problem misspecification, determining the 
optimal hybrid algorithm for each case.  Our results reveal a rich structure of 
how these different computational mechanisms operate and should be balanced in 
different scenarios.  For large systems with low noise and good control, quantum
walk is the best choice, while hybrid strategies can mitigate the effects of 
many shortcomings in hardware and problem misspecification.
\end{abstract}
%%%%%%%%%%%%%%%%%%%%%%%%%%%%%%%%%%%%%%%%%%%%%%%%%%%%%%%%%%%%%%%%%%%%%%%%%%%%%%%

\maketitle

%%%%%%%%%%%%%%%%%%%%%%%%%%%%%%%%%%%%%%%%%%%%%%%%%%%%%%%%%%%%%%%%%%%%%%%%%%%%%%%
\section{Introduction\label{sec:intro}}
%-----------------------------------------------------------------------------%

Quantum computing based not on discrete quantum gates, but on continuous-time 
evolution under quantum Hamiltonians, is a promising route towards near-future 
useful quantum computers.  This is in part because of the success of 
experimental quantum  annealing efforts
\cite{brooke99a,johnson11a,denchev16a,lanting14a,boixo16a}, and also because 
of special purpose quantum simulators \cite{georgescu14,nguyen18}
that employ this technique, and are 
potentially useful for a wider range of computations \cite{kendon17a}.  
Problems known to be suitable for continuous-time algorithms are wide-ranging 
across many important areas, including finance \cite{marzec16a}, aerospace 
\cite{coxson14a}, machine learning \cite{amin16a,Benedetti16a,Benedetti16b}, 
theoretical computer science \cite{chancellor16a}, decoding of communications 
\cite{chancellor16b}, mathematics \cite{Li17a,Bian13a}, and computational 
biology \cite{perdomo-ortiz12a}. 

Continuous-time computation is less familiar than the ubiquitous digital 
computation that underpins everything from mobile phones to internet servers. 
There is no classical equivalent of computing via continuous-time manipulation
of digital data to guide our intuition, or provide a source of classical
algorithmic resources that might be adapted to a quantum setting.  A detailed
study based on a well-characterised problem can thus serve to elucidate the
mechanisms in continuous-time quantum computing and build a firm foundation for
further development.  Hence, we focus this work on the unordered search
problem first studied in a quantum setting by Grover in 1997 \cite{grover97a}.
Grover's algorithm provides a quadratic speed up over classical searching,
proved by Bennett at al.~\cite{bennett97a} to be the best possible improvement.

Two further  examples of quantum search algorithms are \emph{quantum walk} (QW) 
searching \cite{shenvi02a} and the \emph{adiabatic quantum computing} (AQC) 
search algorithm \cite{roland02a}, which both obtain the optimal quadratic speed
up.  There remains the questions of which is more efficient in terms of the
prefactors \cite{lovett11a}, or more robust in the face of imperfections.
While the results in \cite{bennett97a} imply that any protocol we develop here 
will not provide better scaling properties, asymptotic scaling factors don't 
give a full account of algorithm performance.  A recent study in which a 
quantum annealer appears to show the same asymptotic scaling as a classical 
algorithm, but with a prefactor advantage \cite{denchev16a} of $\sim10^8$, 
underscores the importance of practical computational advantages beyond 
asymptotic scaling.  This prompts more detailed study of exactly how the quantum
search algorithms work, the topic of many papers since the original algorithms 
were first presented \cite{coulamy17a,tulsi15a,ambainis11a,magniez11a}.

Quantum walk searching has been shown to implement a similar type of rotation 
in Hilbert space to that which Grover's algorithm employs \cite{shenvi02a}.  
On the other hand, adiabatic quantum searching alters the Hamiltonian over time, 
turning on the term for the marked state slowly enough to keep the quantum 
system in its ground state throughout.  On the face of it, these are quite 
different dynamics, as has been highlighted in \cite{wong16a}.  However, both 
use the same Hamiltonians and initial states, and we argue here that both are 
best viewed as extreme cases of possible quantum annealing schedules.  This 
invites consideration of intermediate quantum annealing schedules, and we show 
how to interpolate smoothly between QW and AQC, enabling both mechanisms to contribute
to solving the search problem.  We examine the hybrid algorithms thus created 
using simplified models for the asymptotic scaling, and numerical simulation
to explore smaller systems where more complex finite size effects contribute.
Taking into account realistic factors, such as a finite initialisation time for 
each run of the algorithm,  our results reveal a rich structure of intermediate 
strategies available to optimise  the performance of a practical quantum 
computer.

The paper is structured as follows: In Sec.~\ref{sec:background}, we give the 
background and lay the groundwork for our study in terms of the QW and AQC 
protocols which we interpolate between. In Sec.~\ref{sub:optim_sched} we 
introduce the two AQC schedules 
which we use in this study, and we explain in detail how they arise from the 
dynamics of the quantum search Hamiltonian on a hypercube.  In 
Sec.~\ref{sec:generalized_scheds}, we construct interpolated protocols which can
take advantage of both QW and AQC mechanisms.  We then turn to the performance 
of the interpolated protocols in finite-sized systems.  In 
Sec.~\ref{sec:scaling}, we examine the scaling for larger systems in detail, and
demonstrate that the interpolated protocols also yield a quadratic speed up
over classical searching.  In Sec.~\ref{sec:multi_strats} we incorporate 
strategies which involve performing multiple runs, including in 
Sec.~\ref{sub:open_systems} the effect of adding decoherence, and in 
Sec.~\ref{sec:problem_misspec} we examine the effect of problem 
misspecification.  Finally, in Sec.~\ref{sec:conclusion} we summarise our 
results and their implications for future work.  The calculation of the optimal 
schedule for the hypercube is outlined in appendix \ref{app:sched_analytics}, 
and notes on our numerical methods are in appendix \ref{app:num_meth}.

%%%%%%%%%%%%%%%%%%%%%%%%%%%%%%%%%%%%%%%%%%%%%%%%%%%%%%%%%%%%%%%%%%%%%%%%%%%%%%%
\section{Background\label{sec:background}}
%-----------------------------------------------------------------------------%

We begin with a discussion of how the unstructured search problem may be encoded
into qubit states. From here we show how, with the use of very similar Hamiltonians,
the search problem can be solved with an optimal quantum scaling advantage, via 
both QW and AQC algorithms.

%%%%%%%%%%%%%%%%%%%%%%%%%%%%%%%%%%%%%%%%%%%%%%%%
\subsection{Encoding search into quantum states}
%%%%%%%%%%%%%%%%%%%%%%%%%%%%%%%%%%%%%%%%%%%%%%%%

The search problem can be framed in terms of the $N=2^n$ basis states of an 
$n$-qubit system  $\{\ket{j}\} = \{\ket{0},\ket{1}\}^{\otimes n}$, where 
$\{\ket{0},\ket{1}\}$ is the basis of a single qubit. We are given that one 
of the basis states behaves differently to the others and denote this 
`marked' state as $\ket{m}$, where $m$ is an $n$-digit bitstring identifying 
one of the basis states.  Because of the difference in behaviour, we can easily 
verify whether a given state is the marked state.  One way to implement this is 
for the marked state to have a lower energy than all other states, e.g., using a
Hamiltonian like $\hat{H}_p = \hat\openone - \ketbra{m}{m}$, where $\hat\openone$ 
is the identity operator. In terms of Pauli 
operators,
\begin{equation}
\hat{H}_p = \hat\openone - \frac{1}{2^{n}}\prod_{j=1}^{n}(q_j\hat\sigma^z_j+\hat\openone),
\label{eq:H_Grover}
\end{equation} 
where $q_j \in\{-1,1\}$ define a logical bitstring $m$ via the mapping $1\rightarrow 0$ and 
$-1\rightarrow 1$. The search problem is 
then to determine which of the basis labels $j$ corresponds to the marked state 
label $m$, given that \emph{a priori} we have no knowledge of $m$, apart from it 
being a basis state.  We represent this ignorance of the marked state by starting 
with the system in a uniform superposition over the basis states,
\begin{equation} \label{eq:psi_init}
	\ket{\psi_{\text{init}}} = \frac{1}{\sqrt{N}}\sum_{j=0}^{N-1} \ket{j}.
\end{equation}

The quantum search algorithms considered in this paper solve the search problem 
by evolving the system into a state with a large overlap with the marked state, 
so that a measurement can be made to return the marked state label $m$ with high
probability.  This is achieved by applying a (generally time-dependent) 
Hamiltonian to evolve the system initially in state
$\ket{\psi_{\text{init}}}$ to a final state $\ket{\psi_\text{final}}$.
Performing a measurement of this state in the basis $\{\ket{j}\}$
will yield the marked state label with probability 
$|\braket{\psi_\text{final}}{m}|^2$.  If $|\braket{\psi_\text{final}}{m}|^2 = 1$
then the search is perfect and the problem is solved.  If the search is 
imperfect then the problem can be solved by searching multiple times: since the 
result of each search is checked independently, a single successful search is 
sufficient. As long as $|\braket{\psi_\text{final}}{m}|^2$ is greater than 
$1$/\textsc{poly}$(n)$ this form of amplitude amplification will be efficient. 
Multiple runs have a cost: see Sec.~\ref{sec:multi_strats} for details of the
trade off between multiple runs and the initialization time for each run.

In general, problems with full permutation symmetry, such as the search problem,
are considered to be toy problems from a practical point of view.  A naive 
implementation of such a problem---in this case 
$\hat{H}_p = \openone - \ketbra{m}{m}$ the marked state Hamiltonian---requires 
exponentially many terms of the form $\prod_{j\in m} \hat{\sigma}_z^{(j)}$, 
where $m$ is a binary number with $n$ bits 
(with $j\in m$ indicating the $1$ digits of the number), and $j$ iterates over the bits in 
$m$ that are equal to one. However, it has recently been shown 
\cite{chancellor16a} that the spectrum of such terms in permutation-symmetric 
problems can be reproduced using $n$ extra qubits and a number of extra coupling
terms of the form $\hat{\sigma}_z^{(j)}\hat{\sigma}_z^{(k)}$ which scales as 
$n^2$.  It has also been suggested that such models may be fully realized perturbatively 
\cite{chancellor_AQC2016,chancellor17b}.  Although this approach to construct 
such terms is much closer to the realm of what can be experimentally realized, 
it would still be highly non-trivial to implement.  Nonetheless, the insights 
gained from studying the search problem can be adapted to realistic problems of 
practical interest.

%%%%%%%%%%%%%%%%%%%%%%%%%%%%%%%%%%%%%%%%%%%%%%%%%%%%%%%%%%%%%%%
\subsection{Quantum walk search algorithm}\label{sub:qwsearch}
%%%%%%%%%%%%%%%%%%%%%%%%%%%%%%%%%%%%%%%%%%%%%%%%%%%%%%%%%%%%%%%

A continuous-time quantum walk can be defined by considering the labels $j$ of 
the $n$-qubit basis states $\{\ket{j}\}$ to be the labels of vertices of an 
undirected graph $G$.  The edges of $G$ can be defined through its adjacency 
matrix $A$, whose elements satisfy $A_{jk}=1$ if an edge in $G$ connects 
vertices $j$ and $k$ and $A_{jk}=0$ otherwise.  Since $G$ is undirected, $A$ is 
symmetric, hence it can be used to define a Hamiltonian.  Although we can use 
the adjacency matrix $A$ directly, it is in general more convenient 
mathematically to define the Hamiltonian of the quantum walk using the Laplacian
$L = A-D$, where $D$ is a diagonal matrix with entries $D_{jj} = d_j$ the degree
of vertex $j$ in the graph.   We follow this convention here, but note that in 
this work we use regular graphs for which the degree $d_j = d$ is the same for 
all vertices, so that $D=d\openone$, where $\openone$ is the identity matrix 
(ones on the diagonal) of the same dimension as $A$.  Terms proportional to the 
identity in the Hamiltonian shift the zero point of the energy scale and 
contribute an unobservable global phase, but otherwise don't affect the 
dynamics.  The quantum walk Hamiltonian is then defined as 
$\hat{H}_{\text{QW}} = -\gamma\hat{L}$, where $\hat{L}$ is the Laplacian 
operator, and the prefactor $\gamma$ is the hopping rate of the quantum walk. 
For any regular graph of degree $d$ we thus have 
\begin{equation}\label{eq:Hqw}
	\hat{H}_{\text{QW}} = \gamma \left(d\hat{\openone} 
						- \sum_{jk} A_{jk} \ketbra{j}{k}\right)
                   \equiv \gamma(d\hat{\openone} - \hat{A}),
\end{equation}
where the adjacency operator $\hat{A}$ has matrix elements in the vertex basis 
$\{\ket{j}\}$ given by the adjacency matrix $A$.  The action of 
$\hat{H}_{\text{QW}}$ is to move amplitude between connected vertices, as 
specified by the non-zero entries in $A$.  During a quantum walk, a pure state 
$\ket{\psi(0)}$ evolves according to the Schr\"{o}dinger equation to give 
\begin{equation} \label{eq:QW_evolution}
	\ket{\psi(t)} = \exp(-i\hat{H}_{\text{QW}} t) \ket{\psi(0)}
\end{equation}
after a time $t$, where we have used units in which $\hbar=1$.  

Quantum walk dynamics can be used to solve the search problem by modifying the
energy of the marked state $\ket{m}$ to give a quantum walk search Hamiltonian
\begin{equation}
\hat{H}_{\text{QWS}} = \gamma(d\hat{\openone} - \hat{A}) -\ket{m}\bra{m}.
\end{equation}
In the units we are using, this amounts to giving state $\ket{m}$ an energy of 
$-1$ while all other states have zero energy.  This also makes $\gamma$ a 
dimensionless parameter controlling the ratio of the strengths of the two parts 
of the quantum walk search Hamiltonian.  Applying $\hat{H}_{\text{QWS}}$ to the 
search initial state $\ket{\psi_{\text{init}}}$ in Eq.~(\ref{eq:psi_init}) 
produces a periodic evolution such that the overlap with the marked state 
oscillates.  The frequency of these oscillations depends on the hopping rate 
$\gamma$, which must be chosen correctly, along with the measurement time $t_f$,
to maximize the final success probability 
$\mathsf{P}=|\braket{\psi(t_f)}{m}|^2$, where 
$\ket{\psi(t_f)} = \exp(-i\hat{H}_{\text{QWS}} t_f) \ket{\psi_{\text{init}}}$ 
is the state at time $t_f$.

The performance of quantum walk search algorithms will clearly have some 
dependence on the choice of the graph $G$. Provided the connectivity isn't too 
sparse or low-dimensional \cite{childs03a}, most choices of graph will work,
even random graphs \cite{Chakraborty16a}.  Two convenient choices on which the 
quantum walk is analytically solvable are the complete graph, for which all 
vertices are directly connected, and a graph whose edges form an $n$-dimensional
hypercube.  Moore and Russell \cite{moore01a} first studied quantum walks on
hypercubes, and Hein et al.~\cite{hein09a} perform a detailed analysis of discrete
time quantum walk searching on the hypercube, extending the work of Shenvi et
al.~\cite{shenvi02a}.  We choose to focus our work on a hypercube, rather than a 
fully-connected graph, because it is the more practical graph in terms of 
implementation on a quantum computer.  A hypercube graph is the natural choice 
for a quantum walk encoded into qubits because moving from one vertex to a 
neighbouring vertex corresponds to flipping a qubit.  The techniques and scaling
arguments we give in this work also apply in the case of a fully connected 
graph, and can be easily extended to a more general setting, for example to the 
`typical' random graphs considered in \cite{Chakraborty16a}. 

The adjacency matrix of an $n$-dimensional hypercube graph has elements 
$A_{jk} = 1$ if and only if the vertex labels $j$ and $k$ have a Hamming 
distance of one.  That is, when written as $n$-digit bitstrings, they differ in 
exactly one bit position.  The corresponding adjacency operator can be 
conveniently expressed as
\begin{equation}
\hat{A}^{(h)} = \sum_{j=1}^n  \hat{\sigma}_x^{(j)},
\end{equation}
where the sum is over all $n$ qubits and $\hat{\sigma}_x^{(j)}$ is the Pauli-$X$ 
operator applied to the $j$th qubit with the identity operator on the other 
qubits.  That is, 
\begin{equation} 
\hat{\sigma}_x^{(j)} = 
\left(\bigotimes_{r=1}^{j-1} \hat{\openone}_2\right) 
\otimes \hat{\sigma}_x \otimes
\left(\bigotimes_{r=j+1}^{n} \hat{\openone}_2\right),
\end{equation}
where $\otimes$ denotes the tensor product, and $\hat{\openone}_2$ is the 
identity operator of dimension two.  The Hamiltonian for the quantum walk on the hypercube is thus given by
\begin{equation} \label{eq:QWhyp}
\hat{H}^{(h)}_{\text{QW}} = \gamma\left( n\hat{\openone} 
						   - \sum_{j=1}^{n}\hat{\sigma}_x^{(j)} \right), 
\end{equation}
since an $n$-dimensional hypercube has vertices which have degree $n$.

To construct the quantum walk search Hamiltonian on the hypercube, we include 
two trivial adjustments for later mathematical convenience.  If we make the 
energy of the marked state lower by adding $\openone - \ketbra{m}{m}$ to the 
quantum walk Hamiltonian, this gives the marked state an energy of zero while 
all other states have an energy of one for this part of the Hamiltonian.  We 
also include a factor of a half in gamma, to match 
Refs.~\cite{childs03a,farhi00a,childs02a} and facilitate the mapping to the 
symmetric subspace (appendix \ref{app:sched_analytics}).  Our quantum walk 
search on the hypercube is then
\begin{equation} \label{eq:QWsearch}
\hat{H}^{(h)}_{\text{QWS}} = \frac{\gamma}{2}\left( n\hat{\openone} 
						   - \sum_{j=1}^{n}\hat{\sigma}_x^{(j)} \right)
						   + (\openone - \ketbra{m}{m}).
\end{equation}

Childs and Goldstone \cite{childs03a} analyze the quantum walk search algorithm 
for both the complete and hypercube graphs.  For each graph, they find optimal 
values of $\gamma$ for which the performance of the search matches the quadratic 
quantum speed up achieved by Grover's search algorithm.  The mechanism for 
finding the marked state can be understood intuitively as follows.  Note that 
the initial state $\ket{\psi_{\text{init}}}$ from Eq.~(\ref{eq:psi_init}) is the 
(non-degenerate) ground state of both the complete graph and the hypercube 
Hamiltonians, i.e., $\hat{H}^{(h)}_{\text{QW}}$ of Eq.~(\ref{eq:QWhyp}).
The marked state $\ket{m}$ is, by design, the ground 
state of the marked state component of the search Hamiltonian.  For large values
of $\gamma$, the marked state term is relatively small so the graph Hamiltonian 
dominates, and the ground state of the full search Hamiltonian 
$\hat{H}^{(h)}_{\text{QWS}}$ of Eq.~(\ref{eq:QWsearch}) is approximately 
$\ket{\psi_{\text{init}}}$.  Conversely, for small values of $\gamma$, the 
ground state of $\hat{H}^{(h)}_{\text{QWS}}$ is approximately $\ket{m}$.   Over 
a narrow range of intermediate values of $\gamma$, the ground state switches 
between the two.  By calculating the low level part of the energy spectrum of 
$\hat{H}^{(h)}_{\text{QWS}}$, Childs and Goldstone tune $\gamma$ until both the 
initial state $\ket{\psi_{\text{init}}}$ and the marked state $\ket{m}$ have 
significant overlap with both the ground state $E_0$ and the first excited state
$E_1$ of the search Hamiltonian.  Intuitively, we want the search Hamiltonian to
drive transitions between $\ket{\psi_{\text{init}}}$ and $\ket{m}$ as 
efficiently as possible.  This occurs when the overlaps are evenly balanced, 
which in turn occurs when the gap $g=E_1-E_0$ between the ground and first 
excited state is smallest: $g_{\text{min}}$.  With this optimally chosen value 
of $\gamma$, the time it takes for the transition to occur turns out to be 
proportional to $1/g_{\text{min}}$.  For the hypercube graph, the optimal value 
of $\gamma$ is
\begin{equation} \label{eq:gamma_hopt}
\gamma_{o}^{(h)} = 
    \frac{1}{N}\sum_{r=1}^n\binom{n}{r}\frac{1}{r} \equiv R_1,
\end{equation}
where $\binom{n}{r}$ is the binomial coefficient $n$ choose $r$.
This sum appears many times in the following calculations, so it is convenient 
to abbreviate it by $R_1$.  Note also that it is not always sufficiently 
accurate to use the approximation $R_1\simeq 2/n$ given in \cite{childs03a}.  
The time to reach the first maximum overlap with the marked state is 
$t_{o}^{(h)}\simeq(\pi/2)\sqrt{N}$, providing a quadratic speed up 
equivalent to Grover's original search algorithm.

%%%%%%%%%%%%%%%%%%%%%%%%%%%%%%%%%%%%%%%%%%%%%%%%%%%%%%
\subsection{Adiabatic quantum search algorithm \label{sub:adiabat_sched}}
%%%%%%%%%%%%%%%%%%%%%%%%%%%%%%%%%%%%%%%%%%%%%%%%%%%%%%

Adiabatic quantum computing (AQC), first introduced by Farhi et 
al.~\cite{farhi00a}, works as follows. The problem of interest is encoded into 
an $n$-qubit Hamiltonian $\hat{H}_{p}$ in such a way that the solution can be 
derived from the ground state of $\hat{H}_{p}$. The system is initialized in 
the ground state of a different Hamiltonian $\hat{H}_0$, for which this 
initialization is easy. The computation then proceeds by implementing a 
time-dependent Hamiltonian that is transformed slowly from $\hat{H}_0$ to 
$\hat{H}_{p}$.  In general this adiabatic `sweep' Hamiltonian can be 
parameterized in terms of a time-dependent schedule function $s\in[0,1]$ as
\begin{equation}
 \hat{H}_{\text{AQC}}(s)=(1-s)\hat{H}_0+s\,\hat{H}_{p}, 
\label{eq:H_AQC}
\end{equation}
with $s\equiv s(t)$ such that $s(t=0)=0$ and at the final time $t_f$ we have 
$s(t=t_f)=1$.  It is useful to define a reduced time $\tau=t/t_f$, with 
$0\le\tau\le 1$.  Whereas $\tau$ is linear in $t$, the schedule function 
$s(\tau)$ -- written as a function of $t$ or $\tau$ -- allows for nonlinear 
transformation.  Nonlinear schedules are essential to obtain a quantum speed up.

The adiabatic theorem of quantum mechanics \cite{born28} says that the system 
will stay in the instantaneous ground state of the time-dependent Hamiltonian 
$\hat{H}_\text{AQC}(s)$ provided the following two conditions are satisfied: 
(i) there is at all times an energy gap $g(s)>0$ between the instantaneous 
ground and first excited states, and 
(ii) the Hamiltonian is changed sufficiently slowly. 
Provided these are both true the system will be in the desired ground state of 
$\hat{H}_{p}$ at the end of the computation, thus solving the problem encoded 
in $\hat{H}_{p}$.  In practice, the duration of this adiabatic sweep would be 
prohibitively long, so a realistic sweep will incur some probability of error. 
We discuss this and other subtleties of the adiabatic theorem in 
Sec.~\ref{sub:optim_sched}, after we introduce the adiabatic quantum search 
algorithm.  For a comprehensive overview of AQC, see Albash and Lidar 
\cite{albash16a}.

Roland and Cerf \cite{roland02a} describe how adiabatic quantum computing can be
used to solve the search problem with a quadratic quantum speed up.  Define the 
problem Hamiltonian as
\begin{equation}
\hat{H}_\textsc{p} = \hat{\openone} - \ket{m}\bra{m}, 
\label{eq:H_prob}
\end{equation}
whose non-degenerate ground state is equal to the marked state $\ket{m}$ with 
eigenvalue zero.  We then need to choose our easy Hamiltonian $\hat{H}_0$ such 
that it has $\ket{\psi_\text{init}}$, as defined in Eq.~(\ref{eq:psi_init}), as 
its non-degenerate ground state.  There are many possible choices, Roland and 
Cerf use $\hat{H}_0=\hat{\openone}-\ketbra{\psi_\text{init}}{\psi_\text{init}}$.
With the system initialized in $\ket{\psi_\text{init}}$, the algorithm proceeds 
by implementing the time-dependent Hamiltonian in Eq.~(\ref{eq:H_AQC}), with a 
suitable schedule function $s(\tau)$, so that after a time $t_f$ the final state
of the system is close to the marked state $\ket{m}$.  Roland and Cerf 
demonstrate that a linear schedule function $s^{(l)}(\tau) = \tau = t/t_f$ does 
not produce a quantum speed up.  It is necessary to use a more efficient 
nonlinear $s(\tau)$, whose rate of change is in proportion to the size of the 
gap $g(s)$ at that point in the schedule, in order to produce the quadratic 
speed up of Grover's search algorithm.

It is easy to show that 
$\hat{H}_0 = \hat{\openone}-\ketbra{\psi_\text{init}}{\psi_\text{init}}$ is 
proportional to the adjacency operator of the fully-connected graph with $N=2^n$
vertices.  For the reasons already given in the context of the quantum walk 
search algorithm, a Hamiltonian corresponding to a less connected graph is 
preferable for practical applications.  In order to make direct comparisons 
between adiabatic and quantum walk searching, we use the hypercube graph, since   
this also has $\ket{\psi_\text{init}}$ as its non-degenerate ground state, with 
Hamiltonian (in its Laplacian form) given by 
\begin{equation}\label{eq:Hh}
\hat{H}^{(h)}_0 = \frac{1}{2}\left( n\hat{\openone} 
                - \sum_{j=1}^{n}\hat{\sigma}_x^{(j)} \right)
\end{equation}
where we have again included a factor of a half for mathematical convenience.  
As further motivation for this choice, we note that this corresponds to a 
transverse-field driver Hamiltonian applied to qubits, which is the most common
choice for
quantum annealing hardware and which can be experimentally realized on a large 
scale \cite{LantingAQC2017}.  Combining Eqs.~(\ref{eq:H_prob}) and 
(\ref{eq:Hh}), we have the adiabatic quantum computing Hamiltonian for search on
a hypercube,
\begin{equation}\label{eq:HhAQC}
\hat{H}^{(h)}_{\text{AQC}} = (1 - s) \frac{1}{2}\left(n\hat{\openone} 
    - \sum_{j=1}^{n}\hat{\sigma}_x^{(j)} \right) 
    + s\left( \openone - \ket{m}\bra{m} \right).
\end{equation}
We note that $\hat{H}^{(h)}_{\text{AQC}}$ contains the same terms as 
$\hat{H}^{(h)}_{\text{QWS}}$ in Eq.~(\ref{eq:QWsearch}), only in different, 
time-varying proportions.  It remains to specify the function $s(\tau)$ for the 
optimal performance of this Hamiltonian for searching.  There are several 
subtleties to deriving an optimal $s(\tau)$ for the hypercube, which we address 
in the next section.

%%%%%%%%%%%%%%%%%%%%%%%%%%%%%%%%%%%%%%%%%%%%%%%%%%%%%%
\section{Optimising AQC schedules \label{sub:optim_sched}}
%%%%%%%%%%%%%%%%%%%%%%%%%%%%%%%%%%%%%%%%%%%%%%%%%%%%%%

We have seen that QW and AQC searching may be achieved with Hamiltonians
that have the same terms but different, time-varying, coefficients. Next we 
would like to interpolate these coefficients to generate hybrid search algorithms.
However, we must first determine an optimal schedule $s(\tau)$ for the AQC search.
In fact this is not entirely straightforward: it is possible to find more than one
optimal schedule. In this section we derive two different
schedules via an analytical method and a numerical method, and demonstrate that
these both give optimal quantum scaling advantage for
the unstructured search problem.

%---------------------------------------------%
\subsection{Adiabatic condition and method\label{sub:opt_method}}
%---------------------------------------------%

We now return to the nuances of the adiabatic theorem
and how, in the regime of limited running time, the schedule $s(\tau)$
may be optimized to minimize the error.
A more quantitative statement of the adiabatic theorem \cite{roland02a,farhi00a,lidar09a,rezakhani09a,albash16a}
proceeds as follows: 
Consider a time-dependent Hamiltonian of the form in Eq.~(\ref{eq:H_AQC}), with 
initial and final Hamiltonians $\hat{H}_0$, $\hat{H}_\text{p}$ respectively, 
and parameterized by the schedule function $s(\tau)$ that sweeps from 
$s(0) = 0$ to $s(1) = 1$ over a time $t_f$, the runtime of the sweep. 
Denote by $\ket{E_j(t)}$ the $j$th energy eigenstate of the Hamiltonian at 
time $t$ and its energy by $E_j(t)$, where $j=0,1$ denotes the ground and first 
excited states respectively. Provided that $E_1(t) > E_0(t)$ for $t\in[0,t_f]$ 
and transitions to higher energy eigenstates can be ignored, the final state 
obeys
\begin{equation} \label{eq:aqc_epsilon}
|\braket{\psi(t_f)}{E_0(t_f)}|^2 ~\geq~ 1-\epsilon^2,
\end{equation}
for small parameter $\epsilon \ll 1$, provided that at all times
\begin{equation} \label{eq:adiabat_cond}
\frac{\Big|\left\langle\frac{d\hat{H}}{dt}\right\rangle_{0,1}\Big|}{g^2(t)} 
                                                       \leq\epsilon \ll 1,
\end{equation}
where the matrix element  $\langle d\hat{H}/dt\rangle_{0,1}$ is given by
\begin{equation}
\left\langle \frac{d\hat{H}}{dt}\right\rangle_{0,1} =
\left\langle E_0(t)\Big|\right.\frac{d\hat{H}}{dt}\left.\Big|E_1(t)\right\rangle 
\end{equation}
and the gap $g(t)$ is given by
\begin{equation}
g(t)  = E_1(t) - E_0(t).
\end{equation}

However, adiabatic protocols derived from Eq.~(\ref{eq:adiabat_cond}) are not
always optimal. This equation accounts for probability amplitude leaking from
the ground state into a nearly empty first excited state. Thus it will break down 
in situations where transitions to higher excited states are important, or where
the population of the first excited state is significant.  We can therefore describe Eq.~(\ref{eq:adiabat_cond}) as a 
two-level approximation.  In the context of the search algorithms studied here, 
such an approximation turns out to be good for all but the smallest values of $n$, 
and becomes more accurate for larger search spaces.  We make extensive use of 
this in what follows, especially in Sec.~\ref{sub:SAC_model}.

Equation (\ref{eq:adiabat_cond}) also does not take into account the return of 
probability amplitude which has already entered the excited state.  Such effects
can become the most relevant to the dynamics under two circumstances.  If the 
first excited state is populated significantly, then non-adiabatic dynamics can 
occur such that this amplitude returns and interferes with the ground state 
amplitude.  This is the regime which we primarily study in this work.  Quantum 
walk dynamics are an extreme example of such behaviour as they can be viewed as 
time independent coherent evolution bracketed by instantaneous quenches, which 
are the ultimate non-adiabatic transitions.  The second and more subtle case is 
deep in the adiabatic regime, where the Hamiltonian sweep rate is so slow that 
the rate of excitation formation is very low during the middle of the anneal.  
In these cases, boundary effects become important, which depend in a complicated
way on both the nature of the annealing schedule and the total runtime 
\cite{rezakhani10a,weibe12a,kieferova14a}.  While this regime is very 
interesting, it is outside of the scope of this study, and not relevant 
for practical implementation of algorithms.  For this reason, we limit our 
numerical studies to a maximum runtime of $\sim 5\pi/g_{\min}$, about ten times
the typical runtime derived from the minimum gap.  With runtimes 
$t_f \lesssim 5 \pi/g_{\min}$, we do not observe any appreciable boundary 
effects in our numerical results.

Roland and Cerf \cite{roland02a} derive a schedule $s(\tau)$ for the fully 
connected graph by optimizing Eq.~(\ref{eq:adiabat_cond}), by matching the 
instantaneous rate of change of the schedule function $s(t)$ to the size of the 
gap at that time.  Using
\begin{equation}
\left\langle \frac{d\hat{H}}{dt}\right\rangle_{0,1} = \frac{ds}{dt} \left\langle \frac{d\hat{H}}{ds}\right\rangle_{0,1}
\end{equation}
in the adiabatic condition of Eq.~(\ref{eq:adiabat_cond}) gives
\begin{equation}\label{eq:ac_inst}
\left\lvert\frac{ds}{dt}\right\rvert \le \epsilon 
	\frac{g^2(t)}{\left\lvert \left\langle \frac{d\hat{H}}{ds}\right\rangle_{0,1} \right\rvert}.
\end{equation}
The instantaneous gap $g(t)$ and 
$\langle d\hat{H}/ds\rangle_{0,1}$ can be calculated from the eigensystem of the 
Hamiltonian, which is analytically tractable for the complete graph.  The 
schedule they obtain this way produces the full quadratic quantum speed up for 
the adiabatic quantum search algorithm on the fully connected graph.  

%-----------------------------------------------------------------%
\subsection{Hypercube schedule calculation\label{sub:calc_opt_hyp}}
%-----------------------------------------------------------------%

Since we are using the hypercube graph, we must do the  equivalent calculation 
for the hypercube AQC search Hamiltonian given by Eq.~(\ref{eq:HhAQC}). 
The eigensystem of this Hamiltonian has been solved in 
Refs.~\cite{farhi00a,childs02a,childs03a} by mapping it to the 
symmetric subspace. From here the position and size of the minimum gap can be found
exactly, and the eigenvalue equations expanded about this point. This is combined
with the saturation of Eq.~(\ref{eq:ac_inst}) at the minimum gap point, where the RHS
takes its minimum value. From the resulting expressions it is possible to derive an 
analytical expression for the schedule $s(t)$.
The full calculation is somewhat lengthy 
and is outlined in appendix \ref{app:sched_analytics}.  We find the calculated 
optimal schedule
\begin{equation}\label{eq:analytic_AQC_schedule_1}
s^{(c)}(t) = \frac{2\sqrt{R_2}}{\sqrt{N}(1+R_1)^2}
\tan\left\{\frac{8\epsilon\sqrt{R_2}R_1^2 t}{n\sqrt{N}R_2^2} - c\right\}
+\frac{1}{1+R_1},
\end{equation}
where terms $O(1/N)$ and smaller have been dropped, 
\begin{equation}\label{eq:analytic_AQC_schedule_2}
c = \arctan\left\{\frac{(1+R_1)\sqrt{N}}{2\sqrt{R_2}}\right\},
\end{equation}
the constant $R_1$ is defined in Eq.~(\ref{eq:gamma_hopt}) and $R_2$ by
\begin{equation}\label{eq:R2}
R_2\equiv\frac{1}{N}\sum_{r=1}^n\bin{n}{r}\frac{1}{r^2}.
\end{equation} 
This analytical schedule is guaranteed to satisfy Eq.~(\ref{eq:ac_inst}) only in the region of
the minimum gap, however it is here that transitions to unwanted higher energy levels 
are most rapid, so the net effect is that this schedule still manages to produce optimal
quantum speedup. For $N\gg 1$, the runtime is given by 
\begin{equation}
	\epsilon\, t^{(c)}_f \simeq \frac{\pi\sqrt{N}}{4},
\label{eq:tfTLAoptc}
\end{equation}
where the approximation of the arctans by $\pi/2$ becomes exact as 
$N\rightarrow\infty$. Note that choosing a value for $\epsilon$ -- the accuracy 
with which the system stays in the ground state, see Eq.~(\ref{eq:aqc_epsilon}) --
determines the corresponding runtime $t_f$, and vice versa.  For our numerical 
calculations we have chosen to specify $t_f$, since this enables direct 
comparisons with QW searching to be made.
\begin{figure}
 \begin{centering}
  \includegraphics[width=\columnwidth]{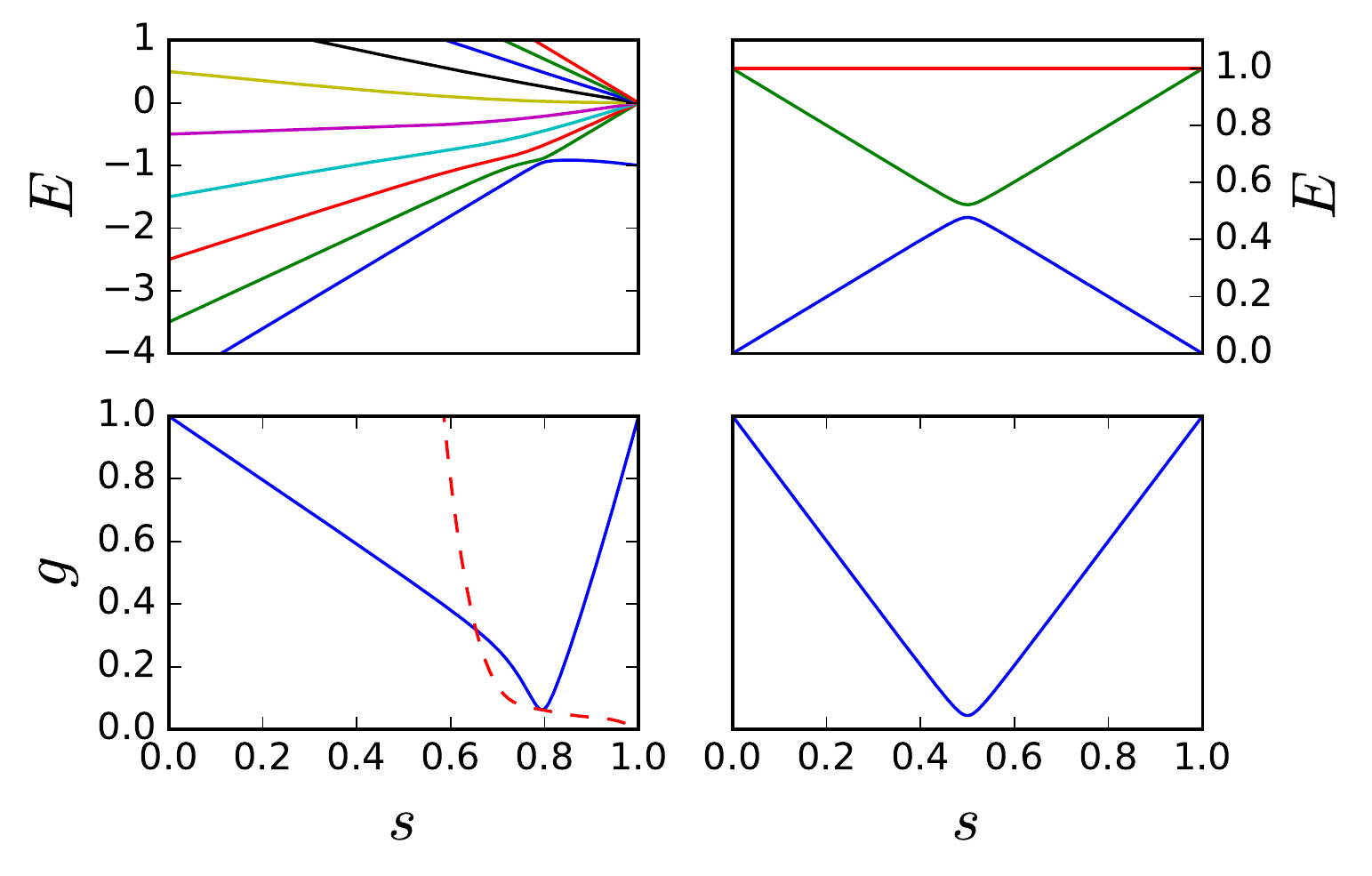}
 \end{centering}
 \caption{%
 (color online) 
 (Top) Energy levels and (bottom) gap for a hypercube of size $n=9$ (left) and 
 complete graph (right). Both the true gap (blue, solid)
 and the approximate, analytical, gap (red, dashed) are shown for the hypercube 
 (bottom left). The analytical gap is only accurate near the true minimum gap,
 however it is here that transitions to higher energy levels are most rapid, 
 so the resulting analytical schedule still yields optimal quantum speedup.
 Energy units defined by Eq.~(\ref{eq:HhAQC}).
 \label{fig:HhAQC_energy}
}
\end{figure}
The energy levels of $\hat{H}^{(h)}_{\text{AQC}}$ are shown in 
Fig.~\ref{fig:HhAQC_energy} (top left) for $n=9$, and for comparison the energy 
levels of the search Hamiltonian for the complete graph (which is the same for 
any size) are shown top right.

We also solve Eq.~(\ref{eq:adiabat_cond}) numerically to obtain $s^{(n)}$ using
an explicit numerical calculation of the gap $g(t)$, and using
the maximum value of $\langle d\hat{H}^{(h)}_{\text{AQC}}/dt\rangle_{0,1}$,
which is shown in appendix \ref{app:sched_analytics} to be $n/4$. 
Our numerical algorithm is described in appendix~\ref{app:num_meth}.  While 
it does not provide a closed form solution, results using $s^{(n)}$ do provide 
insight on the accuracy of $s^{(c)}$.  Provided the numerics are performed to a 
sufficient accuracy, $s^{(n)}$ will always provide an optimal $\sqrt{N}$ speed 
up.  The analytically and numerically calculated gaps are plotted in 
Fig.~\ref{fig:HhAQC_energy} (bottom left) for $n=9$, and the corresponding gap 
for the complete graph is shown bottom right.  For the hypercube, the analytical
and numerical gaps are strikingly different, yet both produce schedules that 
obtain a quantum speed up.  As we will see, this is because for the quantum 
search problem only the position and size of the gap are important.  Elsewhere, 
the transition probabilities are so small it does not matter how fast the
schedule proceeds.

However, note that both of these schedules assume a two-level approximation, as 
they start from Eq.~(\ref{eq:adiabat_cond}).  While in general for large $N$ 
this is a good approximation, for small system sizes the higher energy levels do affect the performance, as we show in the
next subsection.

%-------------------------------------------------------------------%
\subsection{Performance of hypercube schedules\label{sub:hyp_scheds}}

\begin{figure}
 \begin{centering}
  \includegraphics[width=0.9\columnwidth]{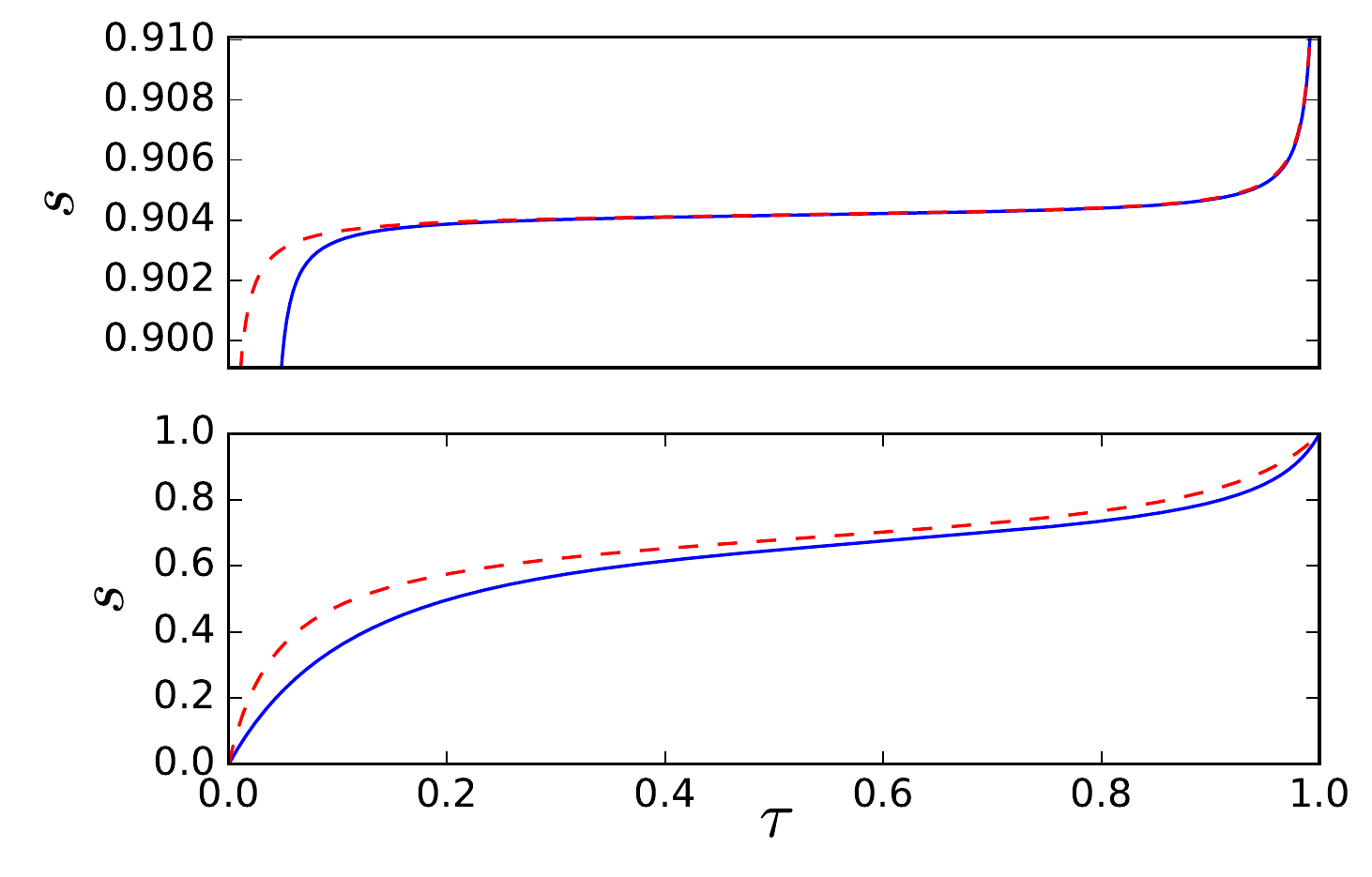}
 \caption{(color online)
 Comparison of annealing schedules from Sec.~\ref{sub:optim_sched} for AQC 
 searching over a hypercube of $n=20$ qubits (top) and $n=5$ qubits (bottom)
 for two-level approximation analytically calculated $s^{(c)}$ (dashed) and 
 numerically calculated $s^{(n)}$ (solid).  Note different scale for $s$ in top 
 figure.
 \label{fig:sched_compare}
 }
 \end{centering}
\end{figure}

Having calculated optimal schedules both analytically and numerically, we now compare their performance for system sizes
up to $n=20$ qubits.
Note that the size of the minimum gap $g_{(\text{min})}$ calculated 
from the two-level approximation in Sec.~\ref{sub:optim_sched} is exactly the 
same for both.  Since both are based only on the 
interactions of the two lowest energy levels, both will find the correct shape 
for the annealing protocol in this region.  Numerical results support this prediction in that 
for $n=20$ the numerically calculated 
optimal schedule $s^{(n)}$ slows down at the same value as $s^{(c)}$ in 
Fig.~\ref{fig:sched_compare}.  For $n=5$ qubits the schedules are distinct over the whole range of $\tau$, 
while for $n=20$, the schedules are almost identical, the only visible difference occurs at $\tau \lesssim 0.1$.
The difference between them around $\tau\lesssim 0.1$ is likely due to interactions with the higher 
excited states of the hypercube Hamiltonian early in the schedule.  In the large
system limit this difference will have little effect on the overall success 
probability, as the overlap with the initial ground state and the manifold of 
states participating in the avoided crossing approaches one exponentially fast 
in the number of qubits $n$ (see table \ref{tab:num_fits}). 

\begin{figure}
 \begin{centering}
  \includegraphics[width=0.9\columnwidth]{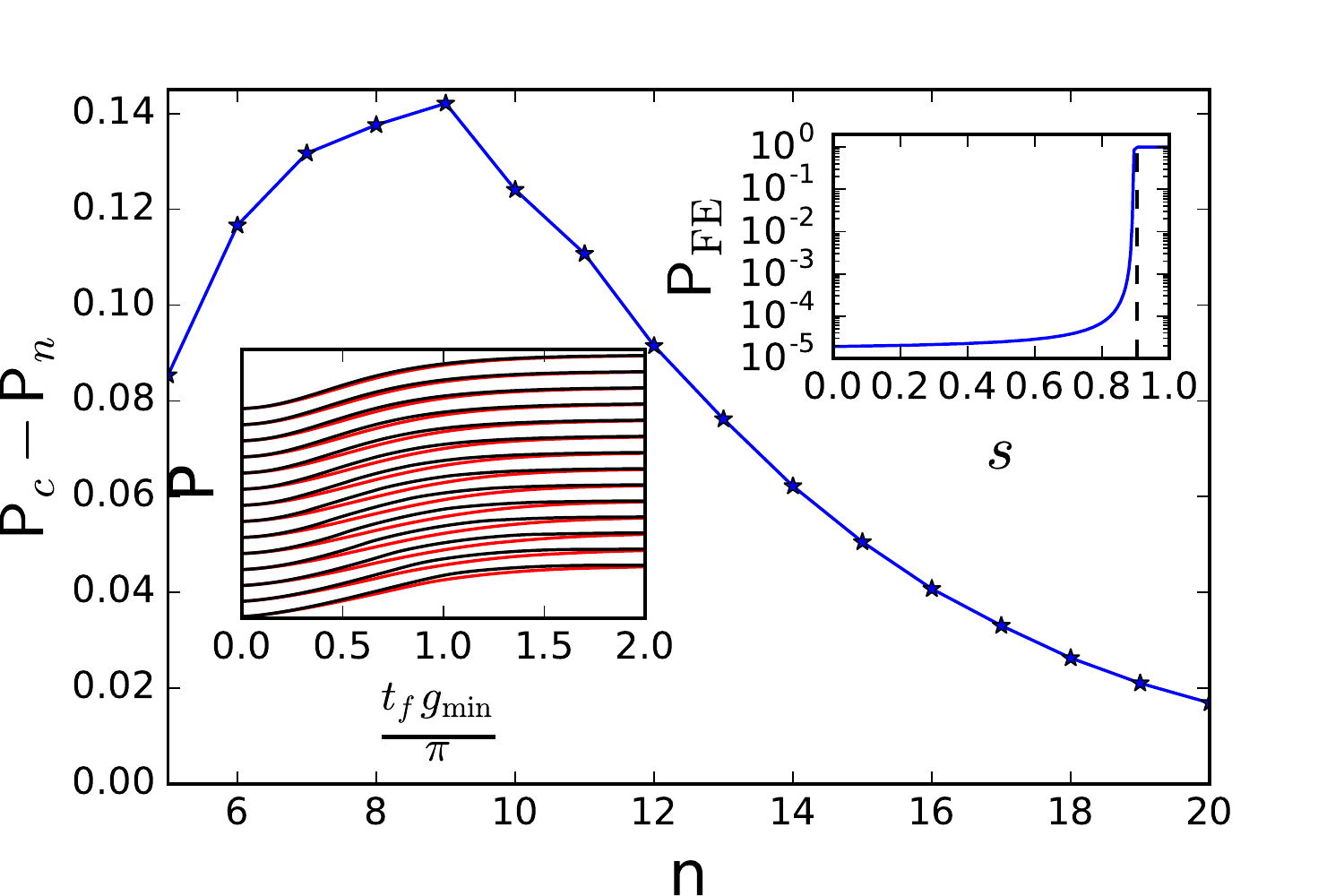}
 \end{centering}
 \caption{(color online) Main figure: Difference in success probabilities 
 $\mathsf{P}_c-\mathsf{P}_n$ between the annealing schedule calculated
 analytically $s^{(c)}$ and the numerically calculated 
 schedule $s^{(n)}$ for a single run over a time $\pi/g_{\text{min}}$.
 Lower Inset: Offset plot of the success probabilities $\mathsf{P}$ versus  
 $t_fg_{\text{min}}/\pi$ for $s^{(c)}$ (red, gray in print) and $s^{(n)}$ (black)
 for $n=5$ to $n=20$. 
 Upper Inset: Sum of the overlaps of $\ket{E_1}$ with $\ket{m}$ and 
 $\ket{\psi_{\text{init}}}$ ($\mathsf{P}_{\mathrm{FE}}=|\braket{E_1}{m}|^2+|\braket{E_1}{\psi_{\text{init}}}|^2$)
 against $s$ for $20$ qubits. Since the relevant avoided crossing is in the space spanned by $
 \ket{m}$ and $\ket{\psi_{\text{init}}}$ a vanishing value $\mathsf{P}_{\mathrm{FE}}$ is indicative of very little rotation of the ground state.  Vertical dashed line is position of $g_{\text{min}}$.
 \label{fig:fig-2_lvl_num}
 }
\end{figure}
For $n<20$, the difference between $s^{(c)}$ and $s^{(n)}$ at early times in 
the run does affect their relative performance, as Fig.~\ref{fig:fig-2_lvl_num}
shows.  Although the numerical schedule $s^{(n)}$ is a more accurate solution 
of the optimization in Eqn.~(\ref{eq:adiabat_cond}), $s^{(c)}$ does better than
$s^{(n)}$.  The reason is that, while the gap is relatively small early in the 
schedule, so is the matrix element between the ground and first excited state of 
the marked state Hamiltonian, as shown in the top inset of 
Fig.~\ref{fig:fig-2_lvl_num}.  As a result, the numerically calculated schedule 
slows down unnecessarily in this region, as can be seen in 
Fig.~\ref{fig:sched_compare}.  The approximate expansion for the gap used 
to derive the schedule $s^{(c)}$ in appendix A grows within this region, see 
Fig.~\ref{fig:HhAQC_energy}.  Hence,  $s^{(c)}$ traverses this part of the schedule 
much faster than $s^{(n)}$.  Effectively, the approximate nature of the expansion 
for the gap used to calculate $s^{(c)}$ partially cancels an unnecessary slowdown 
caused by the approximation that $\langle\frac{d H}{d s}\rangle_{0,1}$ is constant 
for all $s$.  However, as the main figure and lower inset of 
Fig.~\ref{fig:fig-2_lvl_num} show, the difference between the success probabilities 
using the two schedules shrinks as system size increases and the avoided crossing 
becomes more dominant.

%%%%%%%%%%%%%%%%%%%%%%%%%%%%%%%%%%%%%%%%%%%%%%%%%%%%%%%%%%%%%%%%%%%%%%%%%%%%%%%
\section{Hybrid annealing schedules\label{sec:generalized_scheds}}
%-----------------------------------------------------------------------------%

Having arrived at a common Hamiltonian form for QW and AQC searching on an
$n$-dimensional hypercube,
and having derived optimal coefficients for each case, we can now interpolate
the coefficients to generate hybrid search algorithms. In this section we show
how this may be done, and study the resulting dynamics. We begin by looking at
small systems with $n=5$ and $n=8$, and then study the dynamics of systems with
very large $n$ by demonstrating that this limit corresponds to a two-state
single avoided crossing model.

\subsection{Motivation and definition}

We have already noted that QW and AQC search algorithms both use the same terms 
in the Hamiltonian, differing only in the time dependence.  With appropriate 
choice of parameters, both provide a quadratic quantum speed up: a search time 
proportional to $\sqrt{N}$ for a search space of size $N$.  This suggests the
question of whether we can map smoothly between QW and AQC searching, while
maintaining the quantum speed up.

To construct the mapping, we generalize the AQC Hamiltonian of 
Eq.~(\ref{eq:H_AQC}) by defining a time-dependent Hamiltonian
\begin{equation} \label{eq:H_adiabat}
	\hat{H}(\tau) = A(\tau)\hat{H}_0 + B(\tau)\hat{H}_p
\end{equation} 
as a function of the reduced time $\tau = t/t_f$, where the annealing schedules 
$A(\tau)$, $B(\tau)$ satisfy $A(0) \gg B(0)$ and $B(1) \gg A(1)$.  The AQC 
algorithm as described by Eq.~(\ref{eq:H_AQC}) is obtained by setting
\begin{align}
A_\text{AQC}(\tau) &= 1 - s(\tau)\nonumber\\
B_\text{AQC}(\tau) &= s(\tau).  
\end{align}
The QW search Hamiltonian with hopping rate $\gamma$, described by 
Eq~(\ref{eq:QWsearch}), can also be obtained by setting
\begin{align}
A^{(\gamma)}_\text{QW}(\tau) &= 
	\begin{cases}\gamma & \tau<1\\ 0 & \tau=1\end{cases} \nonumber\\
B^{(\gamma)}_\text{QW}(\tau) &= \begin{cases}1 & \tau>0\\ 0 & \tau=0.\end{cases}
\label{eq:QW_AB}
\end{align}
We can make this even closer to the AQC form by defining $\beta = 1/(1+\gamma)$ 
and setting
\begin{align}
A_\text{QW}(\tau) &= 
	\begin{cases}1-\beta & \tau<1\\ 0 & \tau=1\end{cases} \nonumber\\
B_\text{QW}(\tau) &= \begin{cases}\beta & \tau>0\\ 0 & \tau=0.\end{cases}
\label{eq:QW_AB_beta}
\end{align}
For QW search on the hypercube, using Eq.~(\ref{eq:gamma_hopt}) for 
$\gamma^{(h)}_{o}$, to achieve optimal $\sqrt{N}$ scaling we must set 
$\beta$ equal to
\begin{equation}
\beta^{(h)}_{o}=\frac{1}{1 + R_1}.
\end{equation}

For $0<\tau<1$, the re-parameterization of Eq.~(\ref{eq:QW_AB}) in 
Eq.~(\ref{eq:QW_AB_beta}) maintains the ratio of 
$A_\text{QW}(\tau)/B_\text{QW}(\tau)=\gamma$.  However, it also introduces a 
global energy shift 
$A_\text{QW}(\tau) = \beta A^{(\gamma)}_\text{QW}(\tau)$ and
$B_\text{QW}(\tau) = \beta B^{(\gamma)}_\text{QW}(\tau)$. 
The observant reader will note that, because the optimal 
$\gamma^{(h)}_{o}$ is dependent on the size of the system, this 
re-parameterization introduces a weak dependence of the global energy scale on 
system size $N=2^n$.  However, since 
$\beta^{(h)}_{o}  \rightarrow 1$ in the large $N$ limit, this 
weak dependence cannot affect the leading order term in the asymptotic scaling, 
and the re-parameterized quantum walk search algorithm still provides optimal 
$\sqrt{N}$ scaling. 

\begin{figure}
 \begin{centering}
 \includegraphics[width=0.9\columnwidth]{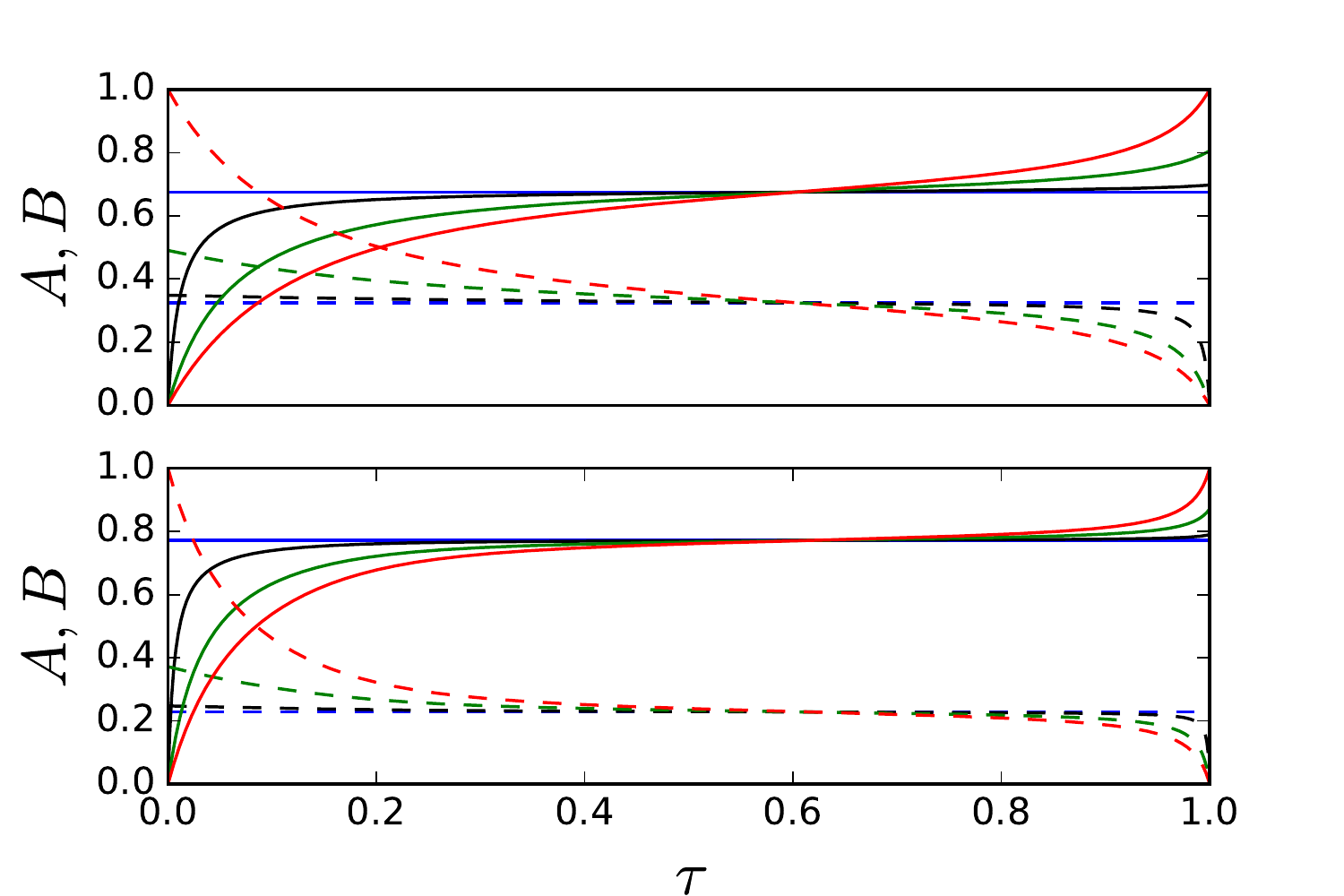}
 \end{centering}
\caption{%
 (color online) Interpolated schedule functions $A(\alpha,\beta,\tau)$ 
 (dashed lines) and $B(\alpha,\beta,\tau)$ (solid lines) as defined by 
 Eq.~(\ref{eq:interp_AB}) for hybrid QW-AQC quantum searching on an $n=5$ (top)
 and $n=8$ (bottom) dimensional hypercube graph.  
 $\alpha=0$ (QW), blue (dark gray in print); 
 $\alpha=0.1$, black; $\alpha=0.5$, green (mid gray in print); and  
 $\alpha=1$ (AQC), red (light gray in print), calculated numerically following 
 the method in Appendix~\ref{app:num_meth}.  
 \label{fig:sched_interp} 
 }
\end{figure}

In the way we have parameterized them above, the AQC and QW protocols differ 
only in the annealing schedules $A(\tau)$ and $B(\tau)$.  Hence, we can use the
QW and AQC schedules as end-points of a smooth interpolation between these two 
search algorithms to define a continuum of hybrid protocols.  Using a parameter 
$\alpha\in[0,1]$, where $\alpha=0$ corresponds to QW and $\alpha=1$ corresponds 
to AQC, we can define
\begin{align} \label{eq:interp_AB}
A(\alpha,\beta,\tau) 
  &= \frac{1-s(\tau)}{\alpha+(1-\alpha)\frac{(1-s(\tau))}{(1-\beta)}}\nonumber\\
B(\alpha,\beta,\tau)
  &= \frac{s(\tau)}{\alpha+(1-\alpha)\frac{s(\tau)}{\beta}}.
\end{align}
giving a family of hybrid quantum algorithms defined by the Hamiltonian
\begin{equation} \label{eq:Hh_AB}
\hat{H}_{\text{AB}} = A(\alpha,\beta,\tau)\hat{H}_0
				+ B(\alpha,\beta,\tau)\hat{H}_p.
\end{equation}
This interpolation is quite general, for well-behaved $\hat{H}_0$ and 
$\hat{H}_p$, with the caveat about the extra dependence of the energy scale on 
the QW hopping rate through $\beta$ mentioned above.  The resulting family of 
functions is illustrated in Fig.~\ref{fig:sched_interp} for search over $5$- 
and $8$-qubit hypercube graphs.

Note that, although it is plausible, it doesn't follow \emph{a priori} from the 
construction that these interpolated AQC-QW schedules will yield a quantum speed
up at all for searching, let alone an optimal $\sqrt{N}$ scaling.  This is 
because the different mechanisms in QW and AQC could be incompatible in 
combination.  We return to this important question in Sec.~\ref{sec:scaling}, 
where we show that properly specified interpolations can indeed achieve the 
theoretical optimum $\sqrt{N}$ scaling.

%%%%%%%%%%%%%%%%%%%%%%%%%%%%%%%%%%%%%%%%%%%%%%%%%%%%%%%%%%%%%%%%%%%%%%%%%%%%%%%
\subsection{Small size examples} \label{sub:small_sys_ex}
%%%%%%%%%%%%%%%%%%%%%%%%%%%%%%%%%%%%%%%%%%%%%%%%%%%%%%%%%%%%%%%%%%%%%%%%%%%%%%%

To gain intuition for how our interpolated schedules behave, we study small 
systems of five and eight qubits.  These have been simulated using the full 
Hamiltonian on the hypercube; for numerical methods, see 
appendix~\ref{app:num_meth}.  Fig.~\ref{fig:5_comp_evo} shows how the final 
success probability varies with the search duration $t_f$ for QW, AQC and an 
intermediate $\alpha=0.5$ search over the $5$-qubit hypercube graph.  Note that,
because the schedules $A$ and $B$ are in general nonlinear functions of time, in
all plots against $t_f$ each point represents a separate run of the quantum 
search algorithm for that value of $t_f$; the plots do not also represent the 
time evolution $0\le t\le t_f$, except for $\alpha=0$ when the schedule 
functions are constant ($A=1-\beta$ and $B=\beta$).  Plots of the time evolution
for a single search can be seen in Ref.~\cite{morley17a} and in Sec. 
\ref{sub:open_systems}.  Also plotted in Fig.~\ref{fig:5_comp_evo} are the 
annealing schedules $A$ and $B$ as a function of the reduced time $\tau$, 
illustrating how the shape of the functions $A(\alpha,\tau)$ and 
$B(\alpha,\tau)$ changes for different values of $\alpha$, from flat for a 
quantum walk to a curving AQC annealing schedule for $\alpha=1$.
\begin{figure}
 \begin{centering}
  \includegraphics[width=0.9\columnwidth]{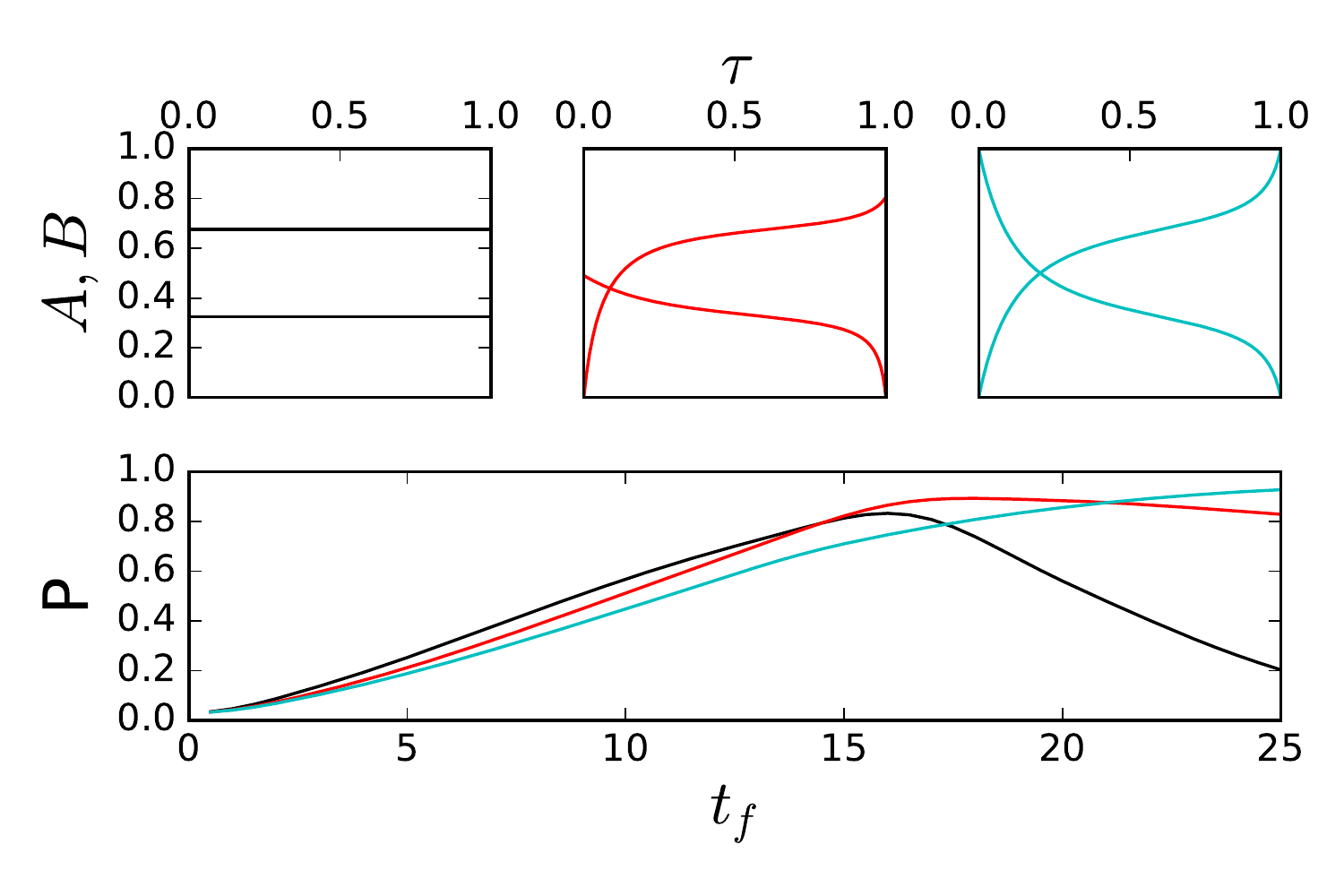}
 \end{centering}
 \caption{%
 (color online) (Top) Numerically calculated hybrid schedules $A$ and $B$ 
 against runtime $\tau$ for quantum search on a 5-qubit hypercube graph for 
 $\alpha=0$ QW, (black, top left), 
 $\alpha=0.5$ (red, mid gray in print, top middle), and 
 $\alpha=1$ AQC, (cyan, light gray in print, top right). 
 (Bottom) Success probability of the corresponding searches (indicated by 
 matching colour/shade of gray in print) against total search time $t_f$,
 in units given by Eq.~(\ref{eq:H_adiabat}).  Note 
 that this does not show time evolution against $t$ or $\tau$.
 \label{fig:5_comp_evo} 
 } 
\end{figure}
We see that the qualitative behaviour of adiabatic evolution is fundamentally 
different from that of the quantum walk search. For the optimal AQC schedule the
success probability increases monotonically to a value very close to one.  In 
contrast, QW shows oscillatory behaviour, and although the success probability 
does not approach one, it does show a faster initial increase than for AQC.  The
intermediate schedule shows a mix of both behaviours, with a locally oscillating
but globally increasing success probability that shows an initial increase rate 
between that of QW and AQC. 

\begin{figure}
 \begin{centering}
  \includegraphics[width=0.9\columnwidth]{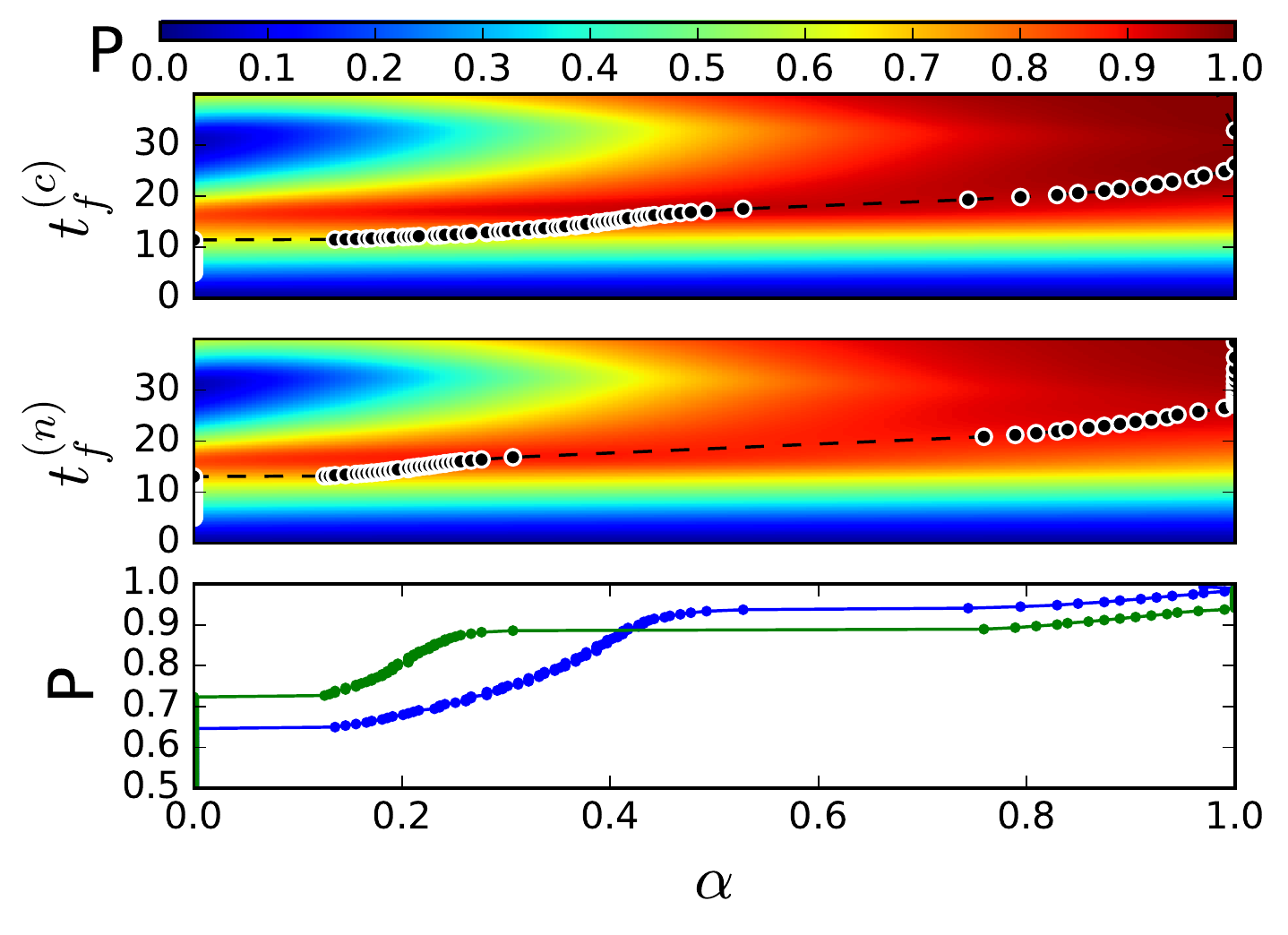}
 \end{centering}
 \caption{(color online)
 Successs probabilities $\mathsf{P}$ of hybrid QW-AQC quantum search on a 
 5-qubit hypercube graph plotted against the interpolation parameter $\alpha$ 
 and total runtime $t_f$ using optimal schedules (top) $s^{(c)}$ analytical, and
 (middle) $s^{(n)}$ from numerical solution.   Dotted lines with black points 
 indicate the optimal protocol at a given runtime $t_f$.  Probability
 corresponding to the optimal protocol (bottom) for analytical (blue, dark gray 
 in print) and
 numerical (green, mid gray in print). Time units given by Eq.~(\ref{eq:H_adiabat}).
 \label{fig:5_two_sched} 
 } 
\end{figure}
\begin{figure}
 \begin{centering}
  \includegraphics[width=0.9\columnwidth]{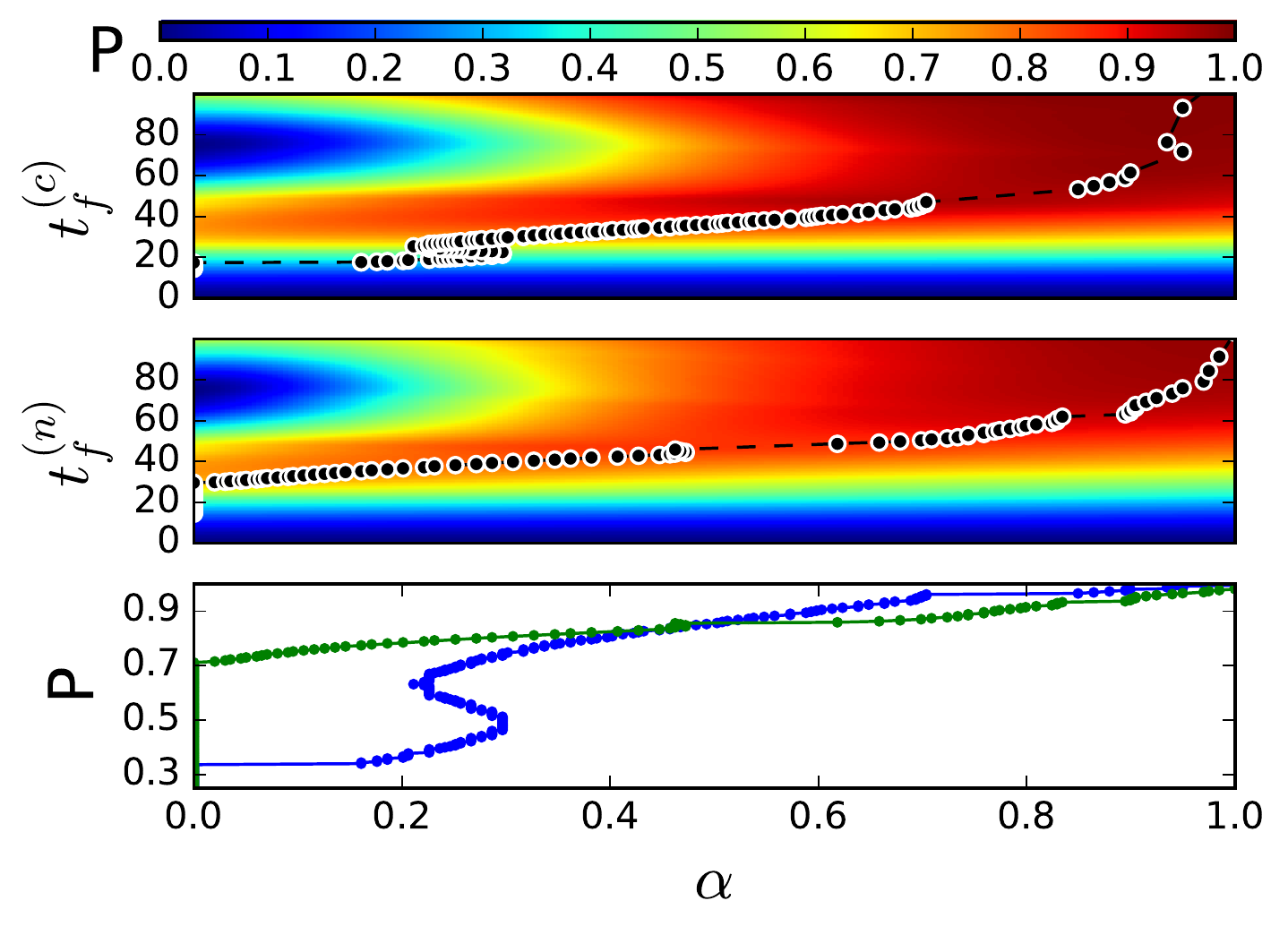}
 \end{centering}
 \caption{(color online)
 Successs probabilities $\mathsf{P}$ of hybrid QW-AQC quantum search on an 
 8-qubit hypercube graph plotted against the interpolation parameter $\alpha$ 
 and total runtime $t_f$ using optimal schedules (top) $s^{(c)}$ analytical, and
 (middle) $s^{(n)}$ from numerical solution.   Dotted lines with black points 
 indicate the optimal protocol at a given runtime $t_f$.  Probability
 corresponding to the optimal protocol (bottom) for analytical (blue, dark gray 
 in print) and
 numerical (green, mid gray in print). Time units given by Eq.~(\ref{eq:H_adiabat}).
 \label{fig:8_two_sched} 
 } 
\end{figure}

We now turn to the probability $\mathsf{P}$ of finding the marked state that is 
obtained for different choices of $\alpha$ and $t_f$.  For a continuum of 
$\alpha$ values, Figs.~\ref{fig:5_two_sched} and \ref{fig:8_two_sched} 
illustrate the same qualitative behaviour for 5-qubit and 8-qubit quantum 
searches.  The oscillatory behaviour associated with a QW slowly fades away as 
the interpolation approaches the respective AQC schedule, at which point the 
success probability $\mathsf{P}$ increases monotonically with $t_f$.  If a 
relatively low success probability is sufficient, only a short total runtime 
$t_f$ is needed, and quantum walk is the best strategy.  As $t_f$ is increased, 
the best strategy is to increase $\alpha$ and start adding some adiabatic 
character into the protocol. Finally, if a high success probability is required 
and a long runtime $t_f$ is possible, then AQC becomes the best strategy.  We 
also see that, for these system sizes, the hybrid protocols maintain the quantum
speed up for the search algorithm runtime.

We now consider the differences between the calculated and numerical annealing 
schedules $s^{(c)}$ and $s^{(n)}$ for these small systems.  
Fig.~\ref{fig:5_two_sched} depicts results for $n=5$ qubits.  The main 
difference for five qubits is that the numerically calculated optimal schedule 
$s^{(n)}$ is able to perform  substantially better for $\alpha < 0.4$, where 
``better'' means a higher probability of success for a given runtime $t_f$ and 
value of $\alpha$. 
Figure~\ref{fig:8_two_sched} shows the same comparisons for the slightly 
larger value of $n=8$ qubits.  The optimal $\alpha$ moves away from $\alpha=0$ 
at a smaller value of $t_f$ and $\mathsf{P}$ for $s^{(c)}$ than it does for 
$s^{(n)}$.  There is also more structure in the optimal $\alpha$ line (black 
dashes) for $s^{(c)}$ than for $s^{(n)}$, with a range of $\alpha$ values that
are optimal for more than one value of $\mathsf{P}$.  Otherwise, the two behave
quite similarly for these small sizes, suggesting that both $s^{(c)}$ and $s^{(n)}$ 
are able to provide a quantum speed up for hybrid protocols.  To confirm this in
general, not just for small $n$, further analysis and simulations of larger 
systems are required, which we tackle in the following subsections.

%%%%%%%%%%%%%%%%%%%%%%%%%%%%%%%%%%%%%%%%%%%%%%%%%%%%%%%%%%%%%%%%%%%%%%%%%%%%%%%
\subsection{Performance of hybrid algorithms\label{sec:scaling}}
%-----------------------------------------------------------------------------%

Our strategy for analyzing the scaling of the hybrid quantum search algorithms 
is to show that the performance is dominated by a single, low energy, avoided 
crossing, see Fig.~\ref{fig:HhAQC_energy}, which is present at the same position
in all our hybrid algorithms.  We then show that the essential features of the
behavior are captured by a simple, two-state single avoided crossing model which
all the hybrid algorithms map to in the large size limit.  For this simple avoided
crossing model we can easily show that the hybrid algorithms all provide an optimal
quantum speed up.  It then follows that our full-size hybrid algorithms have the
same asymptotic scaling.

\begin{figure}
 \begin{centering}
  \includegraphics[width=0.9\columnwidth]{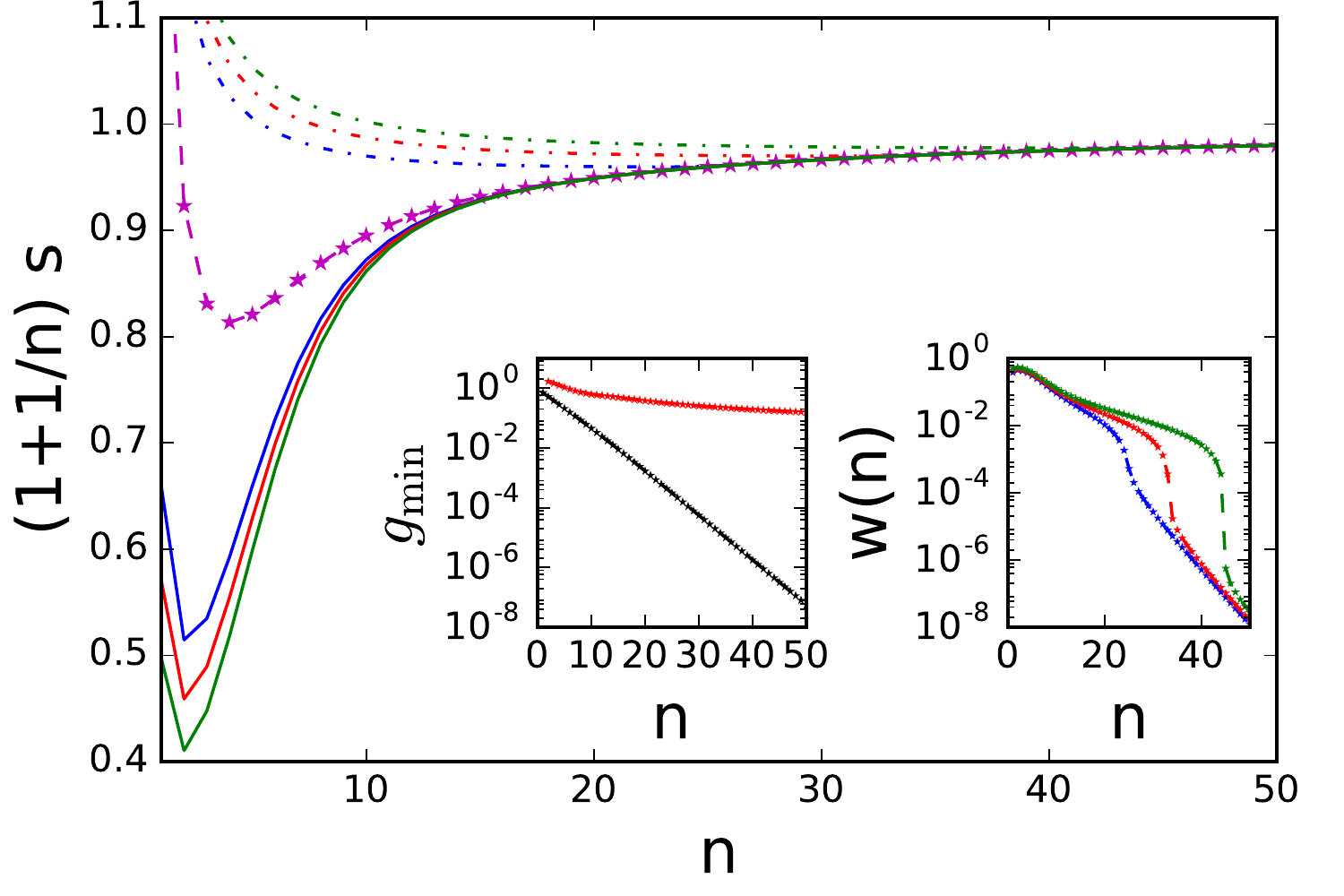} 
 \end{centering}
 \caption{(color online) 
 Main figure: $s(\tau)$ scaled by $(1+1/n)$ against number of qubits $n$ for 
 $90\%$ (blue, dark gray in print), 
 $93\%$ (red, light gray in print), 
 $95\%$ (green, mid gray in print)  
 overlap of $\ket{\psi(t)}$ with $\ket{m}$ (solid) and with 
 $\ket{\psi_{\text{init}}}$ (dot-dashed).
 Magenta stars are the transition point, the value of $s(\tau)$ when the minimum 
 gap $g_{\text{min}}$ occurs.
 Left inset: $g_{\text{min}}=\min(E_1-E_0)$ (lower black stars) and 
 $\min(E_2-E_0)$ (upper red stars, light gray in print). 
 Energy units given by Eq.~(\ref{eq:H_adiabat}).
 Right inset: width of the transition $w(n)=\Delta s(\tau)$, the difference 
 between solid and dot-dashed curves of the same color in the main figure.
 Calculated using the AQC search hypercube Hamiltonian mapped to the line,
 see Appendix \ref{app:num_meth}.
 \label{fig:hyp_scaling} 
 } 
\end{figure}

We first consider the end points of the interpolation, QW and AQC search.  For 
AQC search, the optimal schedule $s^{(c)}(\tau)$ or $s^{(n)}(\tau)$ is derived 
directly from the functional form of the lowest avoided crossing, ensuring that 
the Hamiltonian is changed slowly enough to avoid transitions to higher energy 
levels.  We only need to show that the low energy structure of the Hamiltonian 
is dominated by a single avoided crossing throughout the process.  This is shown
numerically in Fig.~\ref{fig:hyp_scaling}.  The width $w(n)$ of the avoided 
crossing decreases rapidly with $n$.  Even for a modest size of $n=50$ qubits, 
the switch from $95\%$ overlap with the hypercube Hamiltonian ground state to 
$95\%$ overlap with the marked state occurs in less than $10^{-6}$ of the total 
dynamic range of the protocol, which runs from $s(\tau)=0$ to $s(\tau)=1$\footnote{
While the dynamical range $w(n)=\Delta s(\tau)$ in which the state rotates between 
the two nearly-orthogonal states $\ket{\psi_\text{init}}$ and $\ket{m}$ becomes
exponentially small, the total runtime $t_f\sim\sqrt{N}=\exp(n/2)$ grows even more
quickly so that the time taken $w(n)/t_f$ increases with $n$.}.
In contrast, for QW search, transitions to higher 
energy levels are a necessary part of the evolution to the marked state, so we 
need to determine the scaling of several related quantities to show that a 
single avoided crossing dominates in determining the behavior.

%%%%%%%%%%%%%%%%%%%%%%%%%%%%%%%%%%%%%%%%%%%%%%%%%%%%%%%%%%%%%%%%%%%%%%%%%%%%%%%
\subsection{Minimum gap scaling in QW search}\label{sub:min_gap}
%%%%%%%%%%%%%%%%%%%%%%%%%%%%%%%%%%%%%%%%%%%%%%%%%%%%%%%%%%%%%%%%%%%%%%%%%%%%%%%

For QW search, to show numerically that the lowest avoided crossing is the only 
relevant feature in the large $N$ limit, we must demonstrate two things.  First,
that the minimum gap $g_{\text{min}}=(E_1-E_0)$ between the ground state and the
first excited state becomes much smaller than the minimum gap between the ground
state and the second excited state.  Second, that the lowest avoided level 
crossing, where $g(\tau)=g_{\text{min}}$, dominates the transition between the 
ground state of the hypercube Hamiltonian $\hat{H}^{(h)}_0$ and the ground state
of the marked state Hamiltonian $\hat{H}_\textsc{p}$, and becomes more dominant 
as system size increases.  Noting that, as illustrated in 
Fig.~\ref{fig:sched_interp}, around the minimum gap, where all the schedules 
cross, we have $(1-s(\tau)) \simeq \gamma^{(h)}_{o}$, 
Figure \ref{fig:hyp_scaling} shows that both of these do, in fact, occur.  The 
left inset shows that at the avoided crossing, the gap between the ground state 
and first excited state shrinks exponentially faster in $n$ than the gap between
the ground state and second excited state.  The main figure and right inset of 
Fig.~\ref{fig:hyp_scaling} show how the transition between the two ground states
becomes dominated by the dynamics at $g_{\text{min}}$ as $n$ increases.

\begin{figure}
 \begin{centering}
 \includegraphics[width=0.9\columnwidth]{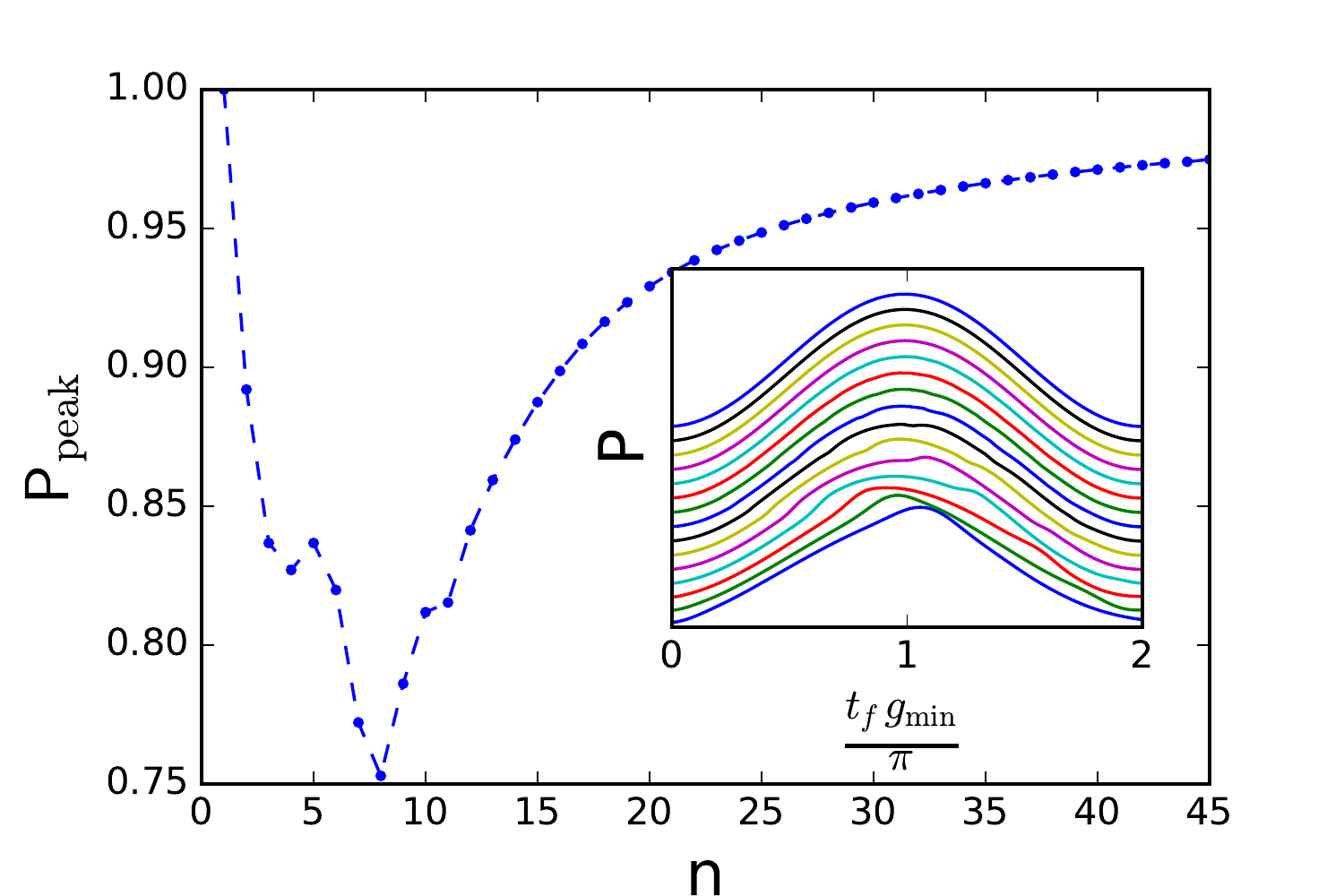}
 \end{centering}
 \caption{(color online) Main figure: Search success probability $\mathsf{P}$ at
 the first peak for a quantum walk search against qubit number $n$ up to $n=50$. 
 Inset: Rescaled offset plot of $\mathsf{P}$ against $t$ starting at the bottom 
 with $n=5$ qubits and going to $n=20$.  Calculated using the hypercube QW 
 search mapped to the line. 
 \label{fig:QWpeak_scale}
 }
\end{figure}
For a pure quantum walk search, this convergence to behaviour dominated by a 
single avoided crossing can be seen in Fig.~\ref{fig:QWpeak_scale}, which shows 
that not only does the search success probability $\mathsf{P}$ approach one in 
the large system limit (main figure), but also that the time evolution of 
$\mathsf{P}$ (inset) approaches the functional form for the single avoided 
crossing $\mathsf{P}(\tau)=\sin^2(g_{\text{min}}t_f/2)$. The non sinusoidal 
shapes of these curves at low qubit number are due to the influence of excited 
states higher than the first exited state.  In the main figure, these small size 
effects are clearly significant up to about $n=12$ qubits.  This highlights the 
potentially atypical nature of the 5- and 8-qubit examples in Sec. 
\ref{sub:small_sys_ex}, and the importance of examining larger system sizes.  
For $n>12$, the probability $\mathsf{P}$ smoothly approaches one, although 
relatively slowly (polynomially) as a function of $n$.
Based on the data in table \ref{tab:num_fits} we can deduce that this 
effect relates to the fact that the overlap of the manifold where the avoided 
crossing takes place with the marked state only approaches one polynomially in 
$n$ (logarithmically in $N$).  

Since states of higher energy than the first excited state play very little role 
in the QW search dynamics for larger systems, we can approximate the probability 
that the marked state can be reached using only the manifold 
$\mathcal{T}=\{|E_0\rangle,|E_1\rangle\}$ of ground and first excited states
of the full search Hamiltonian $\hat{H}^{(h)}_{\text{QWS}}$.
This can be upper bounded by considering the probability that the dynamics transfers
as much as possible of $\ket{\psi_\text{init}}$ into $\mathcal{T}$, and
then optimally aligns the system state with $\ket{m}$ without leaving $\mathcal{T}$.
Using $\hat{\mathcal{P}}_\mathcal{T}=|E_0\rangle\langle E_0| + |E_1\rangle\langle E_1|$,
the projector onto $\mathcal{T}$, this can be shown to be given by the product of the sums
of the overlaps, 
\begin{align} \label{eq:PQWmax}
\mathsf{P}_{\max}^{\text(QW)} &=|\hat{\mathcal{P}}_\mathcal{T}\ket{\psi_\text{init}}|^2
                                \times|\hat{\mathcal{P}}_\mathcal{T}\ket{m}|^2
          \nonumber\\ 
			&=\left(|\braket{\psi_{\text{init}}}{E_0}|^2
		  + |\braket{\psi_{\text{init}}}{E_1}|^2\right)\nonumber\\
							  &\times\left(|\braket{E_0}{m}|^2
		  + |\braket{E_1}{m}|^2\right),
\end{align}
when single avoided crossing behaviour dominates. 
\begin{table*}[t]
 \begin{tabular}{|c|c|c|}
 \hline
 Quantity & Scaling & $1-r^2$  \\ 
 \hline
 $\mathsf{P}_{\max}^{\text(QW)}$ & $1-1.734\times n^{-1.112}$ & $7.820\times10^{-5}$ \\ 
 \hline
 $|\braket{E_0}{m}|^2+|\braket{E_1}{m}|^2$ &  $1-1.734\times n^{-1.112}$ & $7.820 \times 10^{-5}$  \\ 
 \hline
 $|\braket{\psi_{\text{init}}}{E_0}|^2+|\braket{\psi_{\text{init}}}{E_1}|^2$ &
 $1-4.292\times 2^{-1.186\,n}$ & $0.00143$ \\ 
 \hline 
$\gamma^{(h)}_o$ & $1-1.233\times n^{-1.0425}$ & $1.120\times 10^{-5}$ 
 \\ 
 \hline
 \end{tabular}
 \caption{Numerical fits for various quantities related to quantum walks and 
 adiabatic protocols.  These fits were 
 performed using linear fitting on either logarithmic or semi-logrithmic axes in
 the range $n=40$ to $n=70$, except for 
 $|\braket{\psi_{\text{init}}}{E_0}|^2+|\braket{\psi_{\text{init}}}{E_1}|^2$
 which was fit over the range $n=11$ to $n=40$ due to numerical precision 
 issues. The coefficient of determination 
 $r^2\equiv1-\frac{\sum_i(y_i-f_i)^2}{\sum_i(y_i-\bar{y})^2}$, where $f_i$ are 
 the data and $y$ is the fitting function, is calculated 
 against the linear function on the logarithmic or semi-logarithmic axes.  
 These fits are plotted along with the data used to produce them in 
 Fig.~\ref{fig:gamma_and_ovls}.  The slight difference from $-1$ in the scaling 
 exponent for $\gamma^{(h)}_o$ is due to numerical finite size effects.  
 \label{tab:num_fits}}
\end{table*}
\begin{figure}
 \begin{centering}
 \includegraphics[width=0.9\columnwidth]{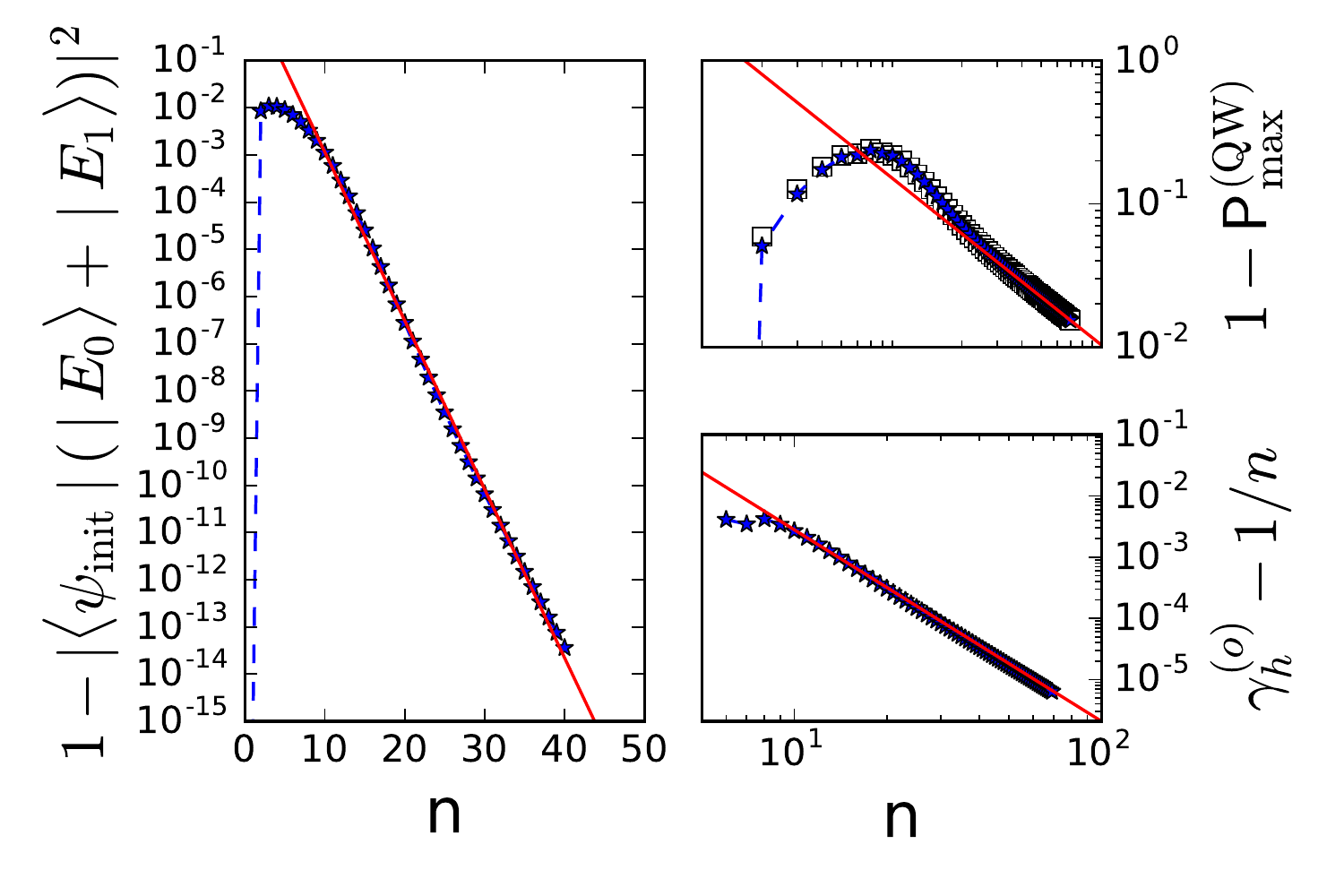}
 \caption{(color online) Scaling of various quantities related to QW searching. 
 Left: difference from one of the overlap of $\ket{\psi_{\text{init}}}$ with 
 $\ket{E_0}$ and $\ket{E_1}$ against number of qubits $n$. 
 Top Right: difference from one of marked state with $\ket{E_0}$ and $\ket{E_1}$
 (stars) and $\mathsf{P}_{\max}^{\text(QWS)}$ (squares), against $n$.
 Bottom right $\gamma^{(h)}_o-1/n$ versus $n$. 
 Solid lines (red online) are numerical fits, summarized in table 
 \ref{tab:num_fits}.  Calculated using the hypercube QW search mapped to the 
 line.
 \label{fig:gamma_and_ovls} 
 }
 \end{centering}
\end{figure}

Figure~\ref{fig:gamma_and_ovls} shows how $\mathsf{P}_{\max}^{\text(QW)}$ 
approaches one as $n$ increases, by plotting the difference from one on a log or
log-log scale.  The top right figure shows that 
$\mathsf{P}_{\max}^{\text(QW)}\rightarrow 1$ only happens relatively slowly, 
with a polynomial scaling in $n$, and therefore logarithmic in $N$.  By plotting
the first overlap in Eq.~(\ref{eq:PQWmax}) separately, the left figure shows 
that the overlap of $\ket{\psi_{\text{init}}}$ with $\ket{E_0}$ and $\ket{E_1}$ 
rapidly approaches one.  Hence, the scaling of $\mathsf{P}_{\max}^{\text(QW)}$ 
shown top right is dominated by the overlap of the marked state with the lowest 
energy states $\ket{E_0}$ and $\ket{E_1}$ at the gap.  We can quantify how 
slowly $\mathsf{P}$ approaches one by doing numerical fits to determine the 
scaling of the relevant quantities: these are summarized in 
table~\ref{tab:num_fits}.  In particular, we note that $\gamma^{(h)}_o$ only 
approaches $1/n$ linearly in $n$, consistent with the analytical results in 
Ref.~\cite{childs03a}.   

The fact that $\mathsf{P}_{\max}^{\text(QW)}\rightarrow 1$ suggests that the 
optimal protocol for all success probabilities should approach QW 
($\alpha=0$) for large system size, because QW does not slow down at the 
minimum gap like AQC does.  However, 
$\mathsf{P}_{\max}^{\text(QW)}\rightarrow 1$ only happens relatively slowly: 
the maximum $\mathsf{P}_{\max}^{\text(QW)}$ which a QW search obtains only 
reaches $99\%$ by around $100$ qubits.  Brute force classical techniques will 
become computationally non-trivial beyond around 30 bits, where 
$\mathsf{P}_{\max}^{\text(QW)}\approx 95\%$.  The finite size effects we study 
here are thus relevant to real world applications. 

%%%%%%%%%%%%%%%%%%%%%%%%%%%%%%%%%%%%%%%%%%%%%%%%%%%%%%%%%%%%%%%%%%%%%%%%%%%%%%%
\subsection{Single avoided crossing model\label{sub:SAC_model}}
%-----------------------------------------------------------------------------%

We have shown that a single avoided crossing dominates for large $N$ for 
both QW and AQC search algorithms on the hypercube.  Dominance of a single 
avoided crossing is the method used to solve analytically for all 
Hamiltonian-based quantum search algorithms treated to date, including the 
complete graph \cite{roland02a} and Cartesian lattices (which provide a quantum 
speed up for $d\ge 4$ dimensions) \cite{childs03a}.  It is also the typical 
behavior for a broad class of random search graphs \cite{Chakraborty16a}. 
We now introduce a simple, two state, single avoided crossing model for 
quantum search which provides the quadratic quantum speed up.  We will then show
how all of our hybrid protocols can be mapped onto it.   

There are several ways to parameterize a two-state single avoided 
crossing model.  If we designate the marked state to be the $\ket{0}$ state of
a qubit, this will be 
the end point of the schedule.  The initial state needs to be orthogonal to 
$\ket{0}$, i.e., it has to be $\ket{1}$.  These two states are the lowest energy
eigenstates of $\frac{1}{2}(\openone + \hat{\sigma}_z)$ and 
$\frac{1}{2}(\openone - \hat{\sigma}_z)$ respectively, where the factor of 
$\frac{1}{2}$ makes the eigenenergies zero and one in our units.  We also need 
a hopping Hamiltonian term $\hat{\sigma}_x$, to drive transitions between 
$\ket{1}$ and $\ket{0}$.  The relative strength of the hopping Hamiltonian is 
$g_{\text{min}}$, the minimum gap at the avoided crossing.  The single avoided 
crossing AQC search Hamiltonian is
\begin{align}\label{eq:H_ac_s}
\hat{H}^{(\text{AC})}(s) &= (1-s)\hat{H}^{\text{(AC)}}_0 + s\hat{H}^{\text{(AC)}}_p \nonumber \\
                         &= (1-s)\left\{\frac{1}{2}(\openone + \hat{\sigma}_z) 
- g_{\text{min}}\hat{\sigma}_x\right\} + s\frac{1}{2}(\openone - \hat{\sigma}_z).
\end{align}
The initial state $\ket{1}$ is only an approximate eigenstate of $\hat{H}^{\text{(AC)}}_0$
but the approximation improves as $g_{\text{min}}$ decreases.
Solving the eigensystem for this Hamiltonian gives
\begin{equation}
g^{(\text{AC})}(s) = \{(1-2s)^2 + 4g^2_{\text{min}}(1-s)^2\}^{\frac{1}{2}}
\end{equation}
for the gap between the two energy levels.  In the limit of small $g_{\text{min}}$ the minimum gap is $g_\text{min}$ and occurs for
$s=\frac{1}{2}$.  We can then apply the method of \cite{roland02a} to find the 
optimal schedule $s(t)$ for this system.  Calculating $d\hat{H}/ds$ we find
\begin{equation}
\frac{d\hat{H}}{ds}^{(\text{AC})} = -\hat{\sigma}_z + g_{\text{min}}\hat{\sigma}_x
\end{equation}
giving a maximum value of one%\!
\footnote{Strictly the maximum value is $\sqrt{1+g_{\text{min}}^2} = 1 + \mathcal{O}(g_
{\text{min}}^2)$, however this correction simply modifies $\epsilon$ in what follows, 
and disappears altogether when terms of order $g_{\text{min}}^2$ are dropped.}
for 
$|\average{\frac{d\hat{H}^{(\text{AC}})}{ds}}_{0,1}|$ in the large-size limit.
Using Eq.~(\ref{eq:ac_inst}) to find the optimal schedule, we need to solve
\begin{equation}
\frac{ds}{dt} = \frac{\epsilon [g^{(\text{AC})}(s)]^2}{|\average{\frac{d\hat{H}_{\text{AC}}}{ds}}_{0,1}|} = \epsilon\{(1-2s)^2 + 4g^2_{\text{min}}(1-s)^2\},
\end{equation}
where the maximum value is used for $|\average{\frac{d\hat{H}^{(\text{AC}})}{ds}}_{0,1}|$.
This can be integrated straightforwardly to give
\begin{equation}
\arctan\left\{2g_{\text{min}}(s-1)+\frac{2s-1}{g_{\text{min}}}\right\} 
    = 2g_{\text{min}}\epsilon t + c
\end{equation}
with
\begin{equation}
c = - \arctan\left\{2g_{\text{min}} + \frac{1}{g_{\text{min}}}\right\}.
\end{equation}
From this we find for $s=1$ that the runtime $t_f^{(AC)}$ is given by
\begin{equation}\label{eq:tf_ac}
\epsilon\, t_f^{(AC)} = \frac{\frac{\pi}{2} - \arctan(g_{\text{min}})}{g_{\text{min}}}
    \simeq \frac{\pi}{2\,g_{\text{min}}} - 1,
\end{equation}
where the approximate expression uses 
$\arctan(1/g_{\text{min}}) \simeq \frac{\pi}{2}-g_{\text{min}}$ for 
$g_{\text{min}}\ll 1$ and terms of order $g^2_{\text{min}}$ have been dropped.  
The runtime of the optimal schedule thus depends inversely on the size of the 
minimum gap, as expected.  Solving for $s(t)$ and dropping terms of order 
$g^2_{\text{min}}$ gives
\begin{equation}\label{eq:s_as_t}
s(t) \simeq \frac{1}{2}\left\{1 - g_{\text{min}}\cot\left[g_{\text{min}} (2\epsilon t + 1) \right] \right\}.
\end{equation}

In this limit where $g_{\text{min}}\ll 1$, an equivalent way to parameterize 
$\hat{H}^{(\text{AC})}$ is
\begin{equation}
\hat{H}^{(\text{AC})}=\frac{g_{\text{min}}}{2}\left[f(t)\hat{\sigma}_z 
    - \hat{\sigma}_x \right], 
\label{eq:H_ac}
\end{equation}
where $-\infty < f(t) < \infty$.  This form is obtained by taking 
$(1-2s(t))/g_{\text{min}}\rightarrow f(t)$ and shifting the zero point of the 
energy scale to the middle of the avoided crossing.  As $f(t)$ changes from 
$-\infty$ to $\infty$ it passes through zero as the sign of the $\hat{\sigma}_z$
term changes, when the $\hat{\sigma}_x$ term drives the transition from 
$\ket{1}$ to $\ket{0}$.  Although the $\hat{\sigma}_x$ term is no longer turned 
off at the end of the schedule, it becomes negligible in comparison to the 
$\hat{\sigma}_z$ term and does not significantly alter the dynamics.  This can 
be intuitively thought of as scaling all features of 
$\hat{H}^{(\text{AC})}$ other than the avoided crossing to $\pm \infty$.

The QW form of the single avoided crossing search Hamiltonian is also simple to 
analyze.  We deduce the optimal value of $\gamma_o=1$ from the value of 
$s=\frac{1}{2}$ at the avoided crossing.  We then use Eqns.~(\ref{eq:QW_AB_beta}) 
in which $\beta_o=1/(1+\gamma_o) = \frac{1}{2}$, whence
\begin{align}
\hat{H}^{\text{AC}}_{(\text{QWS})} &= (1-\beta_o)\hat{H}^{\text{(AC)}}_0 
     + \beta_o\hat{H}^{\text{(AC)}}_p \nonumber \\
    &= \frac{1}{2}\left\{\frac{1}{2}(\openone 
     + \hat{\sigma}_z) - g_{\text{min}}\hat{\sigma}_x 
     + \frac{1}{2}(\openone - \hat{\sigma}_z)\right\} \nonumber\\
    &= \frac{1}{2}(\openone - g_{\text{min}}\hat{\sigma}_x)
\end{align}
The $\hat{\sigma}_x$ term causes deterministic transitions between the two 
states regardless of their energies, at a rate determined by $g_{\text{min}}$.
By solving for the dynamics, the time for the input state $\ket{1}$ to evolve to
the marked state $\ket{0}$ can be shown to be $t_f^{(qw)}=\pi/g_{\text{min}}$.

\begin{figure}
 \begin{centering}
  \includegraphics[width=0.9\columnwidth]{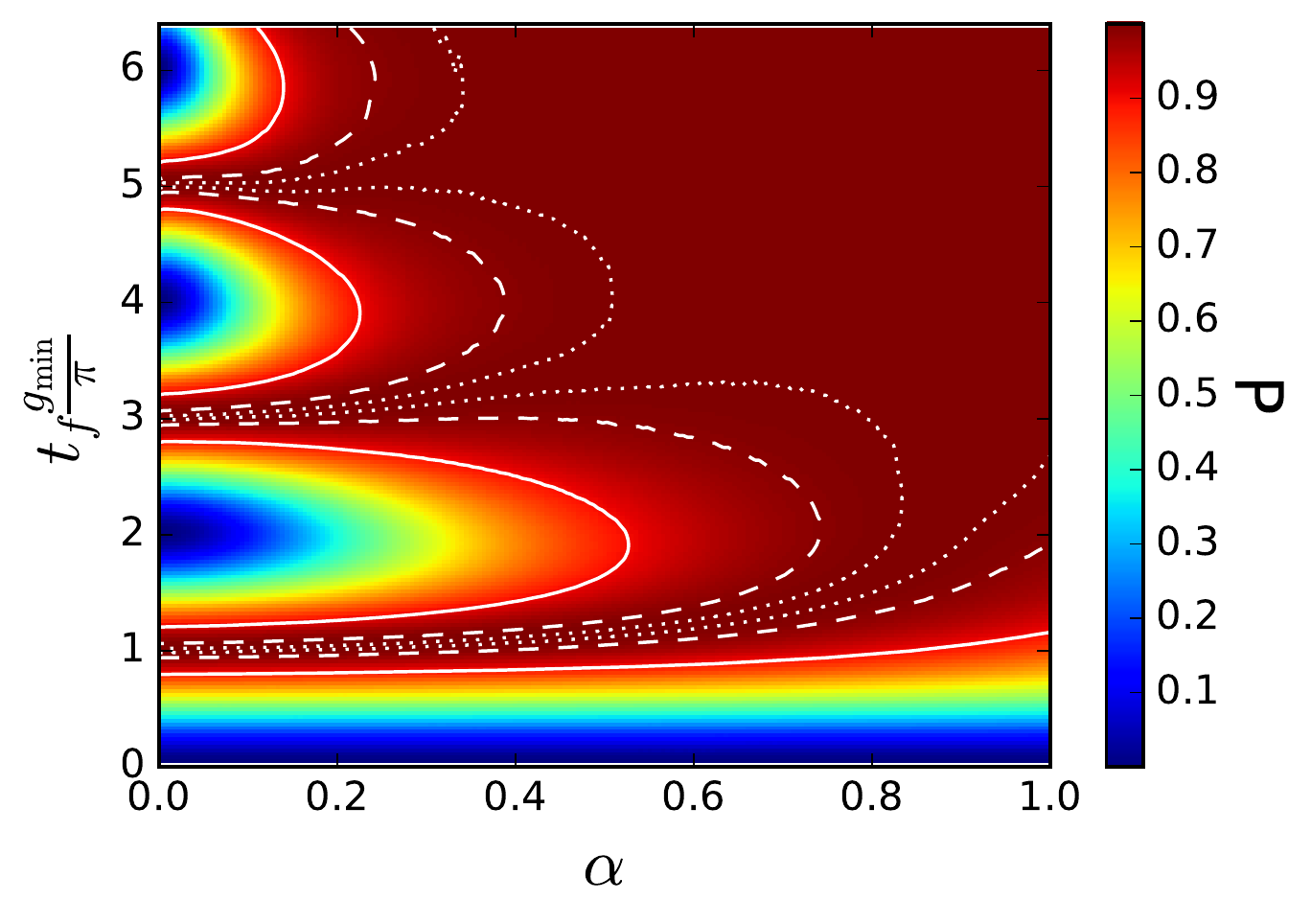}
 \end{centering}
 \caption{Probability $\mathsf{P}$ of finding the marked state versus runtime
 $t_f$ and interpolation parameter $\alpha$ for the single avoided
 crossing model.  White contours show solid=0.9, dashed=0.99, dotted=0.999 
 success probability $\mathsf{P}$.
 \label{fig:ac_plot_noErr}
 }
\end{figure}

We can now map between QW and AQC in the avoided crossing model using 
Eqs.~(\ref{eq:interp_AB}) for $A(\alpha,\beta,\tau)$ and $B(\alpha,\beta,\tau)$.
Using $\beta=\frac{1}{2} = 1/(1+\gamma_o)$, for $s(t)$ from 
Eq.~(\ref{eq:s_as_t}) we have hybrid schedules
\begin{align}\label{eq:interp_AB_AC}
A_{\text{AC}}(\alpha,t) 
  &= \frac{1-s(t)}{\alpha+2(1-\alpha)(1-s(t))}\nonumber\\
B_{\text{AC}}(\alpha,t)
  &= \frac{s(t)}{\alpha+2(1-\alpha)s(t)}.
\end{align}
We can easily show numerically that all the hybrid algorithms defined by 
Eqs.~(\ref{eq:interp_AB_AC}) find the marked state with high probability
(given by $\epsilon$) in a runtime $\lesssim \epsilon t_f^{(AC)}$ given 
by Eqn.~(\ref{eq:tf_ac}), the runtime 
required by the optimal AQC $s(t)$ used to define the hybrid schedules.  
Figure \ref{fig:ac_plot_noErr} shows this is indeed the case.  The white
contours highlight the difference between the pure QW search, which 
succeeds with certainty, and the AQC and hybrid algorithms, which always 
have a probability of error $\epsilon^2$ that can be traded against the 
runtime $t_f$.  The shallow upward curve of these contours towards the AQC 
end of the hybrid protocols shows in what sense the QW search is better 
than AQC in the large size limit.

The hybrid algorithms on the full hypercube map onto the hybrid single
avoided crossing model algorithms for large $n$.  This follows from the
solution methods for the  end points, QW and AQC searching, which all use
the two-level approximation to  prove the quadratic speed up.  Since the
full hypercube hybrid algorithms are defined from these in the same way as
the single avoided crossing model hybrid algorithms are defined, the hybrid
algorithms also map to the corresponding single avoided crossing hybrid
algorithm.  They therefore also obtain the quantum speed up for large $n$, 
which is what we set out to show.

%%%%%%%%%%%%%%%%%%%%%%%%%%%%%%%%%%%%%%%%%%%%%%%%%%%%%%%%%%%%%%%%%%%%%%%%%%%%%%%
\subsection{Optimal hybrid algorithm for a single run} \label{sub:approx_scale}
%%%%%%%%%%%%%%%%%%%%%%%%%%%%%%%%%%%%%%%%%%%%%%%%%%%%%%%%%%%%%%%%%%%%%%%%%%%%%%%
Having shown that hybrid protocols between QW and AQC maintain the quadratic 
quantum speed up, the next question is how to optimize over this continuum of 
hybrid schedules for finite size systems.  The single avoided crossing model 
gives the large size limit in which QW is the optimal strategy.  However, this
limit is only reached in a polynomial scaling with $n$, as described in 
Sec.~\ref{sub:min_gap}.

For a single run of a search algorithm, we can trade off between the magnitude 
of the success probability and the runtime of the search.  For QW searches, 
there is a maximum probability $\mathsf{P}_{\max}^{\text(QW)}$ that can be 
obtained; shorter runtimes reach lower success probabilities, and so do longer 
runtimes.  For AQC searches a longer runtime always reaches a higher success 
probability.  We can thus specify the success probability we require and ask 
which hybrid algorithm attains this success probability with the shortest runtime.
We consider multiple run strategies in Sec.~\ref{sec:multi_strats}.

\begin{figure}
 \begin{centering}
 \includegraphics[width=0.9\columnwidth]{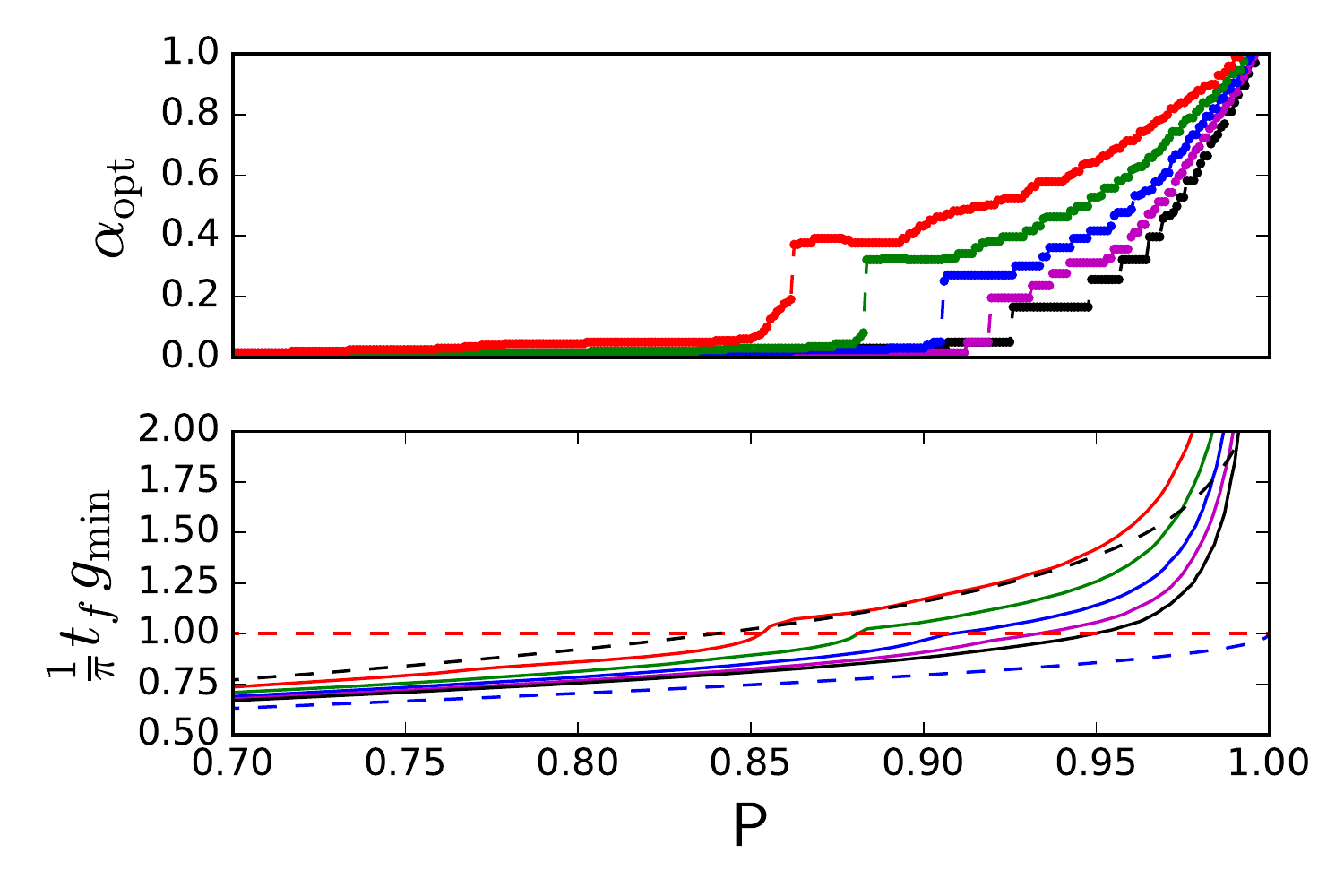}
 \caption{(color online)
 Top: Value of interpolation parameter $\alpha$ giving the shortest runtime for
 a fixed success probability $\mathsf{P}$ for a single search, using numerically 
 calculated optimal schedules $s^{(n)}$ for hypercube dimensions (listed from
 top line to bottom line):
 $n=12$ (red); $n=14$ (green); $n=16$ (blue); $n=18$ (magenta); $n=20$ (black). 
 Bottom: Normalized runtime versus $\mathsf{P}$ for corresponding $\alpha$ and
 hypercube dimension as above (solid lines, same ordering as above). 
 Dashed lines: single avoided crossing model (large $N$ limit) for
 $t_f=g_\text{min}/\pi$, the time at which a QW reaches a success probability of
 one (red), time for QW to reach $\mathsf{P}$ (blue), time for AQC to reach 
 $\mathsf{P}$ (black).
 \label{fig:ac_approach}
 }
 \end{centering}
\end{figure}
As Fig.~\ref{fig:ac_approach} (top) illustrates for sizes from $n=12$ to $n=20$, 
the optimal protocol jumps from QW to hybrid at $\mathsf{P} \approx 
\mathsf{P}_{\max}^{\text(QW)}$, and the optimal hybrid strategy it jumps to
becomes more QW-like (smaller $\alpha$) as the system size increases.  As 
$\mathsf{P}$ is increased further, the optimal hybrid strategy becomes steadily
more AQC-like (larger $\alpha$).  Figure \ref{fig:ac_approach} (bottom) shows
that the hybrid strategies require runtimes $t_f$ larger than $g_\text{min}/\pi$ 
to achieve higher success probabilities in a single run.

%%%%%%%%%%%%%%%%%%%%%%%%%%%%%%%%%%%%%%%%%%%%%%%%%%%%%%%%%%%%%%%%%%%%%%%%%%%%%%%
\section{Multiple runs for one search\label{sec:multi_strats}}
%-----------------------------------------------------------------------------%

In the previous sections we derived hybrid search Hamiltonians for the hypercube,
and studied their dynamics. However this doesn't yet give us a full picture of
the relative usefulness of the different dynamics. In this section we study the
relative performance of the different searches when we allow for the possibility of
multiple searches, and when the system suffers from decohering interactions with its
environment.

%%%%%%%%%%%%%%%%%%%%%%%%%%%%%%%%%%%
\subsection{Motivation}
%%%%%%%%%%%%%%%%%%%%%%%%%%%%%%%%%%%

In a realistic setting of the search problem we can easily check whether the 
result of a search is the correct answer or not.  Hence, we must consider not 
only single run strategies, but also multi-run strategies, where the success 
probability is defined as the probability of succeeding in at least one of 
several runs.  In the context of quantum search on the hypercube, we measure 
which site of the hypercube our state is on, and then determine the energy of 
this state with respect to the search Hamiltonian.  If this energy is zero,
then we have found the state we are looking for, otherwise, we should 
re-initialize and run the search again.  However, we also need to account for a 
non-zero `initialization' time $t_\text{init}$ associated with each run of the 
search.  Such an initialization time is mathematically as well as physically 
necessary.  The fidelity between the initial state and marked state 
$|\braket{\psi_{\text{init}}}{m}|^2=\frac{1}{N}$ is non-zero.  An arbitrarily 
short run is equivalent to making a random guess.  Therefore, without an 
additional penalty per run, it would be possible to guess an arbitrarily large 
number of times for free, thus finding the marked state in a total arbitrarily 
short time.  Any physical device will take a significant amount of time both to 
setup the initial state and to measure the final state.  For the purposes of our
study, the effects on the total search time of initialization and readout times 
are the same, therefore the quantity we call $t_{\text{init}}$ should be taken to 
include all of the time associated with a single run other than the actual 
runtime of the algorithm $t_f$, i.e., as including both initialization and 
measurement.

\iffalse
\begin{itemize}
  \item \textcolor{red}{In realistic setting we can check our answer}
  \item \textcolor{red}{We therefore need to talk not of individual searches as discussed
  above, but 
overall 'strategies'(/algorithms) that can reach a given target success probability in 
the least amount of time}
  \item \textcolor{red}{We have to, therefore, consider the initialization time of our 
  system (otherwise random guessing is best!)}
  \item \textcolor{red}{This leads to multi-anneal strategies, which can be formulated 
                        like so ...}
  \end{itemize}
\fi

%%%%%%%%%%%%%%%%%%%%%%%%%%%%%%%%%%%%%%%%%%%%%%%%%%%%%%%%%%%%%%%%%
\subsection{Multiple run searching}
%%%%%%%%%%%%%%%%%%%%%%%%%%%%%%%%%%%%%%%%%%%%%%%%%%%%%%%%%%%%%%%%%

%
\begin{figure}
 \begin{centering}
  \includegraphics[width=0.9\columnwidth]{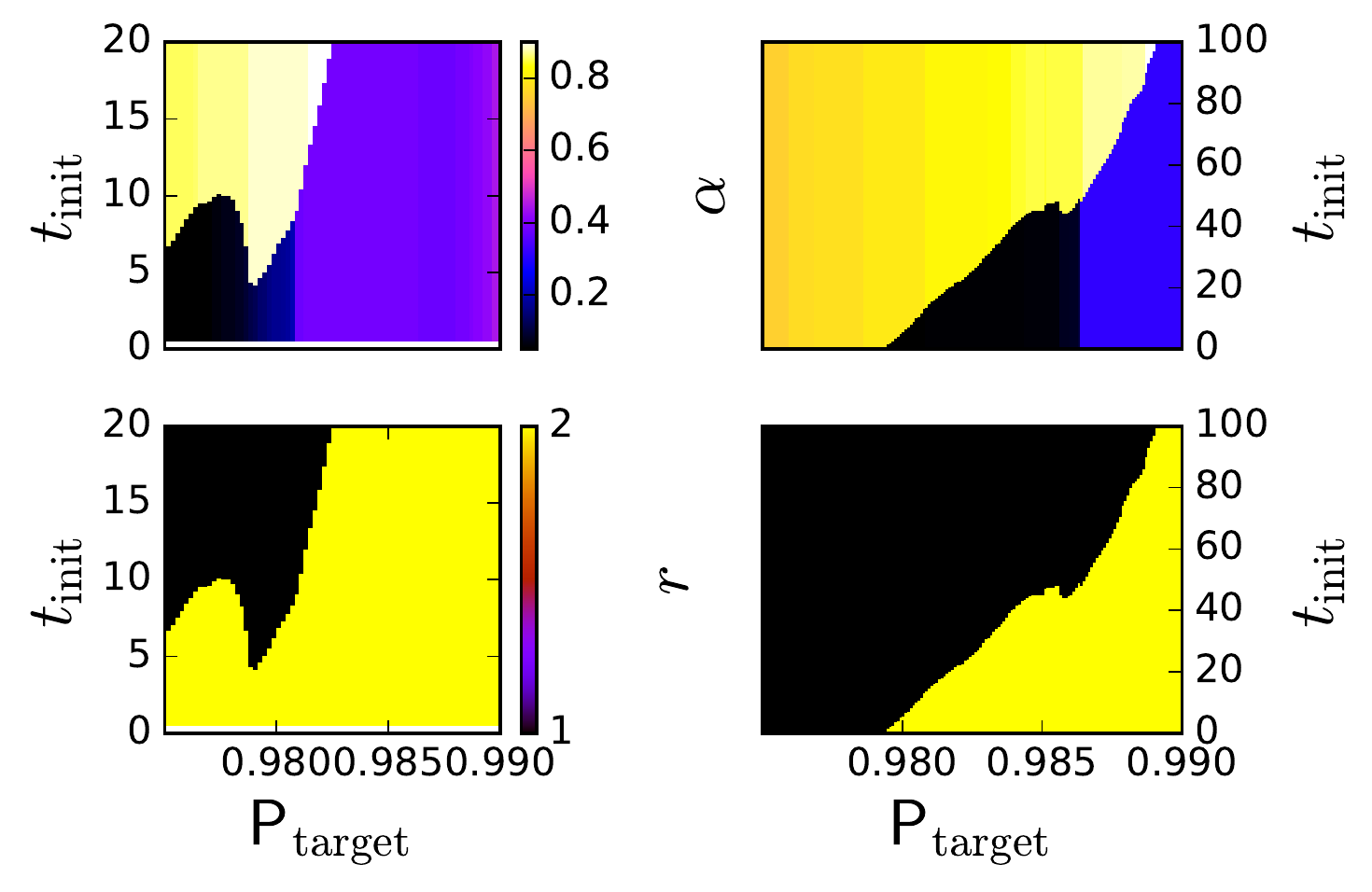}
 \end{centering}
 \caption{(color online)
 Optimal number of runs $r$ (bottom) and optimal $\alpha$ (top) for the 
 numerically optimized strategy $s^{(n)}$ with $n=12$ (left) and $n=14$ (right) 
 qubits versus $t_\text{init}$ and search success probability $\mathsf{P}$. 
 $t_\text{init}$ is in inverse energy units, the same as $t_f$ on other figures.
 \label{fig:num_mult_run} 
 } 
\end{figure}
As examples, we consider $n=12$ and $n=14$ qubits using the numerically 
calculated optimal strategy $s^{(n)}$.  Referring to Fig.~\ref{fig:QWpeak_scale}, 
$n=12$ still shows finite size effects, while $n=14$ is just into the smoothly 
scaling regime.  We find for chosen success probabilities 
in the range $0.95-0.99$, the optimal strategy depends on both $t_{\text{init}}$ 
and $\mathsf{P}_{\text{target}}$ as shown in Fig.~\ref{fig:num_mult_run}. 
For the range of $t_{\text{init}}$ we examine, both sizes show a transition from 
a single run able to reach the required success probability to a region requiring 
two runs.  The single runs are hybrid, becoming progressively more AQC-like as the
required probability increases.  At the point where two runs can do better than 
one AQC run, the two run strategy is much closer to quantum walk, but becomes 
progressively more hybrid as target success probability increases further.  Finite 
size effects are visible for $n=12$ in the non-monotonic shape of the boundary 
between one run and two runs in Fig.~\ref{fig:num_mult_run} (left).  For smaller 
$n<12$, these effects become more complicated, there is no single ``best strategy'' 
for a small search space.  Indeed, we also found that the optimal strategy changes 
significantly when any of the parameters are varied.  The complexity in the optimal 
search strategy for small $n$ is because the two-level approximation does not hold 
well in this regime, and interactions with higher excited states have a 
non-negligible effect.  This suggests that a similarly complex situation will likely 
be present in more sophisticated optimization Hamiltonians, whenever a two-level 
approximation is not valid.

%%%%%%%%%%%%%%%%%%%%%%%%%%%%%%%%%%%%%%%%%%%
\subsection{Noisy quantum searching} \label{sub:open_systems}
%%%%%%%%%%%%%%%%%%%%%%%%%%%%%%%%%%%%%%%%%%%%

Another realistic situation where multiple runs can be helpful is when there is
a significant level of unwanted decoherence or other forms of noise acting on 
the quantum hardware.  In this case, shorter runs that end before decoherence 
effects are too strong, but consequently have lower success probabilities and 
hence need more repeats, may be able to maintain a quantum speed up.  
Decoherence effects on the different AQC and QW mechanisms are analysed in more 
detail in related work \cite{morley17a}, and the effects of noise in AQC 
search have been studied in \cite{vega10a,wild16a}.  Here we focus on hybrid 
algorithms, and the extra options these provide for optimizing the search.

We choose a simple model of decoherence by adding a Lindblad term to the 
von-Neumann equation for the system density operator $\hat\rho(t)$,
\begin{equation} \label{eq:decoherence_rate}
 \frac{\partial\hat\rho(t)}{\partial t} = 
 -\frac{i}{\hbar} [ \hat{H}(t), \hat\rho(t)] + \kappa \mathds{P}[ \hat\rho(t)],
\end{equation}
where $\hat{H}(t)$ is the search Hamiltonian and $\kappa\mathds{P}[\rho(t)]$ is
a decoherence term tuned by a rate $\kappa$.  We choose a form for $\mathds{P}$ 
that uniformly reduces the coherences between states corresponding to vertices of
the hypercube (the computational basis).  This type of decoherence has been 
well-studied in the context of quantum walks 
\cite{Alagic05a,Richter06a,kendon08a} and, for high decoherence rate 
$\kappa\gg\gamma$, can be thought of as continuous measurement in the search 
space resulting in a quantum Zeno effect \cite{misra77a}.  It is equivalent to 
coupling with an infinite temperature bath.

%%%%
Since we now have five parameters to optimize over for a given search size $n$, 
($\mathsf{P},t_f,\alpha,\kappa$ and number of runs $r$), we first consider single 
run searches with success probability $\mathsf{P}(t_f,\alpha,\kappa)$.  
This is the final success probability of a hybrid search specified by $\alpha$ of 
duration $t_f$ in the decoherence model of Eq.~(\ref{eq:decoherence_rate}) with 
decoherence rate $\kappa$.  
We simulate the searches for durations $0\leq t_f\leq 200$, and define the search 
duration $t_o$ that maximizes $\mathsf{P}$ for a particular choice of $\alpha$ and 
$\kappa$.  We also define $\alpha_o$ as the value of $\alpha$ 
which maximizes $\mathsf{P}(t_o,\kappa,\alpha)$, this corresponds to the search 
that reaches highest success probability for a given decoherence rate $\kappa$.  
Note that, for computational reasons, we limited $\alpha$ to the values 
$0.0, 0.1, 0.2\dots 0.9, 1.0$ when performing the maximizations; intermediate values
are of course possible.

\begin{figure}
 \includegraphics[width=\columnwidth,trim =-20 -10 -40 -20]{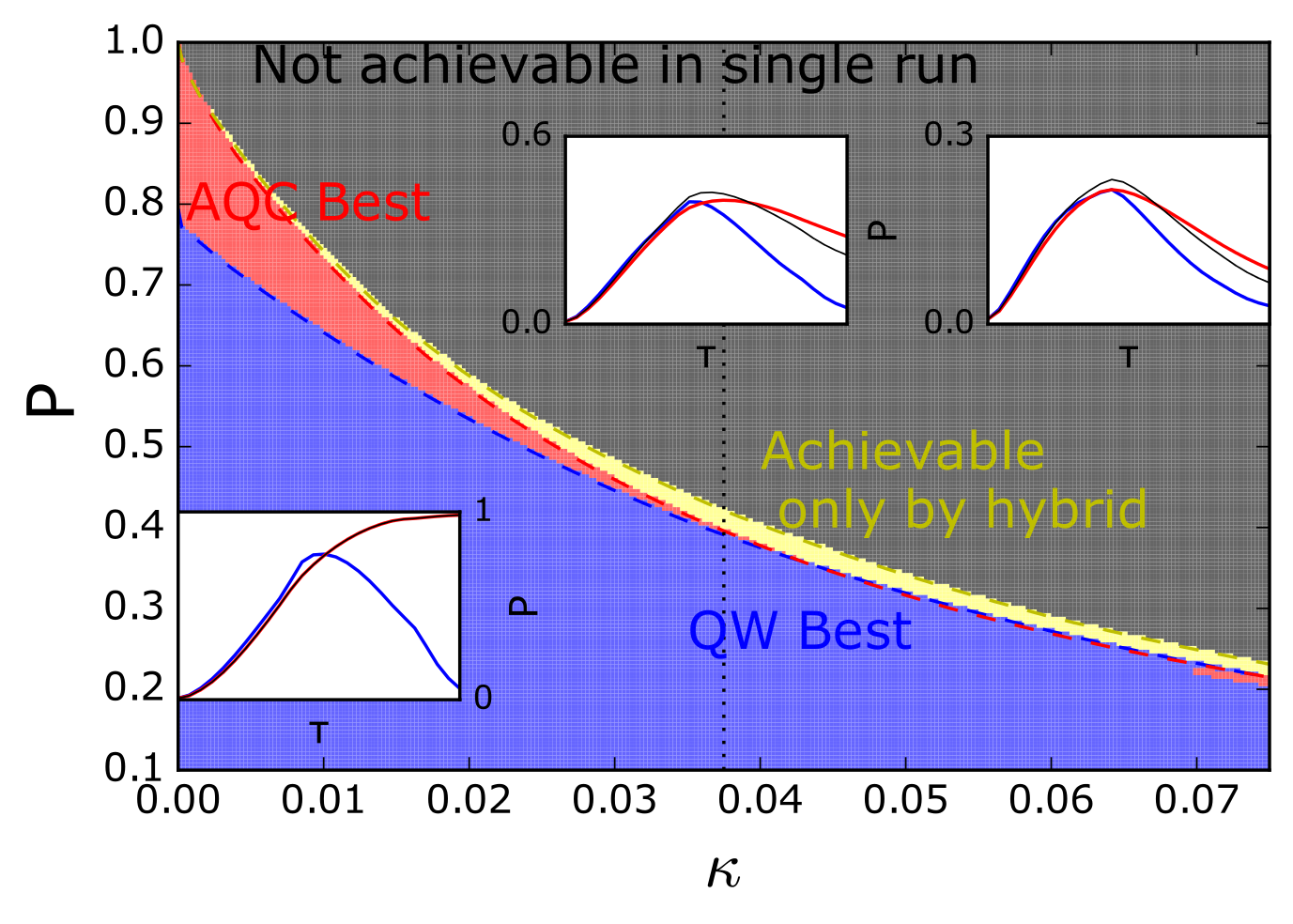}
 \caption{
 (color online)  Main: Shaded regions show fastest protocol: QW (blue), AQC 
 (red), hybrid advantage (yellow), and not achievable in a single run (black). 
 Maximum success probability $\mathsf{P}$ as dashed lines in matching colours, 
 versus decoherence rate $\kappa$, for $n=7$. 
 Insets left to right: $\mathsf{P}$ against reduced time 
 $\tau$ for QW (blue), AQC (red) and the optimal hybrid strategy (thin, black)
 for $\kappa=0$, $0.0385$ (vertical dotted line), $0.075$.  
 \label{fig:evols_and_best_single_run}
 }
\end{figure}

We begin by looking at how the instantaneous success probability 
$\mathsf{P}(t)=\sandwich{m}{\rho(t)}{m}$ evolves during a search, where $m$ 
denotes the marked site. Inset in Fig.~\ref{fig:evols_and_best_single_run} 
are plots of the evolution of 
$\mathsf{P}$ during a search over a $7$-qubit hypercube graph for varying 
decoherence rates $\kappa$, in terms of reduced time $\tau=t/t_f$, for
a $t_f$ that shows the first peak of QW search. The broad effect of the decoherence 
is to reduce the instantaneous success 
probability towards a value of $1/N$, equivalent to classical guessing.
The QW, AQC and hybrid search algorithms retain their characteristics up to an 
overall decoherence damping, which is independent of $\alpha$. As can be seen for
the $\kappa=0$ subplot (left) in Fig.~\ref{fig:evols_and_best_single_run}, the QW
search spreads out more quickly over the search space and therefore exhibits
a more rapid initial increase in $\mathsf{P}$. On the other hand, AQC searching 
can reach higher values of $\mathsf{P}$ for sufficiently
small values of $\kappa$, albeit at later times

The main plot in Fig.~\ref{fig:evols_and_best_single_run} shows which is fastest out
of individual QW, AQC and hybrid searches for a single search,
for a given value of $\kappa$ and of $\mathsf{P}$: from AQC through to hybrid when 
maximal success probability
is required, with QW performing best for slightly lower values of $\mathsf{P}$.
This indicates a remarkably large range of situations where QW dynamics
is desirable - either as part of hybrid algorithms that hit the highest success probabilities
for all but the smallest decoherence rates, or alone in the form of a static Hamiltonian,
if a marginally smaller success probability can be tolerated.
%%%%

Another way to compare the different searches under decoherence is to ask whether a
QW, AQC, or hybrid search will give the maximum possible success probability $\mathsf{P}$
for a given value $\kappa$. The bottom of Fig 15 shows how this maximum $\mathsf{P}$ varies
for the three cases, as well as the value of interpolation parameter $\alpha^{(o)}$ for
the best-case hybrid search. For small values of $\kappa$, $\alpha_o=1$, i.e. AQC gives
the highest peak success probability. As $\kappa$ is increased, the highest-scoring 
search changes and $\alpha_o$ decreases monotonically, indicating hybrid 
searches perform the best overall for intermediate levels of decoherence.  In 
the limit of very high decoherence we are in a quantum Zeno effect regime which 
keeps the search in the initial superposition over all possible states.  This 
means all searches will succeed with the same probability 
$\mathsf{P}=|\langle\psi_\text{init}|m\rangle|^2 = 1/N$, equivalent to classical
guessing. The usefulness of a search is also determined by how quickly it can 
be performed, and so the search time $t_f$ is shown at the top of Fig.~15,
showing that while QW never has the highest success probability in the range we examine, it can be 
substantially quicker.  This helps to explain why
hybrid schedules take on more QW character  as $\kappa$ is increased, and soon begin 
to achieve higher success probabilities than AQC in shorter search times.

%%%%

%

\begin{figure}
  \includegraphics[width=\columnwidth]{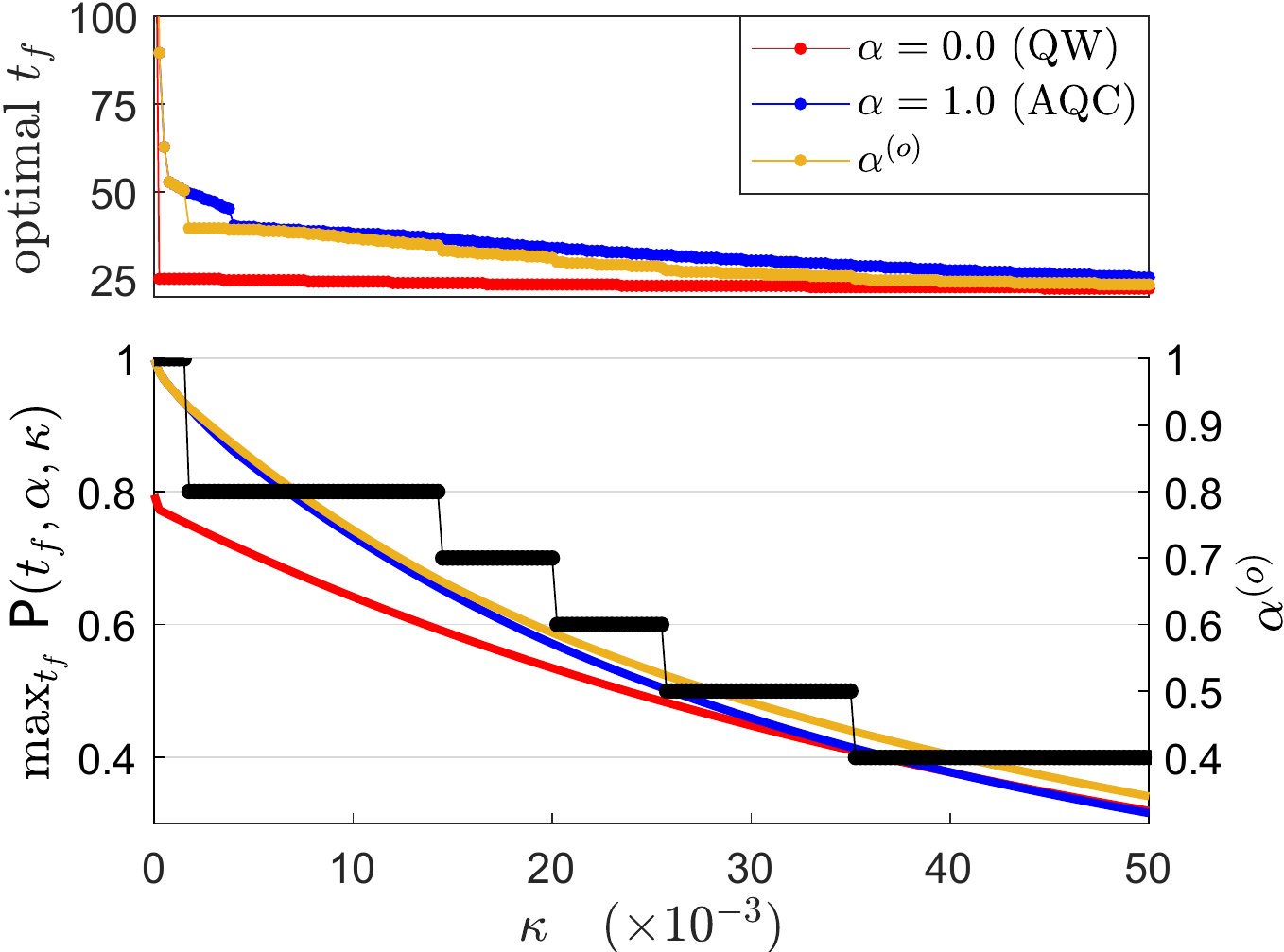}
 \caption{(color online)
 Quantum searching on the $n=7$ hypercube for QW (red, mid gray in print)), AQC (blue, dark gray in print), 
 and the hybrid search which yields the maximum $\mathsf{P}$ (orange, light gray in print) given by
 $\alpha_o$.
 Top: search time $t_o$ which maximizes $\mathsf{P}$ versus $\kappa$. The 
 first data point (not shown) for the AQC and $\alpha_o$ series
 exceeds $t_f = 200$, the upper limit of search times sampled.
 Bottom: search probability $\mathsf{P}(t_f,\alpha,\kappa)$ versus 
 decoherence rate $\kappa$ maximized over
 search times $0\leq t_f\leq 200$ (left axis).  $\alpha_o$ as 
 $\kappa$ varies (black, right axis label). The $\alpha$
 values sampled are $0.0, 0.1,\dots, 1.0$.
 Analytic 
 expression (\ref{eq:analytic_AQC_schedule_1}) used for AQC schedule.
 Time and rate units given by Eq.~(\ref{eq:H_adiabat}).
 \label{fig:singlerun_n=7}
 }
\end{figure}

Having characterized the effects of decoherence on a single run, we now consider
multiple-run search strategies where each search is of the same duration $t_f$.  
We define the optimal annealing schedule as that which minimizes the time taken 
to reach a given success probability, optimized over all equal duration 
multiple-run hybrid search strategies, with durations of individual searches in the 
range $0<t_f\leq 200$.  There are three variables to optimize over: the success 
probability $\mathsf{P}$, the initialization time between searches 
$t_\text{init}$, and the decoherence rate $\kappa$.  We denote the number of runs 
by $r$, so the combined search time is $r t_f$, the combined initialization time is
$r t_\text{init}$, and the total time taken is $r(t_f + t_\text{init})$.

\begin{figure}
 \includegraphics[width = 1.0\columnwidth]{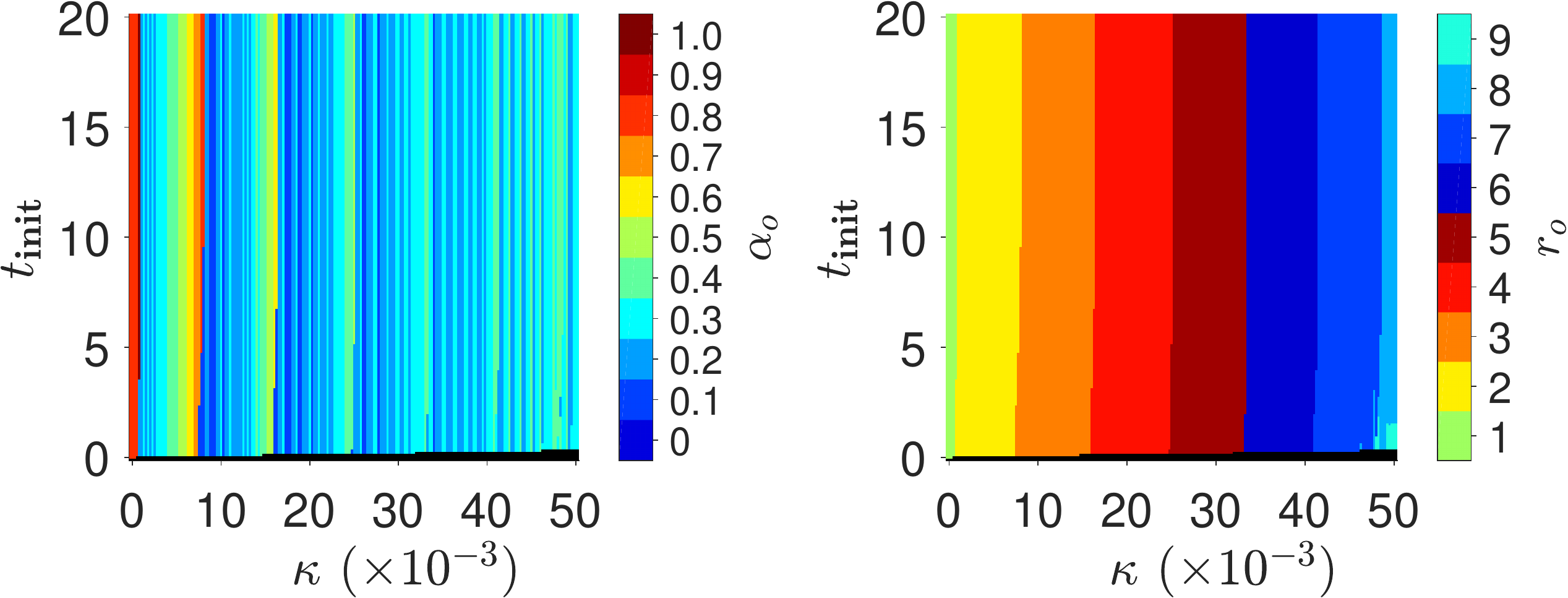}
 \includegraphics[width = 1.0\columnwidth]{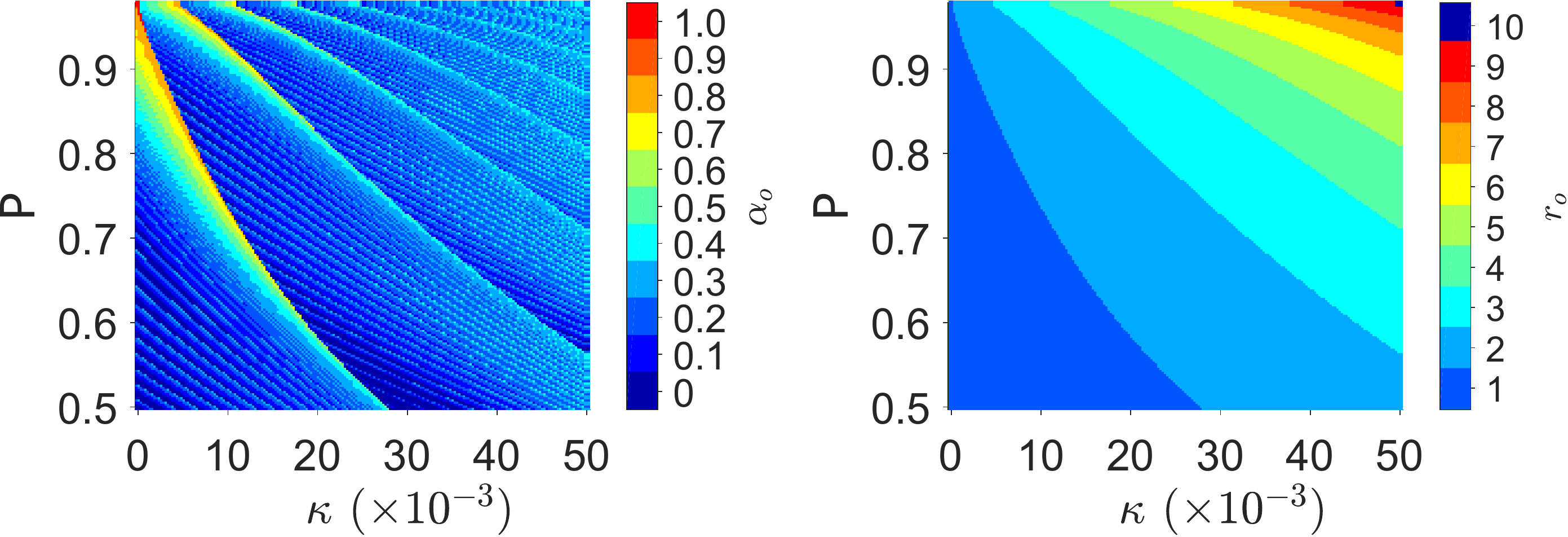}
 \caption{(color online)
 Optimal hybrid search parameter $\alpha_o$ and number of runs $r_o$ for multiple-run
 searching on an $n=7$ hypercube. The optimal search is that which achieves the target
 success probability $\mathsf{P}$ in the shortest total time $r(t_\text{init} + t_f)$,
 where $t_\text{init}$ and $t_f$ are the initialization and run times respectively.
 Top row: dependence on $t_\text{init}$ and decoherence rate $\kappa$ when $\mathsf{P}$
 is fixed to $0.95$. Black indicates the region of instantaneous measuring with $t_f=0$, 
 where $r=382$. Bottom row: dependence on 
 $\mathsf{P}$ and $\kappa$ when $t_\text{init}$ is fixed to 10. LHS plots show $\alpha_o$, RHS
 plots show $r_o$. Analytic 
 expression (\ref{eq:analytic_AQC_schedule_1}) used for AQC schedule. Time and rate units given by Eq.~(\ref{eq:H_adiabat}).
 \label{fig:opt_alpha_1}
 }
\end{figure}

To make this multiple parameter optimization tractable, we considered a discrete
set of values for $\alpha\in\{0.0,0.1,\dots,0.9,1.0\}$, and then minimized the 
total time $r(t_f + t_\text{init})$ while varying $\mathsf{P}$,
$t_\text{init}$ and $\kappa$.
The results can be seen for a $7$-dimensional hypercube in 
Fig.~\ref{fig:opt_alpha_1}, which shows the optimal hybrid schedule $\alpha$ and 
number of runs $r$ taken by the best performing 
multiple-run hybrid search algorithm, as a function of $\kappa$, $t_\text{init}$,
and $\mathsf{P}$.

There is a small threshold initialization time
below which the best strategy is to take multiple measurements of the system state 
as soon as it is 
prepared at a small cost $r t_\text{init}$, indicating that our device can do no better 
than classical random guessing.  Other than this threshold, there is little 
dependence on initialization time. There is a broad tendency towards AQC-like searches
as $\mathsf{P}$ is increased, however for larger values of $\kappa$ an AQC 
search ceases to ever be optimal and hybrid or QW searches are preferred.
As $\kappa$ is increased, there is a localized trend for more AQC-like searches to be
optimal, however this is punctuated with discontinuous changes to a more QW-like search. 
The reason for these discontinuous changes can be seen in the right plots of
Fig.~\ref{fig:opt_alpha_1}. The boundaries where another run is required 
correspond exactly to the regions where the optimal value 
of $\alpha$ suddenly drops.  This transition arises when the decoherence rate
$\kappa$ and/or target success probability $\mathsf{P}$ have increased such that the 
best performing strategy with $r$ searches drops below $\mathsf{P}$,
and another run is required. In this case the target can be reached by $r+1$
lower quality searches. This drop in the quality required of the single search means a
faster, more QW-like search can be used to succeed,
and therefore the optimal value of $\alpha$ drops.

Our numerical results for hybrid algorithms in the presence of noise can be 
understood intuitively by considering how $\mathsf{P}$ 
scales with a small amount of noise in the AQC and QW edge cases.
For noise rate $\kappa$ per unit time, the success probability for a 
single run reduces as $\mathsf{P}\simeq \exp(-\kappa t_f)$, 
where $t_f$ is the time taken for one run of the search algorithm.
For $\mathsf{P}\sim 1$ we thus require $\kappa t_f\ll 1$, 
i.e., $\kappa \ll 1/t_f$.
For QW searching on the hypercube, we have 
$t_f^{(\text{QW})}\simeq\frac{\pi}{2}\sqrt{N}$, hence we obtain 
$\kappa_{\text{QW}}\ll 2/(\pi\sqrt{N})$ for tolerable noise rates.
For AQC on the other hand, from Eqn.~(\ref{eq:tfTLAoptc}) we have 
$t_f^{(\text{AQC})}\simeq \frac{\pi}{4\epsilon}\sqrt{N}$ for large $N$.  
For high success probability, since 
$\mathsf{P}\sim 1-\epsilon$, the adiabatic condition requires $\epsilon\ll 1$ 
and we obtain $\kappa_{\text{AQC}}\ll 4\epsilon/(\pi\sqrt{N})$.
The extra factor of $\epsilon$ implies $\kappa_{\text{AQC}}\ll 
\kappa_{\text{QW}}$.  
Hence, QW search will be more robust to disturbance by noise, as we have found 
numerically for the single run case.
For our $n=7$ example, $\kappa_{\text{QW}}\ll 0.056$ and 
$\kappa_{\text{AQC}}\ll 0.11\epsilon$ = 0.011 for $\mathsf{P}=0.99$,
 and indeed we see in 
Fig.~\ref{fig:evols_and_best_single_run} that performance drops below 
$\mathsf{P}=0.5$ for $\kappa_{\text{QW}}\gtrsim 0.025$.
However, when multiple runs are included, hybrid strategies with significant 
adiabatic character can still outperform QW, depending on hardware 
characteristics determining the initialization and measurement time required 
per run.

%%%%%%%%%%%%%%%%%%%%%%%%%%%%%%%%%%%%%%%%%%%%%%%%%%%%%%%%%%%%%%%%%%%%%%%%%%%%%%%
\section{Problem misspecification\label{sec:problem_misspec}}
%-----------------------------------------------------------------------------

So far we have studied the dynamics of the hybrid search Hamiltonians, as part of 
single and multiple run algorithms, and in the presence of noise. In the following 
section we consider misspecification of the problem, for which the 
dynamics remain coherent, but some parameters are changed in unknown ways.

%%%%%%%%%%%%%%%%%%%%%%%%%%%%%%%%%
\subsection{Motivation}
%%%%%%%%%%%%%%%%%%%%%%%%%%%%%%%%%

\begin{figure}
 \begin{centering}
  \includegraphics[width=0.9\columnwidth]{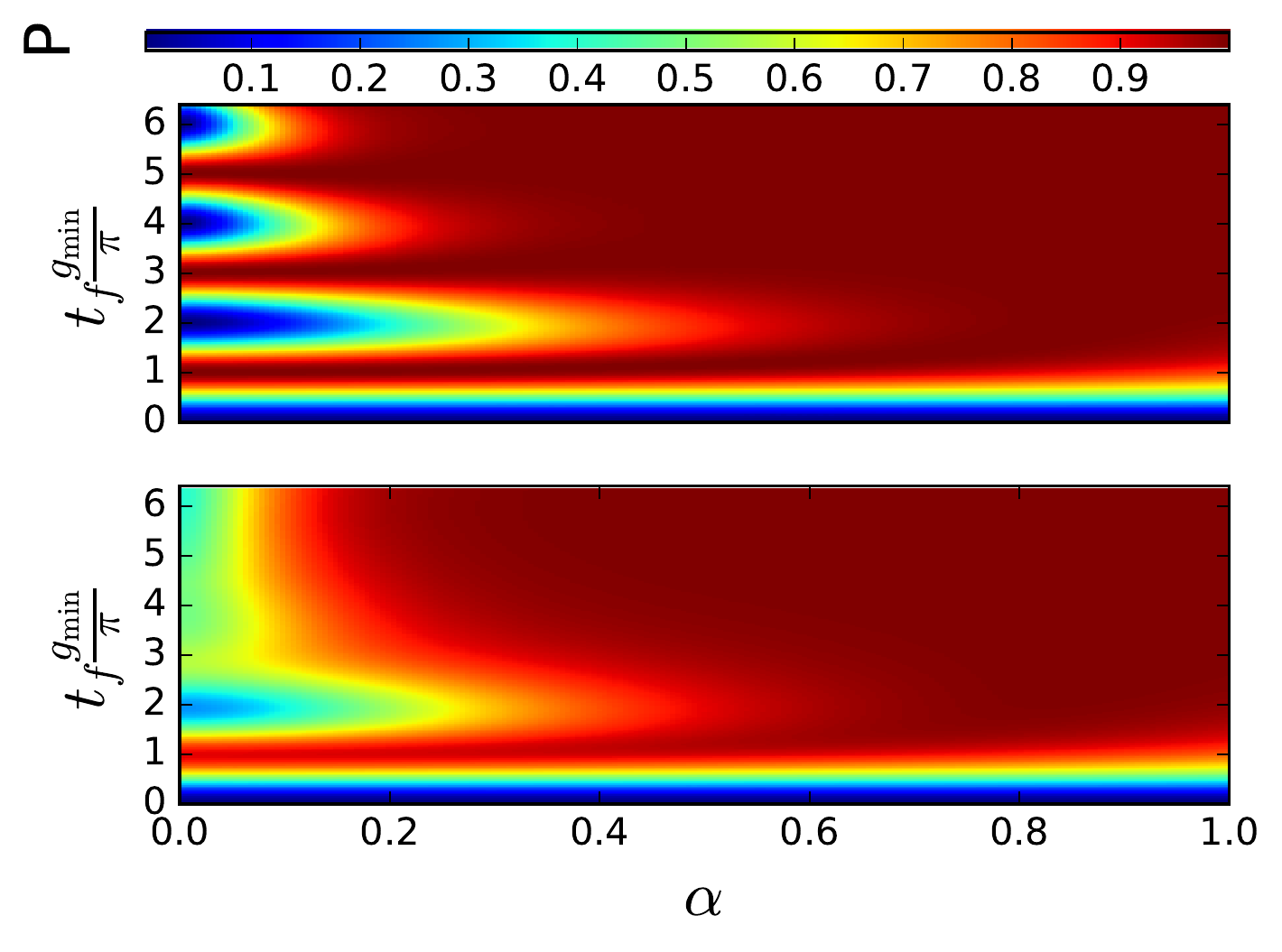}
 \end{centering}
 \caption{Top: Probability $\mathsf{P}$ of finding the marked state versus 
 runtime and $\alpha$ for the single avoided crossing model, same as 
 Fig.~\ref{fig:ac_plot_noErr}.   
 Bottom: as above with a $30\%$ misspecification of the energy 
 $\Delta g_{\text{min}}$. 
 \label{fig:ac_plots}
 }
\end{figure}
Studying the effects of problem misspecification is particularly relevant given 
the critical difficulties which many 
classical analog computing efforts have faced due to propagation of errors 
\cite{bissell04a}.  Misspecifications can come about in a variety of ways, such 
as limited precision for setting the controls in the computer, ignorance of 
what the optimal parameters should be, or noise which is at a much lower frequency 
than the rate of the relevant quantum dynamics.  An important example of the latter 
is so-called $\frac{1}{f}$ noise in superconducting qubit devices 
\cite{koch1983,koch2007}, such as the quantum annealers constructed by D-Wave 
Systems Inc.  It has been shown, for instance, that such misspecifications can 
cause AQC to give an incorrect solution on Ising spin systems \cite{young13a,albash18}, 
and it effectively limits the maximum useful size of such devices.  For an example of
the effects of problem misspecification on a real experiment, see \cite{chancellor16b}.

For this work, we will consider simple misspecification models in the large 
system limit, where the Hamiltonian can be mapped to a single avoided crossing 
in the form of Eq.~(\ref{eq:H_ac_s}) or (\ref{eq:H_ac}). For the purpose of 
studying problem misspecification, it is most convenient to work with the form 
in Eq.~(\ref{eq:H_ac}), which we use for the duration of this section. Because the initial
and marked states are orthogonal in this limit, considering multiple runs which 
can be performed with negligible initialization time is not mathematically 
pathological.  Furthermore, physically, we expect initialization and readout time
to scale, at worst, polynomially with $n$, while runtime will scale as 
$\sqrt{N}\propto 2^{\frac{n}{2}}$.  Therefore, in the large $N$ limit, it is a
natural physical assumption that $t_f \gg t_{init}$.  We first examine the 
effect of having the size of the minimum gap be misspecified, so that we do not 
know when to measure for QW protocols, and then examine the effect of not knowing the 
position of the avoided crossing, which will cause QW protocols to use the wrong
value of $\gamma$ and AQC protocols to slow down at the wrong point.

%%%%%%%%%%%%%%%%%%%%%%%%%%%%%%%%%%%%%%%%%%%%%%%%%
\subsection{Error in gap size}\label{sub:gap_err}
%%%%%%%%%%%%%%%%%%%%%%%%%%%%%%%%%%%%%%%%%%%%%%%%%

The effect of misspecifying the size of the minimum gap can be modelled 
as an uncertainty in the total energy scale $\Delta g_{\text{min}}$, which 
is equivalent to a misspecification of the total runtime $t_f$ through 
Eq.~(\ref{eq:H_ac}).  The effect of uncertainty can be modelled by performing a convolution of
the success probability versus runtime with a distribution describing the uncertainty. An example
result of such a convolution is depicted in Fig.~\ref{fig:ac_plots} bottom.
Assuming that the misspecification is distributed in a Gaussian manner around 
the intended runtime, the new success probability for a given anneal time 
$t_f$ and $\alpha$ becomes 
\begin{align}
&\mathsf{P}(t_f,\alpha,\Delta g_{\text{min}}) = \nonumber\\
&\int_{-\infty}^{\infty} dt_f' 
\frac{\mathsf{P}(|t_f'|,\alpha)}{\Delta g_{\text{min}}\sqrt{2\,\pi}
}\exp\left\{-\frac{(t_f'-t_f)^2}{2(\Delta g_{\text{min}}t_f)^2}\right\},
\label{eq:gap_mis}
\end{align}
where $\Delta g_{\text{min}}$ is the (unitless) fractional uncertainty in 
$g_{\text{min}}$, and the absolute value in the argument of $\mathsf{P}$ within 
the integral is included to avoid negative time arguments.  For reasonable values 
of $\Delta g_{\text{min}}$, it will be rare for $t_f'<0$ and the effect of taking 
the absolute value will be negligible.

Fig.~\ref{fig:ac_plots} shows how the evolution makes a smooth transition 
between the characteristically sinusoidal behavior of success probability 
versus runtime for QW, and the characteristically  monotonic 
behavior of AQC.  As the comparison between the perfect and misspecified cases
demonstrates, gap misspecification causes a large reduction in the success 
probability of QW protocols, but has  almost no effect on the monotonic 
AQC search.
\begin{figure}
 \begin{centering}
  \includegraphics[width=0.9\columnwidth]{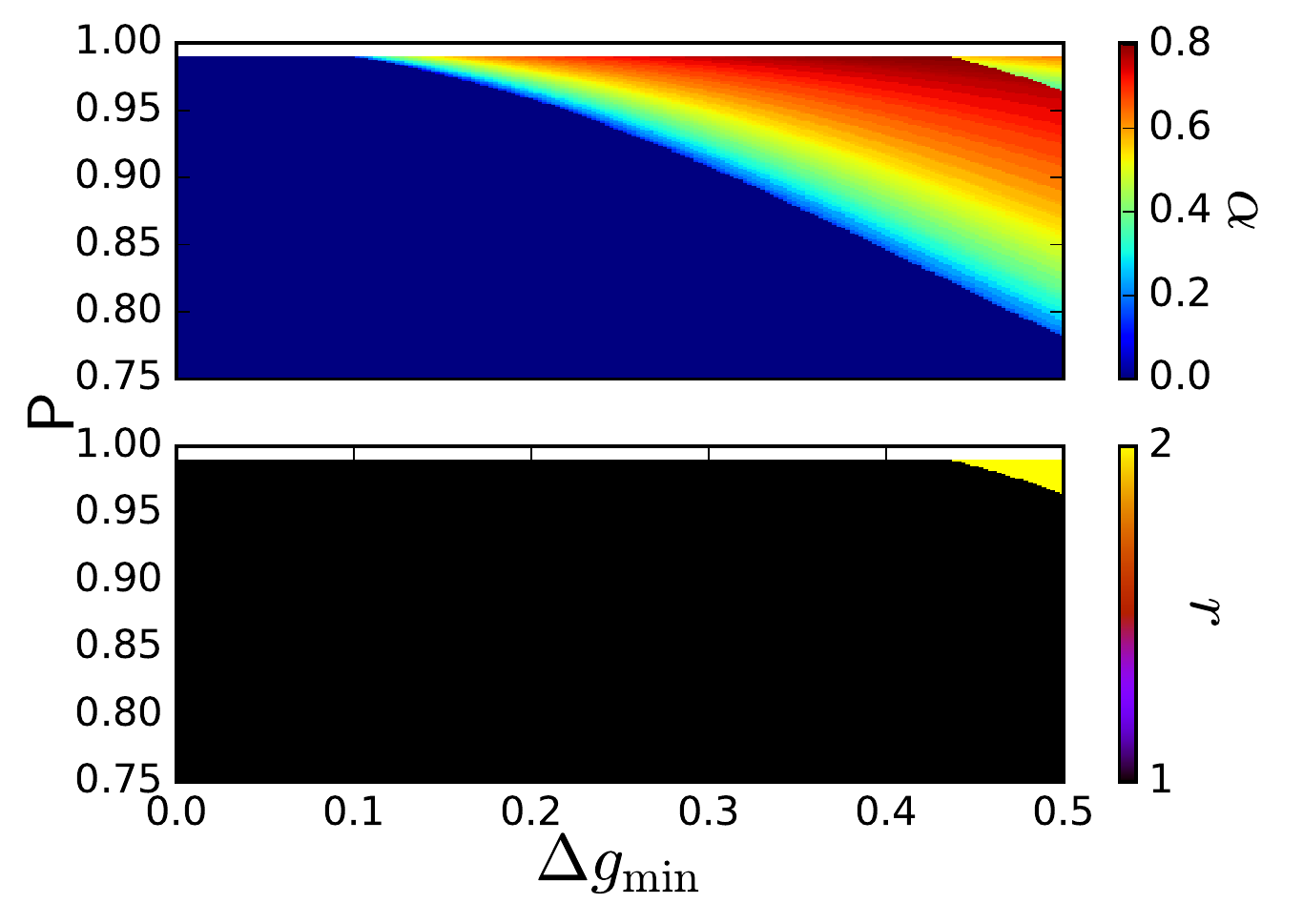}
 \end{centering}
 \caption{(color online)
 Top: Optimal value of $\alpha$ versus success probability $\mathsf{P}$ and 
 $\Delta g_{\text{min}}$ from Eq.~(\ref{eq:gap_mis}). 
 Bottom: number of repeats $r$ in optimal strategy versus $\mathsf{P}$ and 
 $\Delta g_{\text{min}}$. 
 \label{fig:gap_mis}
 }
\end{figure}
Fig.~\ref{fig:gap_mis} illustrates that, for moderately high success 
probability and moderate amounts of misspecification of $\Delta g_{\text{min}}$, the best
protocol is no longer QW, but lies in between the optimal AQC 
schedule and QW.  For large gap misspecification where a high 
success probability is required, the best approach is to run an intermediate 
strategy twice.

The reason that gap size misspecification makes hybrid protocols ($\alpha>0$) 
outperform QW for a large range of parameter space is because a QW can only 
succeed with a probability approaching one if $t_f\,g_{\text{min}}$ is an odd 
multiple of $\pi$.  The misspecification smears out these peaks and implies that
the success probability of a QW will not approach one for any value of  $t_f$. 
For protocols with some adiabatic character, however, the maximum success 
probability will still approach one as $t_f$ becomes larger, as the adiabatic 
theorem holds for any finite gap. In cases where the misspecification overstates
the size of the gap the success probability of AQC will actually improve.

%%%%%%%%%%%%%%%%%%%%%%%%%%%%%%%%%%%%%%%%%%%%%%%%%
\subsection{Error in avoided crossing location}
%%%%%%%%%%%%%%%%%%%%%%%%%%%%%%%%%%%%%%%%%%%%%%%%%

\begin{figure}
 \begin{centering}
  \includegraphics[width=0.9\columnwidth]{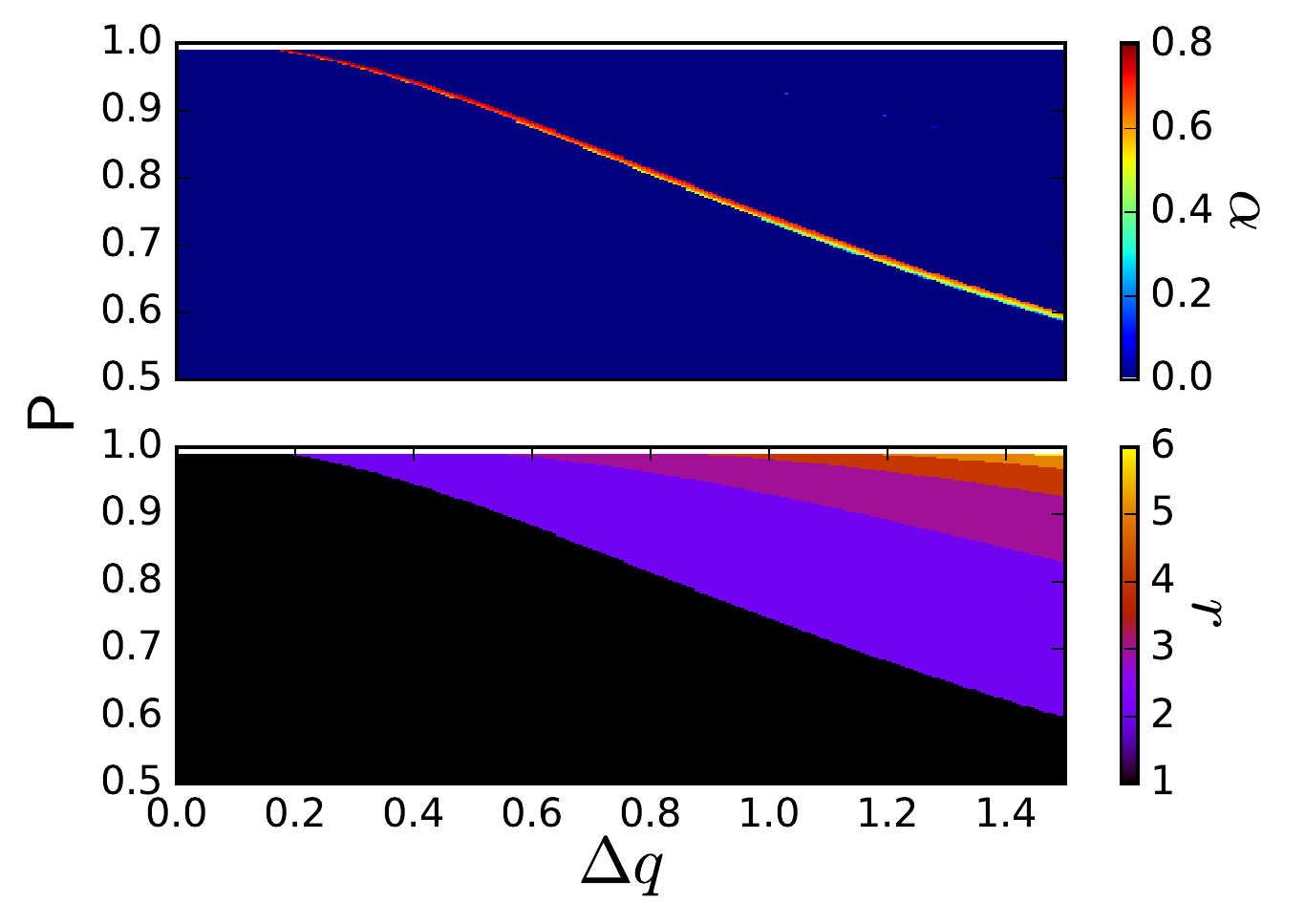}
 \end{centering}
 \caption{(color online)
 Top: Optimal value of $\alpha$ versus success probability $\mathsf{P}$ and 
 $\Delta q$ from Eq.~(\ref{eq:pos_mis}). 
 Bottom: number of repeats $r$ in optimal strategy versus $\mathsf{P}$ and 
 $\Delta q$. 
 \label{fig:pos_mis}
 }
\end{figure}

Another type of problem misspecification  is incorrectly specifying the position
of the avoided crossing.  To model this, we consider a modification of the 
problem Hamiltonian 
\begin{equation}
 \hat{H}^{(\text{AC})\prime}(t,q) = \hat{H}^{(\text{AC})}(t)+\frac{q}{2} g_{\text{min}}
  \hat{\sigma}_z.
\end{equation}
This addition to the problem Hamiltonian provides a shift in the avoided crossing 
position $f(t)\rightarrow f(t)+q$ in Eq.~(\ref{eq:H_ac}). Effectively introducing this shift causes 
the schedule to slow down at the wrong point, reducing the success probability. 
As we did for the case of gap mis-specification, we can model the effect of this error as a 
convolution of the success probability distribution with $q$ with a Gaussian of width $\Delta q$. 
We define the success probability with misspecified avoided 
crossing position as
\begin{equation}
\mathsf{P}(t,\alpha,\Delta q)=
\int_{-\infty}^{\infty} dq \frac{\mathsf{P}(t,\alpha,q)}{\Delta q\,\sqrt{2\,\pi
}}\,\exp\left(-\frac{q^2}{2\,(\Delta q)^2 }\right),
\label{eq:pos_mis}
\end{equation}
where $\Delta q$ is the (unitless) fractional uncertainty in $q$, that controls 
the degree of misspecification.  Figure \ref{fig:pos_mis} illustrates that, in 
contrast to gap misspecification, the best strategy is almost always QW.  
Intermediate strategies only become the superior method briefly, at the edge of 
the regime where single runs are the best way to reach the desired probability.  
At higher misspecification, multiple repeated QW become the best strategy.  

Misspecification in the avoided crossing position does 
significant harm to both AQC and QW protocols. The success probability of a QW 
protocol performed with an incorrectly chosen $\gamma$ does not approach one. 
Similarly, an AQC protocol with a poorly chosen schedule will require a much 
longer runtime for the success probability to approach one.  The faster runtime
of QW then means it beats AQC for multiple runs.

%%%%%%%%%%%%%%%%%%%%%%%%%%%%%%%%%%%%%%%%%%%%%%%%%%%%%%%%%%%%%%%%%%%%%%%%%%%%%%%%%%%%%%%%%
\section{Summary and outlook\label{sec:conclusion}}
%---------------------------------------------------------------------------------------%

In this paper we provide a detailed study of the scaling of continuous-time 
quantum search algorithms on a hypercube graph.  Noting that both quantum walk 
and adiabatic quantum search algorithms can be expressed as two extremes of 
quantum annealing schedules, we define a family of quantum search algorithms 
that are hybrids between QW and AQC.  By mapping the algorithms to a one qubit 
single avoided crossing model, we show that the whole family achieves the 
maximum possible $\sqrt{N}$ quantum speed up.  There are a number of subtleties 
in the scaling behavior on the hypercube that we treat in detail for short 
search times, complementing the work by Weibe and Babcock \cite{weibe12a} on 
long timescales. 

Our hybrid QW-AQC schedules are an example of the advantages we gain by 
treating both QW and AQC as part of the same method of continuous-time 
quantum computing \cite{kendon17a}.  We find that hybrid strategies 
intermediate between QW and AQC provide the best quantum search algorithm 
under a range of realistic conditions.  The techniques we use here can easily 
be extended to hybrid quantum search on other graphs, and to other quantum 
walk or adiabatic quantum computing algorithms.

This work focused on the search problem due to its relative simplicity, 
and the fact that annealing schedules can be derived analytically -- which we 
do for the hypercube graph in appendix \ref{app:sched_analytics}.  The core 
ideas and methods are quite general and can easily be extended to more complex 
and realistic problems, such a `fixed point search', where multiple states 
are marked.  Fixed point search algorithms have been studied in both the QW 
\cite{yoder14a} and AQC \cite{dalzell17a} regimes, so interpolation to 
generate hybrid algorithms should be straightforward.  The quantum walk 
search on random graphs solved in \cite{Chakraborty16a} is based on the same
kind of single avoided crossing arguments which appear in this work, meaning 
that these are also natural for hybrid QW-AQC protocols.

Hybrid algorithms such as the ones we present here can be viewed as particular 
instances of quantum control techniques applied to solving optimization and 
search problems.  Another application of quantum control to quantum algorithms 
is based on the Pontryagin minimum principle of optimal control: that optimal 
control protocols for solving these problems will follow a bang-bang scheme, 
with successive applications of the extreme values of the controls 
\cite{Yang17a}.  An algorithm based on such controls, called the Quantum 
Approximate Optimization Algorithm (QAOA), was first proposed by Farhi, 
Goldstone, and Gutmann \cite{Farhi14a,Farhi14b}.  This protocol can be 
implemented either through digital quantum circuits, or by successively applied 
Hamiltonians.  It has been shown that the QAOA can obtain an optimal $\sqrt{N}$ 
scaling in solving the search problem using a transverse field search 
unitary \cite{Jiang17a}, essentially the problem we consider in this paper.

However, there are two caveats worth noting in terms of the optimality of QAOA 
type bang-bang protocols.  Firstly, when viewed as an application of successive 
Hamiltonians, these protocols require infinitely fast switching time, which is 
generally unphysical.  Secondly, while the optimal control scheme to find the 
solution is mathematically always of a bang-bang form, this solution may exhibit
Fuller's phenomenon \cite{Borisov00a,Fuller60a}, in which the optimal solution 
involves switching back and forth between the two extremal Hamiltonians an 
infinite number of times in a finite time window.  While mathematically valid, 
such a control scheme is clearly not physically realizable.  It is an open 
question what happens to Hamiltonian-based QAOA when finite switching time is 
added as a constraint.  Our result that intermediate protocols between quantum 
walk and adiabatic protocols are still able to obtain an optimal speed up 
provide an encouraging sign that QAOA may remain effective with realistic 
constraints applied.

Recent studies by Muthukrishnan et al.~\cite{muthukrishnan15a,Muthukrishnan2016} 
on a class of permutation symmetric problems related to, but distinct from, 
search, have found that, deep in the diabatic regime, the problem can be 
solved by dynamics which are effectively classical through `diabatic cascades'.
Muthukrishnan et al.~focus only on changing the rate of evolution of an AQC 
algorithm;  in contrast, we examine both the shape of the schedule and the rate
of evolution.  Furthermore, since all of the qubits need to align to interact 
meaningfully with the energy landscape of the search problem, it is unlikely 
that a similar classical diabatic cascade regime exists in our study.

As well as problem size, the performance of a quantum search in a realistic 
setting will depend on many other factors.  By performing a fairly general and 
multi-faceted analysis of such factors, we uncover a landscape where no single 
protocol dominates.  In asymptotically large systems with perfectly specified 
problems, a straightforward QW approach is best.  However, this limit is 
approached slowly, since the success probability for QW scales only as $n$, 
i.e., logarithmically in problem size $N$.   A rich structure exists for 
computationally interesting, non-asymptotic sizes.  On the other hand, for 
asymptotically large systems with some degree of problem misspecification, 
interpolated protocols can outperform the QW approach.  A simple open systems 
analysis reveals another layer of structure that can be exploited in realistic
settings.  For more discussion on the effects of noise and the competition 
between the mechanisms, see our related work \cite{morley17a}.  In future work
we will apply these techniques to algorithms with useful applications which can 
be run on near-future quantum hardware \cite{callison18a}.

%%%%%%%%%%%%%%%%%%%%%%%%%%%%%%%%%%%%%%%%%%%%%%%%%%%%%%%%%%%%%%%%%%%%%%%%%%%%%%%
\acknowledgments
JGM is supported by the UK Engineering and Physical Sciences Research Council 
Grant EP/L015242/1.
NC and VK were supported by the UK Engineering and Physical Sciences Research 
Council Grant EP/L022303/1.  
SB has received funding for this research from the European Research Council 
under the European Union's Seventh Framework Programme (FP7/2007-2013)/ERC 
Grant agreement No.~308253 PACOMANEDIA.

%%%%%%%%%%%%%%%%%%%%%%%%%%%%%%%%%%%%%%%%%%%%%%%%%%%%%%%%%%%%%%%%%%%%%%%%%%%%%%%
\appendix
%%%%%%%%%%%%%%%%%%%%%%%%%%%%%%%%%%%%%%%%%%%%%%%%%%%%%%%%%%%%%%%%%%%%%%%%%%%%%%%
\section{Hypercube optimal schedule calculation}\label{app:sched_analytics}
%-----------------------------------------------------------------------------%

Starting from the Hamiltonian for the AQC search on a hypercube,
Eq:~(\ref{eq:HhAQC})
\begin{equation}
\hat{H}(s) = (1-s)\sum_{j=1}^{n}\frac{1}{2}(1-\hat{\sigma}^{(j)}_x) 
           + s(\openone-\ketbra{m}{m}), \nonumber
\end{equation}
we first apply a gauge transformation (a swap of the $1\leftrightarrow 0$ labels
on a subset of the qubits) to map the marked state $\ket{m}$ to the state 
$\ket{0}$.  We then express the Hamiltonian in the symmetric subspace in terms 
of total spin operators
\begin{equation}
\hat{S}_a = \frac{1}{2}\sum_{j=1}^n \hat{\sigma}^{(j)}_a
\end{equation}
for $a\in\{x,y,z\}$, which have eigenstates $\ket{\frac{n}{2}-r}_a$ for 
$r\in\{0,\dots,n\}$.  In this representation, the marked state is
$\ket{\frac{n}{2}}_z$, and the AQC search Hamiltonian becomes
\begin{equation}\label{eq:aqc_line}
\hat{H}(s) = (1-s)(\tfrac{n}{2}-\hat{S}_x) 
           + s(\openone-\ket{\tfrac{n}{2}}_z\bra{\tfrac{n}{2}}).
\end{equation}

Following Farhi et al \cite{farhi00a} to analyze the eigensystem, we obtain the 
eigenvalue equation
\begin{equation}\label{eq:eigen}
\frac{1-s}{s} = \frac{1}{N}\sum_{r=0}^n \bin{n}{r}\frac{1}{r-\lambda}
\end{equation}
for the energy eigenvalues $E_k=s +(1-s)\lambda_k$.  Farhi et al \cite{farhi00a}
solve this at the minimum gap, which occurs at $s=s_m$ for
\begin{equation}\label{eq:s_gap}
\frac{1-s_m}{s_m} = \frac{1}{N}\sum_{r=1}^n \bin{n}{r}\frac{1}{r}\equiv R_1,
\end{equation}
and show that $\lambda^{(\text{gmin})}_{1,0} \simeq \pm n/(2\sqrt{N})$ for the 
two lowest eigenvalues corresponding the the ground state $E_0$ and first 
excited state $E_1$.

To obtain the optimal schedule following the method in Roland and Cerf 
\cite{roland02a}, we need an expression for the gap as a function of $s(t)$, 
not just at the minimum gap.  We expand the eigenvalue equation (\ref{eq:eigen})
for $\lambda \ll 1$
\begin{equation}
\frac{1-s}{s} = \frac{-1}{N\lambda} 
    + \frac{1}{N} \sum_{r=1}^n \bin{n}{r} \frac{1}{r} (1+\lambda/r)
    + O(\lambda^2).
\end{equation}
Using $R_1$ and $R_2$ from Eqs.~(\ref{eq:gamma_hopt}), (\ref{eq:s_gap}) and 
(\ref{eq:R2}) we obtain
\begin{equation}
\frac{1-s}{s} = \frac{-1}{N\lambda} + R_1 + \lambda R_2.
\end{equation}
This quadratic equation in $\lambda$ has roots
\begin{align}
\lambda = &\frac{1}{2R_2}\left\{\frac{1-s}{s} -R_1\right\}\nonumber \\
    &\pm \frac{1}{2}\left\{\frac{1}{R_2^2}\left(\frac{1-s}{s} -R_1\right)^2
        + \frac{4}{NR_2}\right\}^{\frac{1}{2}}
\end{align}
and gives for the gap $g(s) = (1-s)(\lambda_1-\lambda_0)$
\begin{equation}\label{eq:g_s}
g(s) = (1-s)\left\{\frac{1}{R_2^2}\left(\frac{1-s}{s} -R_1\right)^2
     + \frac{4}{NR_2}\right\}^{\frac{1}{2}}.
\end{equation}

To optimize the schedule, we need to solve Eq.~(\ref{eq:adiabat_cond})
\begin{equation}
\left\lvert\frac{ds}{dt}\right\rvert \le \epsilon 
	\frac{g^2(s)}{\left\lvert \left\langle \frac{d\hat{H}}{ds}\right\rangle_{0,1} \right\rvert} \nonumber
\end{equation}
using the expression for $g(s)$ in Eq.~(\ref{eq:g_s}).
To obtain a suitable approximate value for 
$\langle \frac{d\hat{H}}{ds} \rangle_{0,1}$, we first calculate 
$\frac{d\hat{H}}{ds}$ in the symmetric subspace representation of 
Eq.~(\ref{eq:aqc_line}),
\begin{equation}
\frac{d\hat{H}}{ds} = -(\tfrac{n}{2} - \hat{S}_x) 
                      +(\openone - \ket{\tfrac{n}{2}}_z\bra{\tfrac{n}{2}}).
\end{equation}
It is sufficient to use the maximum value of 
$\langle \frac{d\hat{H}}{ds} \rangle_{0,1}$, which occurs at $s_m$, where the 
eigenstates 
$\ket{E_{1,0}}\simeq (\ket{\frac{n}{2}}_x \pm \ket{\frac{n}{2}}_z)/\sqrt{2}$,  
giving $\langle \frac{d\hat{H}}{ds} \rangle_{0,1}^{(\text{max})} \le \frac{n}{4}$.  
We then have the following equation to solve for $s(t)$
\begin{equation}
\frac{ds}{dt} = \frac{4\epsilon}{nR_2^2}(1-s)^2
    \left\{\left(\frac{1-s}{s} -R_1\right)^2 + \frac{4R_2}{N}\right\}
\end{equation}
This can be integrated to obtain
\begin{align}
\frac{4\epsilon t}{nR_2^2} +c 
 &= \frac{R_1^2 - 4R_2/N}{2\sqrt{R_2/N}(R_1^2 + 4R_2/N)^2} \nonumber \\
 &  \arctan\left\{\frac{((1+R_1)^2+4R_2/N)s - (1+R_1)}{2\sqrt{R_2/N}}\right\}\nonumber \\
 &+ \frac{1}{(1-s)(R_1^2 + 4R_2/N)}\nonumber \\
 &+ \frac{R_1}{(R_1^2 + 4R_2/N)^2}\ln\left\{\frac{(1-s -R_1s)^2 + \frac{4R_2}{N}s^2}{(1-s)^2}\right\}
\end{align}
where $c$ is the constant of integration.  To obtain the constant, set $s=t=0$, giving
\begin{equation}
c' = \arctan\left\{\frac{(1+R_1)\sqrt{N}}{2\sqrt{R_2}}\right\}
   + \frac{R_1^2 + 4R_2/N}{R_1^2 - 4R_2/N}\frac{2\sqrt{R_2}}{\sqrt{N}}
\end{equation}
where the factors in front of the arctan term have been rearranged to give a 
more convenient form for the constant.  One can then in principle solve for
$s(t)$.  However, the terms on the r.h.s., apart from the arctan, are 
potentially problematic as $s\rightarrow 1$.  Given that we started with the 
approximation $\lambda\ll 1$, which occurs at the position of the minimum gap, 
we can't necessarily expect that the solution will be valid for 
$s\rightarrow 1$.  We first note that taking only the arctan 
term on the r.h.s.~gives a schedule that is valid for all $0\le s\le 1$, and 
it provides a runtime proportional to $\sqrt{N}$.  If we don't discard these 
extra terms, we can show that they can be neglected, provided we stop the anneal
very slightly before $s=1$, but still well past the minimum gap.

To solve for $s(t)$ retaining the full expression, invert the arctan to give
\begin{align}
s(t) = &\frac{2\sqrt{R_2}}{\sqrt{N}\{(1+R_1)^2 +4R_2/N\}} \times \nonumber \\
     & \tan\left\{\frac{8\epsilon t\sqrt{R_2}}{nR_2^2\sqrt{N}}\frac{R_1^2+4R_2/N}{R_1^2-4R_2/N} - c''\right\} \nonumber \\
     &+ \frac{1+R_1}{(1+R_1)^2+4R_2/N},
\end{align}
where $c''$ now contains the awkward extra terms,
\begin{align}
c'' = &c' - \frac{1}{(1-s)}\frac{R_1^2 + 4R_2/N}{R_1^2 - 4R_2/N}\frac{2\sqrt{R_2}}{\sqrt{N}}\nonumber \\
	  &- \frac{R_1}{R_1^2 - 4R_2/N}\frac{2\sqrt{R_2}}{\sqrt{N}}\ln\left\{\frac{(1-s -R_1s)^2 + \frac{4R_2}{N}s^2}{(1-s)^2}\right\} \nonumber \\
	= &\arctan\left\{\frac{(1+R_1)\sqrt{N}}{2\sqrt{R_2}}\right\} \nonumber \\
 	  &+ \frac{s}{(1-s)}\frac{R_1^2 + 4R_2/N}{R_1^2 - 4R_2/N}\frac{2\sqrt{R_2}}{\sqrt{N}}\nonumber \\
 	  &- \frac{R_1}{R_1^2 - 4R_2/N}\frac{2\sqrt{R_2}}{\sqrt{N}}\ln\left\{\frac{(1-s -R_1s)^2 + \frac{4R_2}{N}s^2}{(1-s)^2}\right\}. 
\end{align}
The arctan argument is large, so the arctan is close to $\pi/2$.
We note that the extra terms are small for most values of $s$, and only
become large as $s\rightarrow 1$.  To check when these terms become $O(1)$, for 
the first extra term we solve
\begin{equation}
\frac{s}{(1-s)}\frac{R_1^2 + 4R_2/N}{R_1^2 - 4R_2/N}\frac{2\sqrt{R_2}}{\sqrt{N}} \simeq 1
\end{equation}
to obtain
\begin{equation}
s\simeq \frac{1}{1+2\sqrt{R_2/N}}\simeq \frac{1}{1+4/(n\sqrt{N})}.
\end{equation}
This is well past the minimum gap, which occurs at 
$s= 1/(1+R_1) \simeq 1/(1+2/n)$.
Applying the same procedure to the second extra term gives to leading order
\begin{equation}
s \simeq 1 - e^{-\sqrt{N}/4},
\end{equation}
which is even closer to $s=1$ and further from the minimum gap.  Since the 
transition probabilities are only significant close to the minimum gap, and 
hence all the important slowing down of the schedule occurs around the gap, 
what happens this close to $s=1$ has essentially no effect on the success or 
runtime of the algorithm.

Dropping the extra terms from the solution provides an expression for $s(t)$ 
\begin{equation}
s(t) = \frac{2\sqrt{R_2}}{\sqrt{N}(1+R_1)^2}
\tan\left\{\frac{8\epsilon\sqrt{R_2}R_1^2 t}{n\sqrt{N}R_2^2} - c'''\right\}
+\frac{1}{1+R_1}
\end{equation}
where we have also dropped terms $O(1/N)$, and 
\begin{equation}
c''' = \arctan\left\{\frac{(1+R_1)\sqrt{N}}{2\sqrt{R_2}}\right\}.
\end{equation}
Strictly speaking, this is valid for $s \lesssim \frac{1}{1+4/(n\sqrt{N})}$, 
although in fact it is well-behaved right up to and including $s=1$.  
From this we can obtain the runtime
\begin{equation}
\epsilon\, t_f \simeq \frac{\pi\sqrt{N}}{4},
\end{equation}
where the two arctan terms have each been approximated by $\pi/2$, since their
arguments are large, $O(\sqrt{N})$.

%%%%%%%%%%%%%%%%%%%%%%%%%%%%%%%%%%%%%%%%%%%%%%%%%%%%%%%%%%%%%%%%%%%%%%%%%%%%%%%
\section{Numerical methods \label{app:num_meth}}
%%%%%%%%%%%%%%%%%%%%%%%%%%%%%%%%%%%%%%%%%%%%%%%%%%%%%%%%%%%%%%%%%%%%%%%%%%%%%%%

Our numerical calculations were carried out using the Python programming 
language (both Python 2.7 and Python 3.5), making considerable use of the NumPy,
SciPy and Matplotlib packages \cite{python,numpy,scipy,matplotlib}. 
High performance computing resources were not used in this study, although some 
of the simulations took several days to run on standard desktop workstations.  
Most of the simulations consisted of solving the time evolution of the quantum 
search algorithm by numerically integrating the Schr\"{o}dinger equation using 
the appropriate Hamiltonian.  This was done by diagonalising the Hamiltonian and 
exponentiating it in the diagonal basis, before applying it to the wave function.  
This process was iterated for time dependent Hamiltonians, rotating from one 
instantaneous diagonal basis to the next at small time intervals.  For the 
decoherence studies in Sec.~\ref{sub:open_systems}, the same process was applied 
to the density matrix, with dephasing operators also applied along with the 
unitary time evolution.

For larger simulations, we can take advantage of the symmetry in the
hypercube to map the dynamics to a search on the line with appropriately 
weighted edges, as given by Eq.~(\ref{eq:aqc_line}) in appendix 
\ref{app:sched_analytics}.  Provided 
the initial state is also invariant with respect this symmetry, the evolution 
will be restricted to this symmetric subspace.  This allows us to perform 
simulations for much larger numbers of qubits $n\lesssim 100$, and hence extract 
reliable information about the scaling with $n$ from numerics alone.  This 
provides important checks of the validity of the two-level approximations made 
to facilitate the analytical calculations. 

Optimal AQC schedules $s^{(n)}(\tau)$ were calculated numerically as solutions 
of Eq.~(\ref{eq:ac_inst}), both to check the analytical solutions for the 
hypercube, and because we can solve numerically with less approximations than are 
required to obtain analytical expressions.  Specifically, we calculate the gap 
$g(s)$ directly from the Hamiltonian eigensystem, rather than expanding about 
$g_{\text{min}}$ as was done in appendix \ref{app:sched_analytics}. 
However, we do make the same approximation in the analytics and numerics by 
using the maximum value of $\frac{n}{4}$ for 
$\langle \frac{d\hat{H}}{ds} \rangle_{0,1}$ obtained in 
appendix \ref{app:sched_analytics}.  For the hypercube, the matrix which 
describes these systems is $(n+1)\times (n+1)$, even after taking advantage 
of symmetry by mapping to a line.  A Hermitian $2\times2$ matrix can always be 
diagonalized analytically by finding the roots of the characteristic polynomial, 
as was done in \cite{roland02a}.  For larger matrices this is no longer 
feasible, nor generally possible if the matrix is bigger than $4\times 4$.  
Fortunately, the gap $g(s)$ can easily be calculated numerically using the 
iterative eigensolving modules in Numpy \cite{numpy}, and we are 
thence able to iteratively solve $\left|\frac{d s}{d t}\right|=\epsilon 4g^2(s)/n$.
We first define a normalized function 
\begin{equation} \label{eq:F_normed}
F(s)=\int_{0}^{s}ds'\frac{1}{\epsilon g^2(s')}\times\left[
	 \int_{0}^{1}ds'\frac{1}{\epsilon g^2(s')}\right]^{-1},
\end{equation}
where $s$ is a function of the reduced time $\tau$.  To obtain $s(\tau)$, we 
need to invert this function, $s(\tau)=F^{-1}(\tau)$.  The following method
accomplishes this.

Deliberately using a programming-like notation, we define $\tau \List$ to be a 
linearly spaced list of points between $\tau=0$ and $\tau=1$, and $s\List$ to be
a list of the corresponding values of $s(\tau)$, obtained by applying 
$F^{-1}(\tau)$ to each element of $\tau \List$.  Defining
$j(s)$ equal to the number of elements in $s\List$ which are strictly 
less than $s$, we approximate $F(s)$ 
numerically by $\tilde{F}(s)$, where we replace the integral by a finite sum 
plus linear interpolation.  Writing 
$\tilde{s_j} = \frac{1}{2}(s\List(j(s))+s\List(j(s)+1))$
\begin{align}
\tilde{F}(s) &= \sum_{j'=1}^{j(s)}\frac{s\List(j'+1)-s\List(j')}
		{\mathcal{N}\,g^2\left(\tilde{s_{j'}}\right)}\nonumber \\
&+ \frac{s-s\List(j(s))}{\mathcal{N}\,g^2\left(\tilde{s_j}\right)},
\end{align}
where $\mathcal{N}$ is a normalization factor which is included to 
ensure that $\hat{F}(s=1)=1$. It is straightforward to numerically invert 
$\hat{F}(s)$.  This can be accomplished by first finding $j_{\max}(s)$, the 
largest value of $j(s)$ for which $\hat{F}(s)<\tau$, and then solving 
\begin{equation}
	\hat{F}(s)|_{j(s)=j_{\max}(s)}=\tau
\end{equation}
for $s$. 
Based on this numerical function inversion, we define an iterative method of 
converging on the solution for $s^{(n)}(\tau)$,
\begin{enumerate}
\item set a linearly spaced $s\List \in [0,1]$ and $\tau \List \in [0,1]$ each 
with the same number of elements
\item using the values of $s$ in $s\List$, apply $\hat{F}^{-1}(\tau)$ to each 
corresponding element in $\tau \List$ to generate a new $s\List$
\item repeat step 2.~with the new $s\List$ as input, until it has converged
\end{enumerate}
The advantage of this iterative method is that, at each iteration, more points 
in $s\List$ will concentrate in areas where $1/g^2$ is larger, for instance near
the dominant avoided crossing. By using the previously calculated $s\List$ as a 
mesh in the current iteration, the protocol can continuously improve the quality
of the numerical inverse with a fixed number of points in $s\List$.

%%%%%%%%%%%%%%%%%%%%%%%%%%%%%%%%%%%%%%%%%%%%%%%%%%%%%%%%%%%%%%%%%%%%%%%%%%%%%%%
\bibliography{quantum_search_hybrid_realistic}

%merlin.mbs apsrev4-1.bst 2010-07-25 4.21a (PWD, AO, DPC) hacked
%Control: key (0)
%Control: author (0) dotless jnrlst
%Control: editor formatted (1) identically to author
%Control: production of article title (0) allowed
%Control: page (1) range
%Control: year (0) verbatim
%Control: production of eprint (0) enabled
\begin{thebibliography}{71}%
\makeatletter
\providecommand \@ifxundefined [1]{%
 \@ifx{#1\undefined}
}%
\providecommand \@ifnum [1]{%
 \ifnum #1\expandafter \@firstoftwo
 \else \expandafter \@secondoftwo
 \fi
}%
\providecommand \@ifx [1]{%
 \ifx #1\expandafter \@firstoftwo
 \else \expandafter \@secondoftwo
 \fi
}%
\providecommand \natexlab [1]{#1}%
\providecommand \enquote  [1]{``#1''}%
\providecommand \bibnamefont  [1]{#1}%
\providecommand \bibfnamefont [1]{#1}%
\providecommand \citenamefont [1]{#1}%
\providecommand \href@noop [0]{\@secondoftwo}%
\providecommand \href [0]{\begingroup \@sanitize@url \@href}%
\providecommand \@href[1]{\@@startlink{#1}\@@href}%
\providecommand \@@href[1]{\endgroup#1\@@endlink}%
\providecommand \@sanitize@url [0]{\catcode `\\12\catcode `\$12\catcode
  `\&12\catcode `\#12\catcode `\^12\catcode `\_12\catcode `\%12\relax}%
\providecommand \@@startlink[1]{}%
\providecommand \@@endlink[0]{}%
\providecommand \url  [0]{\begingroup\@sanitize@url \@url }%
\providecommand \@url [1]{\endgroup\@href {#1}{\urlprefix }}%
\providecommand \urlprefix  [0]{URL }%
\providecommand \Eprint [0]{\href }%
\providecommand \doibase [0]{http://dx.doi.org/}%
\providecommand \selectlanguage [0]{\@gobble}%
\providecommand \bibinfo  [0]{\@secondoftwo}%
\providecommand \bibfield  [0]{\@secondoftwo}%
\providecommand \translation [1]{[#1]}%
\providecommand \BibitemOpen [0]{}%
\providecommand \bibitemStop [0]{}%
\providecommand \bibitemNoStop [0]{.\EOS\space}%
\providecommand \EOS [0]{\spacefactor3000\relax}%
\providecommand \BibitemShut  [1]{\csname bibitem#1\endcsname}%
\let\auto@bib@innerbib\@empty
%</preamble>
\bibitem [{\citenamefont {Brooke}\ \emph {et~al.}(1999)\citenamefont {Brooke},
  \citenamefont {Bitko}, \citenamefont {Rosenbaum},\ and\ \citenamefont
  {Aeppli}}]{brooke99a}%
  \BibitemOpen
  \bibfield  {author} {\bibinfo {author} {\bibfnamefont {J.}~\bibnamefont
  {Brooke}}, \bibinfo {author} {\bibfnamefont {D.}~\bibnamefont {Bitko}},
  \bibinfo {author} {\bibfnamefont {T.~F.}\ \bibnamefont {Rosenbaum}}, \ and\
  \bibinfo {author} {\bibfnamefont {G.}~\bibnamefont {Aeppli}},\ }\bibfield
  {title} {\enquote {\bibinfo {title} {Quantum annealing of a disordered
  magnet},}\ }\href {\doibase 10.1126/science.284.5415.779} {\bibfield
  {journal} {\bibinfo  {journal} {Science}\ }\textbf {\bibinfo {volume}
  {284}},\ \bibinfo {pages} {779--781} (\bibinfo {year} {1999})},\ \Eprint
  {http://arxiv.org/abs/http://science.sciencemag.org/content/284/5415/779.full.pdf}
  {http://science.sciencemag.org/content/284/5415/779.full.pdf} \BibitemShut
  {NoStop}%
\bibitem [{\citenamefont {Johnson}\ \emph {et~al.}(2011)\citenamefont
  {Johnson}, \citenamefont {Amin}, \citenamefont {Gildert}, \citenamefont
  {Lanting}, \citenamefont {Hamze}, \citenamefont {Dickson}, \citenamefont
  {Harris}, \citenamefont {Berkley}, \citenamefont {Johansson}, \citenamefont
  {Bunyk}, \citenamefont {Chapple}, \citenamefont {Enderud}, \citenamefont
  {Hilton}, \citenamefont {Karimi}, \citenamefont {Ladizinsky}, \citenamefont
  {Ladizinsky}, \citenamefont {Oh}, \citenamefont {Perminov}, \citenamefont
  {Rich}, \citenamefont {Thom}, \citenamefont {Tolkacheva}, \citenamefont
  {Truncik}, \citenamefont {Uchaikin}, \citenamefont {Wang},\ and\
  \citenamefont {Rose}}]{johnson11a}%
  \BibitemOpen
  \bibfield  {author} {\bibinfo {author} {\bibfnamefont {M.~W.}\ \bibnamefont
  {Johnson}}, \bibinfo {author} {\bibfnamefont {M.~H.~S.}\ \bibnamefont
  {Amin}}, \bibinfo {author} {\bibfnamefont {S.}~\bibnamefont {Gildert}},
  \bibinfo {author} {\bibfnamefont {T.}~\bibnamefont {Lanting}}, \bibinfo
  {author} {\bibfnamefont {F.}~\bibnamefont {Hamze}}, \bibinfo {author}
  {\bibfnamefont {N.}~\bibnamefont {Dickson}}, \bibinfo {author} {\bibfnamefont
  {R.}~\bibnamefont {Harris}}, \bibinfo {author} {\bibfnamefont {A.~J.}\
  \bibnamefont {Berkley}}, \bibinfo {author} {\bibfnamefont {J.}~\bibnamefont
  {Johansson}}, \bibinfo {author} {\bibfnamefont {P.}~\bibnamefont {Bunyk}},
  \bibinfo {author} {\bibfnamefont {E.~M.}\ \bibnamefont {Chapple}}, \bibinfo
  {author} {\bibfnamefont {C.}~\bibnamefont {Enderud}}, \bibinfo {author}
  {\bibfnamefont {J.~P.}\ \bibnamefont {Hilton}}, \bibinfo {author}
  {\bibfnamefont {K.}~\bibnamefont {Karimi}}, \bibinfo {author} {\bibfnamefont
  {E.}~\bibnamefont {Ladizinsky}}, \bibinfo {author} {\bibfnamefont
  {N.}~\bibnamefont {Ladizinsky}}, \bibinfo {author} {\bibfnamefont
  {T.}~\bibnamefont {Oh}}, \bibinfo {author} {\bibfnamefont {I.}~\bibnamefont
  {Perminov}}, \bibinfo {author} {\bibfnamefont {C.}~\bibnamefont {Rich}},
  \bibinfo {author} {\bibfnamefont {M.~C.}\ \bibnamefont {Thom}}, \bibinfo
  {author} {\bibfnamefont {E.}~\bibnamefont {Tolkacheva}}, \bibinfo {author}
  {\bibfnamefont {C.~J.~S.}\ \bibnamefont {Truncik}}, \bibinfo {author}
  {\bibfnamefont {S.}~\bibnamefont {Uchaikin}}, \bibinfo {author}
  {\bibfnamefont {J.}~\bibnamefont {Wang}}, \ and\ \bibinfo {author}
  {\bibfnamefont {B.~Wilsonand~G.}\ \bibnamefont {Rose}},\ }\bibfield  {title}
  {\enquote {\bibinfo {title} {Quantum annealing with manufactured spins},}\
  }\href {\doibase doi:10.1038/nature10012} {\bibfield  {journal} {\bibinfo
  {journal} {Nature}\ }\textbf {\bibinfo {volume} {473}},\ \bibinfo {pages}
  {194--198} (\bibinfo {year} {2011})}\BibitemShut {NoStop}%
\bibitem [{\citenamefont {Denchev}\ \emph {et~al.}(2016)\citenamefont
  {Denchev}, \citenamefont {Boixo}, \citenamefont {Isakov}, \citenamefont
  {Ding}, \citenamefont {Babbush}, \citenamefont {Smelyanskiy}, \citenamefont
  {Martinis},\ and\ \citenamefont {Neven}}]{denchev16a}%
  \BibitemOpen
  \bibfield  {author} {\bibinfo {author} {\bibfnamefont {Vasil~S.}\
  \bibnamefont {Denchev}}, \bibinfo {author} {\bibfnamefont {Sergio}\
  \bibnamefont {Boixo}}, \bibinfo {author} {\bibfnamefont {Sergei~V.}\
  \bibnamefont {Isakov}}, \bibinfo {author} {\bibfnamefont {Nan}\ \bibnamefont
  {Ding}}, \bibinfo {author} {\bibfnamefont {Ryan}\ \bibnamefont {Babbush}},
  \bibinfo {author} {\bibfnamefont {Vadim}\ \bibnamefont {Smelyanskiy}},
  \bibinfo {author} {\bibfnamefont {John}\ \bibnamefont {Martinis}}, \ and\
  \bibinfo {author} {\bibfnamefont {Hartmut}\ \bibnamefont {Neven}},\
  }\bibfield  {title} {\enquote {\bibinfo {title} {What is the computational
  value of finite-range tunneling?}}\ }\href {\doibase
  10.1103/PhysRevX.6.031015} {\bibfield  {journal} {\bibinfo  {journal} {Phys.
  Rev. X}\ }\textbf {\bibinfo {volume} {6}},\ \bibinfo {pages} {031015}
  (\bibinfo {year} {2016})}\BibitemShut {NoStop}%
\bibitem [{\citenamefont {Lanting}\ \emph {et~al.}(2014)\citenamefont
  {Lanting}, \citenamefont {Przybysz}, \citenamefont {Smirnov}, \citenamefont
  {Spedalieri}, \citenamefont {Amin}, \citenamefont {Berkley}, \citenamefont
  {Harris}, \citenamefont {Altomare}, \citenamefont {Boixo}, \citenamefont
  {Bunyk}, \citenamefont {Dickson}, \citenamefont {Enderud}, \citenamefont
  {Hilton}, \citenamefont {Hoskinson}, \citenamefont {Johnson}, \citenamefont
  {Ladizinsky}, \citenamefont {Ladizinsky}, \citenamefont {Neufeld},
  \citenamefont {Oh}, \citenamefont {Perminov}, \citenamefont {Rich},
  \citenamefont {Thom}, \citenamefont {Tolkacheva}, \citenamefont {Uchaikin},
  \citenamefont {Wilson},\ and\ \citenamefont {Rose}}]{lanting14a}%
  \BibitemOpen
  \bibfield  {author} {\bibinfo {author} {\bibfnamefont {T.}~\bibnamefont
  {Lanting}}, \bibinfo {author} {\bibfnamefont {A.~J.}\ \bibnamefont
  {Przybysz}}, \bibinfo {author} {\bibfnamefont {A.~Yu.}\ \bibnamefont
  {Smirnov}}, \bibinfo {author} {\bibfnamefont {F.~M.}\ \bibnamefont
  {Spedalieri}}, \bibinfo {author} {\bibfnamefont {M.~H.}\ \bibnamefont
  {Amin}}, \bibinfo {author} {\bibfnamefont {A.~J.}\ \bibnamefont {Berkley}},
  \bibinfo {author} {\bibfnamefont {R.}~\bibnamefont {Harris}}, \bibinfo
  {author} {\bibfnamefont {F.}~\bibnamefont {Altomare}}, \bibinfo {author}
  {\bibfnamefont {S.}~\bibnamefont {Boixo}}, \bibinfo {author} {\bibfnamefont
  {P.}~\bibnamefont {Bunyk}}, \bibinfo {author} {\bibfnamefont
  {N.}~\bibnamefont {Dickson}}, \bibinfo {author} {\bibfnamefont
  {C.}~\bibnamefont {Enderud}}, \bibinfo {author} {\bibfnamefont {J.~P.}\
  \bibnamefont {Hilton}}, \bibinfo {author} {\bibfnamefont {E.}~\bibnamefont
  {Hoskinson}}, \bibinfo {author} {\bibfnamefont {M.~W.}\ \bibnamefont
  {Johnson}}, \bibinfo {author} {\bibfnamefont {E.}~\bibnamefont {Ladizinsky}},
  \bibinfo {author} {\bibfnamefont {N.}~\bibnamefont {Ladizinsky}}, \bibinfo
  {author} {\bibfnamefont {R.}~\bibnamefont {Neufeld}}, \bibinfo {author}
  {\bibfnamefont {T.}~\bibnamefont {Oh}}, \bibinfo {author} {\bibfnamefont
  {I.}~\bibnamefont {Perminov}}, \bibinfo {author} {\bibfnamefont
  {C.}~\bibnamefont {Rich}}, \bibinfo {author} {\bibfnamefont {M.~C.}\
  \bibnamefont {Thom}}, \bibinfo {author} {\bibfnamefont {E.}~\bibnamefont
  {Tolkacheva}}, \bibinfo {author} {\bibfnamefont {S.}~\bibnamefont
  {Uchaikin}}, \bibinfo {author} {\bibfnamefont {A.~B.}\ \bibnamefont
  {Wilson}}, \ and\ \bibinfo {author} {\bibfnamefont {G.}~\bibnamefont
  {Rose}},\ }\bibfield  {title} {\enquote {\bibinfo {title} {Entanglement in a
  quantum annealing processor},}\ }\href {\doibase 10.1103/PhysRevX.4.021041}
  {\bibfield  {journal} {\bibinfo  {journal} {Phys. Rev. X}\ }\textbf {\bibinfo
  {volume} {4}},\ \bibinfo {pages} {021041} (\bibinfo {year}
  {2014})}\BibitemShut {NoStop}%
\bibitem [{\citenamefont {Boixo}\ \emph {et~al.}(2016)\citenamefont {Boixo},
  \citenamefont {Smelyanskiy}, \citenamefont {Shabani}, \citenamefont {Isakov},
  \citenamefont {Dykman}, \citenamefont {Denchev}, \citenamefont {Amin},
  \citenamefont {Smirnov}, \citenamefont {Mohseni},\ and\ \citenamefont
  {Neven}}]{boixo16a}%
  \BibitemOpen
  \bibfield  {author} {\bibinfo {author} {\bibfnamefont {Sergio}\ \bibnamefont
  {Boixo}}, \bibinfo {author} {\bibfnamefont {Vadim~N.}\ \bibnamefont
  {Smelyanskiy}}, \bibinfo {author} {\bibfnamefont {Alireza}\ \bibnamefont
  {Shabani}}, \bibinfo {author} {\bibfnamefont {Sergei~V.}\ \bibnamefont
  {Isakov}}, \bibinfo {author} {\bibfnamefont {Mark}\ \bibnamefont {Dykman}},
  \bibinfo {author} {\bibfnamefont {Vasil~S.}\ \bibnamefont {Denchev}},
  \bibinfo {author} {\bibfnamefont {Mohammad~H.}\ \bibnamefont {Amin}},
  \bibinfo {author} {\bibfnamefont {Anatoly~Yu}\ \bibnamefont {Smirnov}},
  \bibinfo {author} {\bibfnamefont {Masoud}\ \bibnamefont {Mohseni}}, \ and\
  \bibinfo {author} {\bibfnamefont {Hartmut}\ \bibnamefont {Neven}},\
  }\bibfield  {title} {\enquote {\bibinfo {title} {Computational multiqubit
  tunnelling in programmable quantum annealers},}\ }\href {\doibase
  doi:10.1038/ncomms10327} {\bibfield  {journal} {\bibinfo  {journal} {Nature
  Communications}\ }\textbf {\bibinfo {volume} {7}} (\bibinfo {year} {2016}),\
  doi:10.1038/ncomms10327}\BibitemShut {NoStop}%
\bibitem [{\citenamefont {Georgescu}\ \emph {et~al.}(2014)\citenamefont
  {Georgescu}, \citenamefont {Ashhab},\ and\ \citenamefont
  {Nori}}]{georgescu14}%
  \BibitemOpen
  \bibfield  {author} {\bibinfo {author} {\bibfnamefont {I.~M.}\ \bibnamefont
  {Georgescu}}, \bibinfo {author} {\bibfnamefont {S.}~\bibnamefont {Ashhab}}, \
  and\ \bibinfo {author} {\bibfnamefont {Franco}\ \bibnamefont {Nori}},\
  }\bibfield  {title} {\enquote {\bibinfo {title} {Quantum simulation},}\
  }\href {\doibase 10.1103/RevModPhys.86.153} {\bibfield  {journal} {\bibinfo
  {journal} {Rev. Mod. Phys.}\ }\textbf {\bibinfo {volume} {86}},\ \bibinfo
  {pages} {153--185} (\bibinfo {year} {2014})}\BibitemShut {NoStop}%
\bibitem [{\citenamefont {Nguyen}\ \emph {et~al.}(2018)\citenamefont {Nguyen},
  \citenamefont {Raimond}, \citenamefont {Sayrin}, \citenamefont {Corti\~nas},
  \citenamefont {Cantat-Moltrecht}, \citenamefont {Assemat}, \citenamefont
  {Dotsenko}, \citenamefont {Gleyzes}, \citenamefont {Haroche}, \citenamefont
  {Roux}, \citenamefont {Jolicoeur},\ and\ \citenamefont {Brune}}]{nguyen18}%
  \BibitemOpen
  \bibfield  {author} {\bibinfo {author} {\bibfnamefont {T.~L.}\ \bibnamefont
  {Nguyen}}, \bibinfo {author} {\bibfnamefont {J.~M.}\ \bibnamefont {Raimond}},
  \bibinfo {author} {\bibfnamefont {C.}~\bibnamefont {Sayrin}}, \bibinfo
  {author} {\bibfnamefont {R.}~\bibnamefont {Corti\~nas}}, \bibinfo {author}
  {\bibfnamefont {T.}~\bibnamefont {Cantat-Moltrecht}}, \bibinfo {author}
  {\bibfnamefont {F.}~\bibnamefont {Assemat}}, \bibinfo {author} {\bibfnamefont
  {I.}~\bibnamefont {Dotsenko}}, \bibinfo {author} {\bibfnamefont
  {S.}~\bibnamefont {Gleyzes}}, \bibinfo {author} {\bibfnamefont
  {S.}~\bibnamefont {Haroche}}, \bibinfo {author} {\bibfnamefont
  {G.}~\bibnamefont {Roux}}, \bibinfo {author} {\bibfnamefont {Th.}\
  \bibnamefont {Jolicoeur}}, \ and\ \bibinfo {author} {\bibfnamefont
  {M.}~\bibnamefont {Brune}},\ }\bibfield  {title} {\enquote {\bibinfo {title}
  {Towards quantum simulation with circular rydberg atoms},}\ }\href {\doibase
  10.1103/PhysRevX.8.011032} {\bibfield  {journal} {\bibinfo  {journal} {Phys.
  Rev. X}\ }\textbf {\bibinfo {volume} {8}},\ \bibinfo {pages} {011032}
  (\bibinfo {year} {2018})}\BibitemShut {NoStop}%
\bibitem [{\citenamefont {Kendon}\ \emph {et~al.}(2018)\citenamefont {Kendon},
  \citenamefont {Chancellor}, \citenamefont {Bose},\ and\ \citenamefont
  {Daley}}]{kendon17a}%
  \BibitemOpen
  \bibfield  {author} {\bibinfo {author} {\bibfnamefont {V.M.}\ \bibnamefont
  {Kendon}}, \bibinfo {author} {\bibfnamefont {N.}~\bibnamefont {Chancellor}},
  \bibinfo {author} {\bibfnamefont {S.}~\bibnamefont {Bose}}, \ and\ \bibinfo
  {author} {\bibfnamefont {A.}~\bibnamefont {Daley}},\ }\href@noop {} {\enquote
  {\bibinfo {title} {Developing continuous-time quantum computing},}\ }
  (\bibinfo {year} {2018}),\ \bibinfo {note} {in preparation.}\BibitemShut
  {Stop}%
\bibitem [{\citenamefont {Marzec}(2016)}]{marzec16a}%
  \BibitemOpen
  \bibfield  {author} {\bibinfo {author} {\bibfnamefont {Michael}\ \bibnamefont
  {Marzec}},\ }\enquote {\bibinfo {title} {Portfolio optimization: Applications
  in quantum computing},}\ in\ \href {\doibase 10.1002/9781118593486.ch4}
  {\emph {\bibinfo {booktitle} {Handbook of High-Frequency Trading and Modeling
  in Finance}}}\ (\bibinfo  {publisher} {John Wiley \& Sons, Inc.},\ \bibinfo
  {year} {2016})\ pp.\ \bibinfo {pages} {73--106}\BibitemShut {NoStop}%
\bibitem [{\citenamefont {Russo}(2014)}]{coxson14a}%
  \BibitemOpen
  \bibfield  {author} {\bibinfo {author} {\bibfnamefont {G.~E.Coxson C. R. Hill
  J.~C.}\ \bibnamefont {Russo}},\ }\href@noop {} {\enquote {\bibinfo {title}
  {Adiabatic quantum computing for finding low-peak-sidelobe codes},}\ }
  (\bibinfo {year} {2014}),\ \bibinfo {note} {presented at the 2014 IEEE High
  Performance Extreme Computing conference}\BibitemShut {NoStop}%
\bibitem [{\citenamefont {Melko}(2016)}]{amin16a}%
  \BibitemOpen
  \bibfield  {author} {\bibinfo {author} {\bibfnamefont {M.~H. Amin E.
  Andriyash J. Rolfe B. Kulchytskyy~R.}\ \bibnamefont {Melko}},\ }\href@noop {}
  {\enquote {\bibinfo {title} {Quantum {B}oltzmann machine},}\ } (\bibinfo
  {year} {2016}),\ \Eprint {http://arxiv.org/abs/arXiv:quant-ph:1601.02036}
  {arXiv:quant-ph:1601.02036} \BibitemShut {NoStop}%
\bibitem [{\citenamefont {Benedetti}\ \emph
  {et~al.}(2016{\natexlab{a}})\citenamefont {Benedetti}, \citenamefont
  {Realpe-G\'omez}, \citenamefont {Biswas},\ and\ \citenamefont
  {Perdomo-Ortiz}}]{Benedetti16a}%
  \BibitemOpen
  \bibfield  {author} {\bibinfo {author} {\bibfnamefont {M.}~\bibnamefont
  {Benedetti}}, \bibinfo {author} {\bibfnamefont {J.}~\bibnamefont
  {Realpe-G\'omez}}, \bibinfo {author} {\bibfnamefont {R.}~\bibnamefont
  {Biswas}}, \ and\ \bibinfo {author} {\bibfnamefont {A.}~\bibnamefont
  {Perdomo-Ortiz}},\ }\bibfield  {title} {\enquote {\bibinfo {title}
  {Estimation of effective temperatures in quantum annealers for sampling
  applications: A case study with possible applications in deep learning},}\
  }\href {\doibase 10.1103/PhysRevA.94.022308} {\bibfield  {journal} {\bibinfo
  {journal} {Phys. Rev. A}\ }\textbf {\bibinfo {volume} {94}},\ \bibinfo
  {pages} {022308} (\bibinfo {year} {2016}{\natexlab{a}})}\BibitemShut
  {NoStop}%
\bibitem [{\citenamefont {Benedetti}\ \emph
  {et~al.}(2016{\natexlab{b}})\citenamefont {Benedetti}, \citenamefont
  {Realpe-G{\'o}mez}, \citenamefont {Biswas},\ and\ \citenamefont
  {Perdomo-Ortiz}}]{Benedetti16b}%
  \BibitemOpen
  \bibfield  {author} {\bibinfo {author} {\bibfnamefont {M.}~\bibnamefont
  {Benedetti}}, \bibinfo {author} {\bibfnamefont {J.}~\bibnamefont
  {Realpe-G{\'o}mez}}, \bibinfo {author} {\bibfnamefont {R.}~\bibnamefont
  {Biswas}}, \ and\ \bibinfo {author} {\bibfnamefont {A.}~\bibnamefont
  {Perdomo-Ortiz}},\ }\href@noop {} {\enquote {\bibinfo {title}
  {Quantum-assisted learning of graphical models with arbitrary pairwise
  connectivity},}\ } (\bibinfo {year} {2016}{\natexlab{b}}),\ \Eprint
  {http://arxiv.org/abs/arXiv:1609.02542} {arXiv:1609.02542} \BibitemShut
  {NoStop}%
\bibitem [{\citenamefont {Chancellor}\ \emph
  {et~al.}(2016{\natexlab{a}})\citenamefont {Chancellor}, \citenamefont
  {Zohren}, \citenamefont {Warburton}, \citenamefont {Benjamin},\ and\
  \citenamefont {Roberts}}]{chancellor16a}%
  \BibitemOpen
  \bibfield  {author} {\bibinfo {author} {\bibfnamefont {N.}~\bibnamefont
  {Chancellor}}, \bibinfo {author} {\bibfnamefont {S.}~\bibnamefont {Zohren}},
  \bibinfo {author} {\bibfnamefont {P.}~\bibnamefont {Warburton}}, \bibinfo
  {author} {\bibfnamefont {S.}~\bibnamefont {Benjamin}}, \ and\ \bibinfo
  {author} {\bibfnamefont {S.}~\bibnamefont {Roberts}},\ }\bibfield  {title}
  {\enquote {\bibinfo {title} {A direct mapping of {M}ax k-{SAT} and high order
  parity checks to a chimera graph},}\ }\href {\doibase 10.1038/srep37107}
  {\bibfield  {journal} {\bibinfo  {journal} {Scientific Reports}\ }\textbf
  {\bibinfo {volume} {6}} (\bibinfo {year} {2016}{\natexlab{a}}),\
  10.1038/srep37107},\ \Eprint {http://arxiv.org/abs/arXiv:1604.00651}
  {arXiv:1604.00651} \BibitemShut {NoStop}%
\bibitem [{\citenamefont {Chancellor}\ \emph
  {et~al.}(2016{\natexlab{b}})\citenamefont {Chancellor}, \citenamefont
  {Szoke}, \citenamefont {Vinci}, \citenamefont {Aeppli},\ and\ \citenamefont
  {Warburton}}]{chancellor16b}%
  \BibitemOpen
  \bibfield  {author} {\bibinfo {author} {\bibfnamefont {N.}~\bibnamefont
  {Chancellor}}, \bibinfo {author} {\bibfnamefont {S.}~\bibnamefont {Szoke}},
  \bibinfo {author} {\bibfnamefont {W.}~\bibnamefont {Vinci}}, \bibinfo
  {author} {\bibfnamefont {G.}~\bibnamefont {Aeppli}}, \ and\ \bibinfo {author}
  {\bibfnamefont {P.~A.}\ \bibnamefont {Warburton}},\ }\bibfield  {title}
  {\enquote {\bibinfo {title} {Maximum--entropy inference with a programmable
  annealer},}\ }\href {\doibase doi:10.1038/srep22318} {\bibfield  {journal}
  {\bibinfo  {journal} {Scientific Reports}\ }\textbf {\bibinfo {volume} {6}}
  (\bibinfo {year} {2016}{\natexlab{b}}),\ doi:10.1038/srep22318}\BibitemShut
  {NoStop}%
\bibitem [{\citenamefont {Li}\ \emph {et~al.}(2017)\citenamefont {Li},
  \citenamefont {Dattani}, \citenamefont {Chen}, \citenamefont {Liu},
  \citenamefont {Wang}, \citenamefont {Tanburn}, \citenamefont {Chen},
  \citenamefont {Peng},\ and\ \citenamefont {Du}}]{Li17a}%
  \BibitemOpen
  \bibfield  {author} {\bibinfo {author} {\bibfnamefont {Zhaokai}\ \bibnamefont
  {Li}}, \bibinfo {author} {\bibfnamefont {Nikesh~S.}\ \bibnamefont {Dattani}},
  \bibinfo {author} {\bibfnamefont {Xi}~\bibnamefont {Chen}}, \bibinfo {author}
  {\bibfnamefont {Xiaomei}\ \bibnamefont {Liu}}, \bibinfo {author}
  {\bibfnamefont {Hengyan}\ \bibnamefont {Wang}}, \bibinfo {author}
  {\bibfnamefont {Richard}\ \bibnamefont {Tanburn}}, \bibinfo {author}
  {\bibfnamefont {Hongwei}\ \bibnamefont {Chen}}, \bibinfo {author}
  {\bibfnamefont {Xinhua}\ \bibnamefont {Peng}}, \ and\ \bibinfo {author}
  {\bibfnamefont {Jiangfeng}\ \bibnamefont {Du}},\ }\href@noop {} {\enquote
  {\bibinfo {title} {High-fidelity adiabatic quantum computation using the
  intrinsic hamiltonian of a spin system: Application to the experimental
  factorization of 291311},}\ } (\bibinfo {year} {2017}),\ \Eprint
  {http://arxiv.org/abs/1706.08061} {arXiv:1706.08061} \BibitemShut {NoStop}%
\bibitem [{\citenamefont {Bian}\ \emph {et~al.}(2013)\citenamefont {Bian},
  \citenamefont {Chudak}, \citenamefont {Macready}, \citenamefont {Clark},\
  and\ \citenamefont {Gaitan}}]{Bian13a}%
  \BibitemOpen
  \bibfield  {author} {\bibinfo {author} {\bibfnamefont {Zhengbing}\
  \bibnamefont {Bian}}, \bibinfo {author} {\bibfnamefont {Fabian}\ \bibnamefont
  {Chudak}}, \bibinfo {author} {\bibfnamefont {William~G.}\ \bibnamefont
  {Macready}}, \bibinfo {author} {\bibfnamefont {Lane}\ \bibnamefont {Clark}},
  \ and\ \bibinfo {author} {\bibfnamefont {Frank}\ \bibnamefont {Gaitan}},\
  }\bibfield  {title} {\enquote {\bibinfo {title} {Experimental determination
  of ramsey numbers},}\ }\href {\doibase 10.1103/PhysRevLett.111.130505}
  {\bibfield  {journal} {\bibinfo  {journal} {Phys. Rev. Lett.}\ }\textbf
  {\bibinfo {volume} {111}},\ \bibinfo {pages} {130505} (\bibinfo {year}
  {2013})}\BibitemShut {NoStop}%
\bibitem [{\citenamefont {Perdomo-Ortiz}\ \emph {et~al.}(2012)\citenamefont
  {Perdomo-Ortiz}, \citenamefont {Dickson}, \citenamefont {Drew-Brook},
  \citenamefont {Rose},\ and\ \citenamefont {Aspuru-Guzik}}]{perdomo-ortiz12a}%
  \BibitemOpen
  \bibfield  {author} {\bibinfo {author} {\bibfnamefont {Alejandro}\
  \bibnamefont {Perdomo-Ortiz}}, \bibinfo {author} {\bibfnamefont {Neil}\
  \bibnamefont {Dickson}}, \bibinfo {author} {\bibfnamefont {Marshall}\
  \bibnamefont {Drew-Brook}}, \bibinfo {author} {\bibfnamefont {Geordie}\
  \bibnamefont {Rose}}, \ and\ \bibinfo {author} {\bibfnamefont {Alan}\
  \bibnamefont {Aspuru-Guzik}},\ }\bibfield  {title} {\enquote {\bibinfo
  {title} {Finding low-energy conformations of lattice protein models by
  quantum annealing},}\ }\href@noop {} {\bibfield  {journal} {\bibinfo
  {journal} {Scientific Reports}\ }\textbf {\bibinfo {volume} {2}} (\bibinfo
  {year} {2012})}\BibitemShut {NoStop}%
\bibitem [{\citenamefont {Grover}(1997)}]{grover97a}%
  \BibitemOpen
  \bibfield  {author} {\bibinfo {author} {\bibfnamefont {L.~K.}\ \bibnamefont
  {Grover}},\ }\bibfield  {title} {\enquote {\bibinfo {title} {Quantum
  mechanics helps in searching for a needle in a haystack},}\ }\href@noop {}
  {\bibfield  {journal} {\bibinfo  {journal} {Phys.~Rev.~Lett.}\ }\textbf
  {\bibinfo {volume} {79}},\ \bibinfo {pages} {325} (\bibinfo {year} {1997})},\
  \Eprint {http://arxiv.org/abs/arXiv:quant-ph/9706033}
  {arXiv:quant-ph/9706033} \BibitemShut {NoStop}%
\bibitem [{\citenamefont {Bennett}\ \emph {et~al.}(1997)\citenamefont
  {Bennett}, \citenamefont {Bernstein}, \citenamefont {Brassard},\ and\
  \citenamefont {Vazirani}}]{bennett97a}%
  \BibitemOpen
  \bibfield  {author} {\bibinfo {author} {\bibfnamefont {Charles~H.}\
  \bibnamefont {Bennett}}, \bibinfo {author} {\bibfnamefont {Ethan}\
  \bibnamefont {Bernstein}}, \bibinfo {author} {\bibfnamefont {Gilles}\
  \bibnamefont {Brassard}}, \ and\ \bibinfo {author} {\bibfnamefont {Umesh}\
  \bibnamefont {Vazirani}},\ }\bibfield  {title} {\enquote {\bibinfo {title}
  {Strengths and weaknesses of quantum computing},}\ }\href@noop {} {\bibfield
  {journal} {\bibinfo  {journal} {SIAM J.~Comput.}\ }\textbf {\bibinfo {volume}
  {26}},\ \bibinfo {pages} {151--152} (\bibinfo {year} {1997})}\BibitemShut
  {NoStop}%
\bibitem [{\citenamefont {Shenvi}\ \emph {et~al.}(2003)\citenamefont {Shenvi},
  \citenamefont {Kempe},\ and\ \citenamefont {Whaley}}]{shenvi02a}%
  \BibitemOpen
  \bibfield  {author} {\bibinfo {author} {\bibfnamefont {Neil}\ \bibnamefont
  {Shenvi}}, \bibinfo {author} {\bibfnamefont {Julia}\ \bibnamefont {Kempe}}, \
  and\ \bibinfo {author} {\bibfnamefont {K~Birgitta}\ \bibnamefont {Whaley}},\
  }\bibfield  {title} {\enquote {\bibinfo {title} {A quantum random walk search
  algorithm},}\ }\href@noop {} {\bibfield  {journal} {\bibinfo  {journal}
  {Phys.~Rev.~A}\ }\textbf {\bibinfo {volume} {67}},\ \bibinfo {pages} {052307}
  (\bibinfo {year} {2003})},\ \Eprint
  {http://arxiv.org/abs/arXiv:quant-ph/0210064} {arXiv:quant-ph/0210064}
  \BibitemShut {NoStop}%
\bibitem [{\citenamefont {Roland}\ and\ \citenamefont
  {Cerf}(2002)}]{roland02a}%
  \BibitemOpen
  \bibfield  {author} {\bibinfo {author} {\bibfnamefont {J\'er\'emie}\
  \bibnamefont {Roland}}\ and\ \bibinfo {author} {\bibfnamefont {Nicolas~J.}\
  \bibnamefont {Cerf}},\ }\bibfield  {title} {\enquote {\bibinfo {title}
  {Quantum search by local adiabatic evolution},}\ }\href {\doibase
  10.1103/PhysRevA.65.042308} {\bibfield  {journal} {\bibinfo  {journal}
  {Phys.~Rev.~A}\ }\textbf {\bibinfo {volume} {65}},\ \bibinfo {pages} {042308}
  (\bibinfo {year} {2002})}\BibitemShut {NoStop}%
\bibitem [{\citenamefont {Lovett}\ \emph {et~al.}(2011)\citenamefont {Lovett},
  \citenamefont {Everitt}, \citenamefont {Heath},\ and\ \citenamefont
  {Kendon}}]{lovett11a}%
  \BibitemOpen
  \bibfield  {author} {\bibinfo {author} {\bibfnamefont {Neil~B.}\ \bibnamefont
  {Lovett}}, \bibinfo {author} {\bibfnamefont {Matthew}\ \bibnamefont
  {Everitt}}, \bibinfo {author} {\bibfnamefont {Robert~M.}\ \bibnamefont
  {Heath}}, \ and\ \bibinfo {author} {\bibfnamefont {Viv}\ \bibnamefont
  {Kendon}},\ }\href@noop {} {\enquote {\bibinfo {title} {The quantum walk
  search algorithm: Factors affecting efficiency},}\ } (\bibinfo {year}
  {2011}),\ \bibinfo {note} {to appear in Math.~Struct.~Comp.~Sci.},\ \Eprint
  {http://arxiv.org/abs/1110.4366v1[quant-ph]} {1110.4366v1[quant-ph]}
  \BibitemShut {NoStop}%
\bibitem [{\citenamefont {Coulamy}\ \emph {et~al.}(2017)\citenamefont
  {Coulamy}, \citenamefont {Saguia},\ and\ \citenamefont
  {Sarandy}}]{coulamy17a}%
  \BibitemOpen
  \bibfield  {author} {\bibinfo {author} {\bibfnamefont {Ivan~B.}\ \bibnamefont
  {Coulamy}}, \bibinfo {author} {\bibfnamefont {Andreia}\ \bibnamefont
  {Saguia}}, \ and\ \bibinfo {author} {\bibfnamefont {Marcelo~S.}\ \bibnamefont
  {Sarandy}},\ }\bibfield  {title} {\enquote {\bibinfo {title} {Dynamics of the
  quantum search and quench-induced first-order phase transitions},}\ }\href
  {\doibase 10.1103/PhysRevE.95.022127} {\bibfield  {journal} {\bibinfo
  {journal} {Phys. Rev. E}\ }\textbf {\bibinfo {volume} {95}},\ \bibinfo
  {pages} {022127} (\bibinfo {year} {2017})}\BibitemShut {NoStop}%
\bibitem [{\citenamefont {Tulsi}(2015)}]{tulsi15a}%
  \BibitemOpen
  \bibfield  {author} {\bibinfo {author} {\bibfnamefont {Avatar}\ \bibnamefont
  {Tulsi}},\ }\bibfield  {title} {\enquote {\bibinfo {title} {Postprocessing
  can speed up general quantum search algorithms},}\ }\href {\doibase
  10.1103/PhysRevA.92.022353} {\bibfield  {journal} {\bibinfo  {journal} {Phys.
  Rev. A}\ }\textbf {\bibinfo {volume} {92}},\ \bibinfo {pages} {022353}
  (\bibinfo {year} {2015})}\BibitemShut {NoStop}%
\bibitem [{\citenamefont {Ambainis}\ \emph {et~al.}(2011)\citenamefont
  {Ambainis}, \citenamefont {Backurs}, \citenamefont {Nahimovs}, \citenamefont
  {Ozols},\ and\ \citenamefont {Rivosh}}]{ambainis11a}%
  \BibitemOpen
  \bibfield  {author} {\bibinfo {author} {\bibfnamefont {Andris}\ \bibnamefont
  {Ambainis}}, \bibinfo {author} {\bibfnamefont {Arturs}\ \bibnamefont
  {Backurs}}, \bibinfo {author} {\bibfnamefont {Nikolajs}\ \bibnamefont
  {Nahimovs}}, \bibinfo {author} {\bibfnamefont {Raitis}\ \bibnamefont
  {Ozols}}, \ and\ \bibinfo {author} {\bibfnamefont {Alexander}\ \bibnamefont
  {Rivosh}},\ }\href@noop {} {\enquote {\bibinfo {title} {Search by quantum
  walks on two-dimensional grid without amplitude amplification},}\ } (\bibinfo
  {year} {2011}),\ \Eprint {http://arxiv.org/abs/arXiv:1112.3337}
  {arXiv:1112.3337} \BibitemShut {NoStop}%
\bibitem [{\citenamefont {Magniez}\ \emph {et~al.}(2011)\citenamefont
  {Magniez}, \citenamefont {Nayak}, \citenamefont {Roland},\ and\ \citenamefont
  {Santha}}]{magniez11a}%
  \BibitemOpen
  \bibfield  {author} {\bibinfo {author} {\bibfnamefont {Fr\'{e}d\'{e}ric}\
  \bibnamefont {Magniez}}, \bibinfo {author} {\bibfnamefont {Ashwin}\
  \bibnamefont {Nayak}}, \bibinfo {author} {\bibfnamefont {J\'{e}r\'{e}mie}\
  \bibnamefont {Roland}}, \ and\ \bibinfo {author} {\bibfnamefont {Miklos}\
  \bibnamefont {Santha}},\ }\bibfield  {title} {\enquote {\bibinfo {title}
  {Search via quantum walk},}\ }\href {\doibase 10.1137/090745854} {\bibfield
  {journal} {\bibinfo  {journal} {SIAM Journal on Computing}\ }\textbf
  {\bibinfo {volume} {40}},\ \bibinfo {pages} {142--164} (\bibinfo {year}
  {2011})},\ \Eprint {http://arxiv.org/abs/quant-ph/0608026v4}
  {quant-ph/0608026v4} \BibitemShut {NoStop}%
\bibitem [{\citenamefont {Wong}\ and\ \citenamefont {Meyer}(2016)}]{wong16a}%
  \BibitemOpen
  \bibfield  {author} {\bibinfo {author} {\bibfnamefont {Thomas~G.}\
  \bibnamefont {Wong}}\ and\ \bibinfo {author} {\bibfnamefont {David~A.}\
  \bibnamefont {Meyer}},\ }\bibfield  {title} {\enquote {\bibinfo {title}
  {Irreconcilable difference between quantum walks and adiabatic quantum
  computing},}\ }\href {\doibase 10.1103/PhysRevA.93.062313} {\bibfield
  {journal} {\bibinfo  {journal} {Phys. Rev. A}\ }\textbf {\bibinfo {volume}
  {93}},\ \bibinfo {pages} {062313} (\bibinfo {year} {2016})}\BibitemShut
  {NoStop}%
\bibitem [{\citenamefont {Chancellor}(2016)}]{chancellor_AQC2016}%
  \BibitemOpen
  \bibfield  {author} {\bibinfo {author} {\bibfnamefont {Nicholas}\
  \bibnamefont {Chancellor}},\ }\href@noop {} {\enquote {\bibinfo {title}
  {Max-k-{SAT}, multi-body frustration, \& multi-body sampling on a two local
  {I}sing system},}\ }\bibinfo {howpublished} {AQC 2016
  \url{https://www.youtube.com/watch?v=aC-6hg_h3EA}} (\bibinfo {year}
  {2016})\BibitemShut {NoStop}%
\bibitem [{\citenamefont {Chancellor}\ \emph {et~al.}(2018)\citenamefont
  {Chancellor}, \citenamefont {Dodds},\ and\ \citenamefont
  {Kendon}}]{chancellor17b}%
  \BibitemOpen
  \bibfield  {author} {\bibinfo {author} {\bibfnamefont {Nicholas}\
  \bibnamefont {Chancellor}}, \bibinfo {author} {\bibfnamefont {A.~Ben}\
  \bibnamefont {Dodds}}, \ and\ \bibinfo {author} {\bibfnamefont {Viv}\
  \bibnamefont {Kendon}},\ }\href@noop {} {\enquote {\bibinfo {title}
  {Practical designs for permutation symmetric problem {H}amiltonians on
  hypercubes},}\ } (\bibinfo {year} {2018}),\ \bibinfo {note} {in
  preparation}\BibitemShut {NoStop}%
\bibitem [{\citenamefont {Childs}\ and\ \citenamefont
  {Goldstone}(2004)}]{childs03a}%
  \BibitemOpen
  \bibfield  {author} {\bibinfo {author} {\bibfnamefont {Andrew}\ \bibnamefont
  {Childs}}\ and\ \bibinfo {author} {\bibfnamefont {Jeffrey}\ \bibnamefont
  {Goldstone}},\ }\bibfield  {title} {\enquote {\bibinfo {title} {Spatial
  search by quantum walk},}\ }\href@noop {} {\bibfield  {journal} {\bibinfo
  {journal} {Phys.~Rev.~A}\ }\textbf {\bibinfo {volume} {70}},\ \bibinfo
  {pages} {022314} (\bibinfo {year} {2004})},\ \Eprint
  {http://arxiv.org/abs/quant-ph/0306054} {quant-ph/0306054} \BibitemShut
  {NoStop}%
\bibitem [{\citenamefont {Chakraborty}\ \emph {et~al.}(2016)\citenamefont
  {Chakraborty}, \citenamefont {Novo}, \citenamefont {Ambainis},\ and\
  \citenamefont {Omar}}]{Chakraborty16a}%
  \BibitemOpen
  \bibfield  {author} {\bibinfo {author} {\bibfnamefont {S.}~\bibnamefont
  {Chakraborty}}, \bibinfo {author} {\bibfnamefont {L.}~\bibnamefont {Novo}},
  \bibinfo {author} {\bibfnamefont {A.}~\bibnamefont {Ambainis}}, \ and\
  \bibinfo {author} {\bibfnamefont {Y.}~\bibnamefont {Omar}},\ }\bibfield
  {title} {\enquote {\bibinfo {title} {Spatial search by quantum walk is
  optimal for almost all graphs},}\ }\href {\doibase
  10.1103/PhysRevLett.116.100501} {\bibfield  {journal} {\bibinfo  {journal}
  {Phys.~Rev.~Lett.}\ }\textbf {\bibinfo {volume} {116}} (\bibinfo {year}
  {2016}),\ 10.1103/PhysRevLett.116.100501}\BibitemShut {NoStop}%
\bibitem [{\citenamefont {Moore}\ and\ \citenamefont
  {Russell}(2002)}]{moore01a}%
  \BibitemOpen
  \bibfield  {author} {\bibinfo {author} {\bibfnamefont {Christopher}\
  \bibnamefont {Moore}}\ and\ \bibinfo {author} {\bibfnamefont {Alexander}\
  \bibnamefont {Russell}},\ }\bibfield  {title} {\enquote {\bibinfo {title}
  {Quantum walks on the hypercube},}\ }in\ \href@noop {} {\emph {\bibinfo
  {booktitle} {Proc.~6th Intl.~Workshop on Randomization and Approximation
  Techniques in Computer Science (RANDOM '02)}}},\ \bibinfo {editor} {edited
  by\ \bibinfo {editor} {\bibfnamefont {J.~D.~P.}\ \bibnamefont {Rolim}}\ and\
  \bibinfo {editor} {\bibfnamefont {S.}~\bibnamefont {Vadhan}}}\ (\bibinfo
  {publisher} {Springer},\ \bibinfo {year} {2002})\ pp.\ \bibinfo {pages}
  {164--178},\ \Eprint {http://arxiv.org/abs/quant-ph/0104137}
  {quant-ph/0104137} \BibitemShut {NoStop}%
\bibitem [{\citenamefont {Hein}\ and\ \citenamefont {Tanner}(2009)}]{hein09a}%
  \BibitemOpen
  \bibfield  {author} {\bibinfo {author} {\bibfnamefont {Birgit}\ \bibnamefont
  {Hein}}\ and\ \bibinfo {author} {\bibfnamefont {Gregor}\ \bibnamefont
  {Tanner}},\ }\bibfield  {title} {\enquote {\bibinfo {title} {Quantum search
  algorithms on the hypercube},}\ }\href {\doibase
  10.1088/1751-8113/42/8/085303} {\bibfield  {journal} {\bibinfo  {journal} {J.
  Phys. A: Math. Theor.}\ }\textbf {\bibinfo {volume} {42}},\ \bibinfo {pages}
  {085303} (\bibinfo {year} {2009})}\BibitemShut {NoStop}%
\bibitem [{\citenamefont {Farhi}\ \emph {et~al.}(2000)\citenamefont {Farhi},
  \citenamefont {Goldstone}, \citenamefont {Gutmann},\ and\ \citenamefont
  {Sipser}}]{farhi00a}%
  \BibitemOpen
  \bibfield  {author} {\bibinfo {author} {\bibfnamefont {E.}~\bibnamefont
  {Farhi}}, \bibinfo {author} {\bibfnamefont {J.}~\bibnamefont {Goldstone}},
  \bibinfo {author} {\bibfnamefont {S.}~\bibnamefont {Gutmann}}, \ and\
  \bibinfo {author} {\bibfnamefont {M.}~\bibnamefont {Sipser}},\ }\href@noop {}
  {\enquote {\bibinfo {title} {Quantum computation by adiabatic evolution},}\ }
  (\bibinfo {year} {2000}),\ \Eprint {http://arxiv.org/abs/quant-ph/0001106}
  {quant-ph/0001106} \BibitemShut {NoStop}%
\bibitem [{\citenamefont {Childs}\ \emph {et~al.}(2002)\citenamefont {Childs},
  \citenamefont {Deotto}, \citenamefont {Farhi}, \citenamefont {Goldstone},
  \citenamefont {Gutmann},\ and\ \citenamefont {Landahl}}]{childs02a}%
  \BibitemOpen
  \bibfield  {author} {\bibinfo {author} {\bibfnamefont {A.~M.}\ \bibnamefont
  {Childs}}, \bibinfo {author} {\bibfnamefont {E.}~\bibnamefont {Deotto}},
  \bibinfo {author} {\bibfnamefont {E.}~\bibnamefont {Farhi}}, \bibinfo
  {author} {\bibfnamefont {J.}~\bibnamefont {Goldstone}}, \bibinfo {author}
  {\bibfnamefont {S.}~\bibnamefont {Gutmann}}, \ and\ \bibinfo {author}
  {\bibfnamefont {A.~J.}\ \bibnamefont {Landahl}},\ }\bibfield  {title}
  {\enquote {\bibinfo {title} {Quantum search by measurement},}\ }\href@noop {}
  {\bibfield  {journal} {\bibinfo  {journal} {Phys. Rev. A}\ }\textbf {\bibinfo
  {volume} {66}},\ \bibinfo {pages} {032314} (\bibinfo {year} {2002})},\
  \Eprint {http://arxiv.org/abs/quant-ph/0204013} {quant-ph/0204013}
  \BibitemShut {NoStop}%
\bibitem [{\citenamefont {Born}\ and\ \citenamefont {Fock}(1928)}]{born28}%
  \BibitemOpen
  \bibfield  {author} {\bibinfo {author} {\bibfnamefont {M.}~\bibnamefont
  {Born}}\ and\ \bibinfo {author} {\bibfnamefont {V.}~\bibnamefont {Fock}},\
  }\bibfield  {title} {\enquote {\bibinfo {title} {Beweis des
  adiabatensatzes},}\ }\href@noop {} {\bibfield  {journal} {\bibinfo  {journal}
  {Z. Phys.}\ }\textbf {\bibinfo {volume} {51(3-4)}},\ \bibinfo {pages}
  {165--180} (\bibinfo {year} {1928})}\BibitemShut {NoStop}%
\bibitem [{\citenamefont {Albash}\ and\ \citenamefont
  {Lidar}(2018)}]{albash16a}%
  \BibitemOpen
  \bibfield  {author} {\bibinfo {author} {\bibfnamefont {Tameem}\ \bibnamefont
  {Albash}}\ and\ \bibinfo {author} {\bibfnamefont {Daniel~A.}\ \bibnamefont
  {Lidar}},\ }\bibfield  {title} {\enquote {\bibinfo {title} {Adiabatic quantum
  computing},}\ }\href {\doibase 10.1103/RevModPhys.90.015002} {\bibfield
  {journal} {\bibinfo  {journal} {Rev. Mod. Phys.}\ }\textbf {\bibinfo {volume}
  {90}} (\bibinfo {year} {2018}),\ 10.1103/RevModPhys.90.015002}\BibitemShut
  {NoStop}%
\bibitem [{\citenamefont {Lanting}(2017)}]{LantingAQC2017}%
  \BibitemOpen
  \bibfield  {author} {\bibinfo {author} {\bibfnamefont {Trevor}\ \bibnamefont
  {Lanting}},\ }\href
  {https://www.youtube.com/watch?v=_Y9sVY-XBfI&index=2&list=PLAlerseOylzZw8A-R4lyU3mqfYfQSggiz}
  {\enquote {\bibinfo {title} {The {D-Wave 2000Q Processor}},}\ } (\bibinfo
  {year} {2017}),\ \bibinfo {note} {presented at AQC 2017}\BibitemShut
  {NoStop}%
\bibitem [{\citenamefont {Lidar}\ \emph {et~al.}(2009)\citenamefont {Lidar},
  \citenamefont {Rezakhani},\ and\ \citenamefont {Hamma}}]{lidar09a}%
  \BibitemOpen
  \bibfield  {author} {\bibinfo {author} {\bibfnamefont {Daniel~A.}\
  \bibnamefont {Lidar}}, \bibinfo {author} {\bibfnamefont {Ali~T.}\
  \bibnamefont {Rezakhani}}, \ and\ \bibinfo {author} {\bibfnamefont
  {Alioscia}\ \bibnamefont {Hamma}},\ }\bibfield  {title} {\enquote {\bibinfo
  {title} {Adiabatic approximation with exponential accuracy for many-body
  systems and quantum computation},}\ }\href {\doibase 10.1063/1.3236685}
  {\bibfield  {journal} {\bibinfo  {journal} {Journal of Mathematical Physics}\
  }\textbf {\bibinfo {volume} {50}},\ \bibinfo {pages} {102106} (\bibinfo
  {year} {2009})}\BibitemShut {NoStop}%
\bibitem [{\citenamefont {Rezakhani}\ \emph {et~al.}(2009)\citenamefont
  {Rezakhani}, \citenamefont {Kuo}, \citenamefont {Hamma}, \citenamefont
  {Lidar},\ and\ \citenamefont {Zanardi}}]{rezakhani09a}%
  \BibitemOpen
  \bibfield  {author} {\bibinfo {author} {\bibfnamefont {A.~T.}\ \bibnamefont
  {Rezakhani}}, \bibinfo {author} {\bibfnamefont {W.-J.}\ \bibnamefont {Kuo}},
  \bibinfo {author} {\bibfnamefont {A.}~\bibnamefont {Hamma}}, \bibinfo
  {author} {\bibfnamefont {D.~A.}\ \bibnamefont {Lidar}}, \ and\ \bibinfo
  {author} {\bibfnamefont {P.}~\bibnamefont {Zanardi}},\ }\bibfield  {title}
  {\enquote {\bibinfo {title} {Quantum adiabatic brachistochrone},}\ }\href
  {\doibase 10.1103/PhysRevLett.103.080502} {\bibfield  {journal} {\bibinfo
  {journal} {Phys. Rev. Lett.}\ }\textbf {\bibinfo {volume} {103}},\ \bibinfo
  {pages} {080502} (\bibinfo {year} {2009})}\BibitemShut {NoStop}%
\bibitem [{\citenamefont {Rezakhani}\ \emph {et~al.}(2010)\citenamefont
  {Rezakhani}, \citenamefont {Pimachev},\ and\ \citenamefont
  {Lidar}}]{rezakhani10a}%
  \BibitemOpen
  \bibfield  {author} {\bibinfo {author} {\bibfnamefont {A.~T.}\ \bibnamefont
  {Rezakhani}}, \bibinfo {author} {\bibfnamefont {A.~K.}\ \bibnamefont
  {Pimachev}}, \ and\ \bibinfo {author} {\bibfnamefont {D.~A.}\ \bibnamefont
  {Lidar}},\ }\bibfield  {title} {\enquote {\bibinfo {title} {Accuracy vs run
  time in adiabatic quantum search},}\ }\href {\doibase
  10.1103/PhysRevA.82.052305} {\bibfield  {journal} {\bibinfo  {journal} {Phys.
  Rev. A}\ }\textbf {\bibinfo {volume} {82}},\ \bibinfo {pages} {052305}
  (\bibinfo {year} {2010})}\BibitemShut {NoStop}%
\bibitem [{\citenamefont {Weibe}\ and\ \citenamefont
  {Babcock}(2012)}]{weibe12a}%
  \BibitemOpen
  \bibfield  {author} {\bibinfo {author} {\bibfnamefont {Nathan}\ \bibnamefont
  {Weibe}}\ and\ \bibinfo {author} {\bibfnamefont {Nathan~S}\ \bibnamefont
  {Babcock}},\ }\bibfield  {title} {\enquote {\bibinfo {title} {Improved
  error-scaling for adiabatic quantum evolutions},}\ }\href {\doibase
  10.1088/1367-2630/14/1/013024} {\bibfield  {journal} {\bibinfo  {journal}
  {New Journal of Physics}\ }\textbf {\bibinfo {volume} {14}} (\bibinfo {year}
  {2012}),\ 10.1088/1367-2630/14/1/013024}\BibitemShut {NoStop}%
\bibitem [{\citenamefont {Kieferov\'a}\ and\ \citenamefont
  {Wiebe}(2014)}]{kieferova14a}%
  \BibitemOpen
  \bibfield  {author} {\bibinfo {author} {\bibfnamefont {M\'aria}\ \bibnamefont
  {Kieferov\'a}}\ and\ \bibinfo {author} {\bibfnamefont {Nathan}\ \bibnamefont
  {Wiebe}},\ }\bibfield  {title} {\enquote {\bibinfo {title} {On the power of
  coherently controlled quantum adiabatic evolutions},}\ }\href
  {http://stacks.iop.org/1367-2630/16/i=12/a=123034} {\bibfield  {journal}
  {\bibinfo  {journal} {New Journal of Physics}\ }\textbf {\bibinfo {volume}
  {16}},\ \bibinfo {pages} {123034} (\bibinfo {year} {2014})}\BibitemShut
  {NoStop}%
\bibitem [{\citenamefont {Morley}\ \emph {et~al.}(2018)\citenamefont {Morley},
  \citenamefont {Chancellor}, \citenamefont {Kendon},\ and\ \citenamefont
  {Bose}}]{morley17a}%
  \BibitemOpen
  \bibfield  {author} {\bibinfo {author} {\bibfnamefont {J.~G.}\ \bibnamefont
  {Morley}}, \bibinfo {author} {\bibfnamefont {N.}~\bibnamefont {Chancellor}},
  \bibinfo {author} {\bibfnamefont {V.~M.}\ \bibnamefont {Kendon}}, \ and\
  \bibinfo {author} {\bibfnamefont {S.}~\bibnamefont {Bose}},\ }\href@noop {}
  {\enquote {\bibinfo {title} {Quench vs adiabacity: balancing competing
  mechanisms for quantum search on noisy machines},}\ } (\bibinfo {year}
  {2018}),\ \bibinfo {note} {co-submitted.}\BibitemShut {Stop}%
\bibitem [{\citenamefont {de~Vega}\ \emph {et~al.}(2010)\citenamefont
  {de~Vega}, \citenamefont {Carmen Ba\~{n}uls},\ and\ \citenamefont
  {P\'{e}rez}}]{vega10a}%
  \BibitemOpen
  \bibfield  {author} {\bibinfo {author} {\bibfnamefont {Ines}\ \bibnamefont
  {de~Vega}}, \bibinfo {author} {\bibfnamefont {Mari}\ \bibnamefont {Carmen
  Ba\~{n}uls}}, \ and\ \bibinfo {author} {\bibfnamefont {A.}~\bibnamefont
  {P\'{e}rez}},\ }\bibfield  {title} {\enquote {\bibinfo {title} {Effects of
  dissipation in an adiabatic quantum search algorithm},}\ }\href {\doibase
  10.1088/1367-2630/12/12/123010} {\bibfield  {journal} {\bibinfo  {journal}
  {New J. Phys.}\ }\textbf {\bibinfo {volume} {12}},\ \bibinfo {pages} {123010}
  (\bibinfo {year} {2010})}\BibitemShut {NoStop}%
\bibitem [{\citenamefont {Wild}\ \emph {et~al.}(2016)\citenamefont {Wild},
  \citenamefont {Gopalakrishnan}, \citenamefont {Knap}, \citenamefont {Yao},\
  and\ \citenamefont {Lukin}}]{wild16a}%
  \BibitemOpen
  \bibfield  {author} {\bibinfo {author} {\bibfnamefont {Dominik~S.}\
  \bibnamefont {Wild}}, \bibinfo {author} {\bibfnamefont {Sarang}\ \bibnamefont
  {Gopalakrishnan}}, \bibinfo {author} {\bibfnamefont {Michael}\ \bibnamefont
  {Knap}}, \bibinfo {author} {\bibfnamefont {Norman~Y.}\ \bibnamefont {Yao}}, \
  and\ \bibinfo {author} {\bibfnamefont {Mikhail~D.}\ \bibnamefont {Lukin}},\
  }\bibfield  {title} {\enquote {\bibinfo {title} {Adiabatic quantum search in
  open systems},}\ }\href {\doibase 10.1103/PhysRevLett.117.150501} {\bibfield
  {journal} {\bibinfo  {journal} {Phys. Rev. Lett.}\ }\textbf {\bibinfo
  {volume} {117}},\ \bibinfo {pages} {150501} (\bibinfo {year}
  {2016})}\BibitemShut {NoStop}%
\bibitem [{\citenamefont {Alagic}\ and\ \citenamefont
  {Russell}(2005)}]{Alagic05a}%
  \BibitemOpen
  \bibfield  {author} {\bibinfo {author} {\bibfnamefont {Gorjan}\ \bibnamefont
  {Alagic}}\ and\ \bibinfo {author} {\bibfnamefont {Alexander}\ \bibnamefont
  {Russell}},\ }\bibfield  {title} {\enquote {\bibinfo {title} {Decoherence in
  quantum walks on the hypercube},}\ }\href {\doibase
  10.1103/PhysRevA.72.062304} {\bibfield  {journal} {\bibinfo  {journal} {Phys.
  Rev. A}\ }\textbf {\bibinfo {volume} {72}},\ \bibinfo {pages} {062304}
  (\bibinfo {year} {2005})}\BibitemShut {NoStop}%
\bibitem [{\citenamefont {Richter}(2007)}]{Richter06a}%
  \BibitemOpen
  \bibfield  {author} {\bibinfo {author} {\bibfnamefont {Peter~C.}\
  \bibnamefont {Richter}},\ }\bibfield  {title} {\enquote {\bibinfo {title}
  {Quantum speedup of classical mixing processes},}\ }\href {\doibase
  10.1103/PhysRevA.76.042306} {\bibfield  {journal} {\bibinfo  {journal} {Phys.
  Rev. A}\ }\textbf {\bibinfo {volume} {76}},\ \bibinfo {pages} {042306}
  (\bibinfo {year} {2007})}\BibitemShut {NoStop}%
\bibitem [{\citenamefont {Kendon}\ and\ \citenamefont
  {Maloyer}(2008)}]{kendon08a}%
  \BibitemOpen
  \bibfield  {author} {\bibinfo {author} {\bibfnamefont {Viv}\ \bibnamefont
  {Kendon}}\ and\ \bibinfo {author} {\bibfnamefont {Olivier}\ \bibnamefont
  {Maloyer}},\ }\bibfield  {title} {\enquote {\bibinfo {title} {Optimal
  computation with non-unitary quantum walks},}\ }\href {\doibase
  http://dx.doi.org/10.1016/j.tcs.2007.12.011} {\bibfield  {journal} {\bibinfo
  {journal} {Theoretical Computer Science}\ }\textbf {\bibinfo {volume}
  {394}},\ \bibinfo {pages} {187 -- 196} (\bibinfo {year} {2008})}\BibitemShut
  {NoStop}%
\bibitem [{\citenamefont {Misra}\ and\ \citenamefont
  {Sudarshan}(1977)}]{misra77a}%
  \BibitemOpen
  \bibfield  {author} {\bibinfo {author} {\bibfnamefont {B.}~\bibnamefont
  {Misra}}\ and\ \bibinfo {author} {\bibfnamefont {E.~C.~G.}\ \bibnamefont
  {Sudarshan}},\ }\bibfield  {title} {\enquote {\bibinfo {title} {{The Zeno's
  paradox in quantum theory}},}\ }\href {\doibase 10.1063/1.523304} {\bibfield
  {journal} {\bibinfo  {journal} {Journal of Mathematical Physics}\ }\textbf
  {\bibinfo {volume} {18}},\ \bibinfo {pages} {756--763} (\bibinfo {year}
  {1977})},\ \Eprint {http://arxiv.org/abs/1011.1669v3} {arXiv:1011.1669v3}
  \BibitemShut {NoStop}%
\bibitem [{\citenamefont {C}(2004)}]{bissell04a}%
  \BibitemOpen
  \bibfield  {author} {\bibinfo {author} {\bibfnamefont {Bissell~C}\
  \bibnamefont {C}},\ }\href {http://oro.open.ac.uk/id/eprint/5795} {\enquote
  {\bibinfo {title} {A great disappearing act: the electronic analogue
  computer},}\ }\bibinfo {howpublished} {IEEE Conference on the History of
  Electronics (Bletchley, UK,)} (\bibinfo {year} {2004}),\ \bibinfo {note}
  {pages 28--30, available at http://oro.open.ac.uk/id/eprint/5795}\BibitemShut
  {NoStop}%
\bibitem [{\citenamefont {Koch}\ \emph {et~al.}(1983)\citenamefont {Koch},
  \citenamefont {Clarke}, \citenamefont {Goubau}, \citenamefont {Martinis},
  \citenamefont {Pegrum},\ and\ \citenamefont {van Harlingen}}]{koch1983}%
  \BibitemOpen
  \bibfield  {author} {\bibinfo {author} {\bibfnamefont {Roger~H.}\
  \bibnamefont {Koch}}, \bibinfo {author} {\bibfnamefont {John}\ \bibnamefont
  {Clarke}}, \bibinfo {author} {\bibfnamefont {W.~M.}\ \bibnamefont {Goubau}},
  \bibinfo {author} {\bibfnamefont {J.~M.}\ \bibnamefont {Martinis}}, \bibinfo
  {author} {\bibfnamefont {C.~M.}\ \bibnamefont {Pegrum}}, \ and\ \bibinfo
  {author} {\bibfnamefont {D.~J.}\ \bibnamefont {van Harlingen}},\ }\bibfield
  {title} {\enquote {\bibinfo {title} {Flicker (1/f) noise in tunnel junction
  dc squids},}\ }\href {\doibase 10.1007/BF00683423} {\bibfield  {journal}
  {\bibinfo  {journal} {Journal of Low Temperature Physics}\ }\textbf {\bibinfo
  {volume} {51}},\ \bibinfo {pages} {207--224} (\bibinfo {year}
  {1983})}\BibitemShut {NoStop}%
\bibitem [{\citenamefont {Koch}\ \emph {et~al.}(2007)\citenamefont {Koch},
  \citenamefont {DiVincenzo},\ and\ \citenamefont {Clarke}}]{koch2007}%
  \BibitemOpen
  \bibfield  {author} {\bibinfo {author} {\bibfnamefont {Roger~H.}\
  \bibnamefont {Koch}}, \bibinfo {author} {\bibfnamefont {David~P.}\
  \bibnamefont {DiVincenzo}}, \ and\ \bibinfo {author} {\bibfnamefont {John}\
  \bibnamefont {Clarke}},\ }\bibfield  {title} {\enquote {\bibinfo {title}
  {Model for $1/f$ flux noise in squids and qubits},}\ }\href {\doibase
  10.1103/PhysRevLett.98.267003} {\bibfield  {journal} {\bibinfo  {journal}
  {Phys. Rev. Lett.}\ }\textbf {\bibinfo {volume} {98}},\ \bibinfo {pages}
  {267003} (\bibinfo {year} {2007})}\BibitemShut {NoStop}%
\bibitem [{\citenamefont {Young}\ \emph {et~al.}(2013)\citenamefont {Young},
  \citenamefont {Blume-Kohout},\ and\ \citenamefont {Lidar}}]{young13a}%
  \BibitemOpen
  \bibfield  {author} {\bibinfo {author} {\bibfnamefont {Kevin~C.}\
  \bibnamefont {Young}}, \bibinfo {author} {\bibfnamefont {Robin}\ \bibnamefont
  {Blume-Kohout}}, \ and\ \bibinfo {author} {\bibfnamefont {Daniel~A.}\
  \bibnamefont {Lidar}},\ }\bibfield  {title} {\enquote {\bibinfo {title}
  {Adiabatic quantum optimization with the wrong {H}amiltonian},}\ }\href
  {\doibase 10.1103/PhysRevA.88.062314} {\bibfield  {journal} {\bibinfo
  {journal} {Phys. Rev. A}\ }\textbf {\bibinfo {volume} {88}} (\bibinfo {year}
  {2013}),\ 10.1103/PhysRevA.88.062314}\BibitemShut {NoStop}%
\bibitem [{\citenamefont {Albash}\ \emph {et~al.}(2018)\citenamefont {Albash},
  \citenamefont {Martin-Mayor},\ and\ \citenamefont {Hen}}]{albash18}%
  \BibitemOpen
  \bibfield  {author} {\bibinfo {author} {\bibfnamefont {T.}~\bibnamefont
  {Albash}}, \bibinfo {author} {\bibfnamefont {V.}~\bibnamefont
  {Martin-Mayor}}, \ and\ \bibinfo {author} {\bibfnamefont {Itay}\ \bibnamefont
  {Hen}},\ }\bibfield  {title} {\enquote {\bibinfo {title} {Analog errors in
  ising machines},}\ }\href@noop {} {\  (\bibinfo {year} {2018})},\ \Eprint
  {http://arxiv.org/abs/arXiv:1806.03744} {arXiv:1806.03744} \BibitemShut
  {NoStop}%
\bibitem [{\citenamefont {Yoder}\ \emph {et~al.}(2014)\citenamefont {Yoder},
  \citenamefont {Low},\ and\ \citenamefont {Chuang}}]{yoder14a}%
  \BibitemOpen
  \bibfield  {author} {\bibinfo {author} {\bibfnamefont {Theodore~J.}\
  \bibnamefont {Yoder}}, \bibinfo {author} {\bibfnamefont {Guang~Hao}\
  \bibnamefont {Low}}, \ and\ \bibinfo {author} {\bibfnamefont {Isaac~L.}\
  \bibnamefont {Chuang}},\ }\bibfield  {title} {\enquote {\bibinfo {title}
  {Fixed-point quantum search with an optimal number of queries},}\ }\href
  {\doibase 10.1103/PhysRevLett.113.210501} {\bibfield  {journal} {\bibinfo
  {journal} {Phys. Rev. Lett.}\ }\textbf {\bibinfo {volume} {113}},\ \bibinfo
  {pages} {210501} (\bibinfo {year} {2014})}\BibitemShut {NoStop}%
\bibitem [{\citenamefont {Dalzell}\ \emph {et~al.}(2017)\citenamefont
  {Dalzell}, \citenamefont {Yoder},\ and\ \citenamefont {Chuang}}]{dalzell17a}%
  \BibitemOpen
  \bibfield  {author} {\bibinfo {author} {\bibfnamefont {Alexander~M.}\
  \bibnamefont {Dalzell}}, \bibinfo {author} {\bibfnamefont {Theodore~J.}\
  \bibnamefont {Yoder}}, \ and\ \bibinfo {author} {\bibfnamefont {Isaac~L.}\
  \bibnamefont {Chuang}},\ }\bibfield  {title} {\enquote {\bibinfo {title}
  {Fixed-point adiabatic quantum search},}\ }\href {\doibase
  10.1103/PhysRevA.95.012311} {\bibfield  {journal} {\bibinfo  {journal} {Phys.
  Rev. A}\ }\textbf {\bibinfo {volume} {95}},\ \bibinfo {pages} {012311}
  (\bibinfo {year} {2017})}\BibitemShut {NoStop}%
\bibitem [{\citenamefont {Yang}\ \emph {et~al.}(2017)\citenamefont {Yang},
  \citenamefont {Rahmani}, \citenamefont {Shabani}, \citenamefont {Neven},\
  and\ \citenamefont {Chamon}}]{Yang17a}%
  \BibitemOpen
  \bibfield  {author} {\bibinfo {author} {\bibfnamefont {Zhi-Cheng}\
  \bibnamefont {Yang}}, \bibinfo {author} {\bibfnamefont {Armin}\ \bibnamefont
  {Rahmani}}, \bibinfo {author} {\bibfnamefont {Alireza}\ \bibnamefont
  {Shabani}}, \bibinfo {author} {\bibfnamefont {Hartmut}\ \bibnamefont
  {Neven}}, \ and\ \bibinfo {author} {\bibfnamefont {Claudio}\ \bibnamefont
  {Chamon}},\ }\bibfield  {title} {\enquote {\bibinfo {title} {Optimizing
  variational quantum algorithms using {P}ontryagin's minimum principle},}\
  }\href {\doibase 10.1103/PhysRevX.7.021027} {\bibfield  {journal} {\bibinfo
  {journal} {Phys. Rev. X}\ }\textbf {\bibinfo {volume} {7}},\ \bibinfo {pages}
  {021027} (\bibinfo {year} {2017})}\BibitemShut {NoStop}%
\bibitem [{\citenamefont {Farhi}\ \emph
  {et~al.}(2014{\natexlab{a}})\citenamefont {Farhi}, \citenamefont
  {Goldstone},\ and\ \citenamefont {Gutmann}}]{Farhi14a}%
  \BibitemOpen
  \bibfield  {author} {\bibinfo {author} {\bibfnamefont {Edward}\ \bibnamefont
  {Farhi}}, \bibinfo {author} {\bibfnamefont {Jeffrey}\ \bibnamefont
  {Goldstone}}, \ and\ \bibinfo {author} {\bibfnamefont {Sam}\ \bibnamefont
  {Gutmann}},\ }\href@noop {} {\enquote {\bibinfo {title} {A quantum
  approximate optimization algorithm},}\ } (\bibinfo {year}
  {2014}{\natexlab{a}}),\ \Eprint {http://arxiv.org/abs/arXiv:1411.4028}
  {arXiv:1411.4028} \BibitemShut {NoStop}%
\bibitem [{\citenamefont {Farhi}\ \emph
  {et~al.}(2014{\natexlab{b}})\citenamefont {Farhi}, \citenamefont
  {Goldstone},\ and\ \citenamefont {Gutmann}}]{Farhi14b}%
  \BibitemOpen
  \bibfield  {author} {\bibinfo {author} {\bibfnamefont {Edward}\ \bibnamefont
  {Farhi}}, \bibinfo {author} {\bibfnamefont {Jeffrey}\ \bibnamefont
  {Goldstone}}, \ and\ \bibinfo {author} {\bibfnamefont {Sam}\ \bibnamefont
  {Gutmann}},\ }\href@noop {} {\enquote {\bibinfo {title} {A quantum
  approximate optimization algorithm applied to a bounded occurrence constraint
  problem},}\ } (\bibinfo {year} {2014}{\natexlab{b}}),\ \Eprint
  {http://arxiv.org/abs/arXiv:1412.6062} {arXiv:1412.6062} \BibitemShut
  {NoStop}%
\bibitem [{\citenamefont {Jiang}\ \emph {et~al.}(2017)\citenamefont {Jiang},
  \citenamefont {Rieffel},\ and\ \citenamefont {Wang}}]{Jiang17a}%
  \BibitemOpen
  \bibfield  {author} {\bibinfo {author} {\bibfnamefont {Zhang}\ \bibnamefont
  {Jiang}}, \bibinfo {author} {\bibfnamefont {Eleanor~G.}\ \bibnamefont
  {Rieffel}}, \ and\ \bibinfo {author} {\bibfnamefont {Zhihui}\ \bibnamefont
  {Wang}},\ }\bibfield  {title} {\enquote {\bibinfo {title} {Near-optimal
  quantum circuit for {G}rover's unstructured search using a transverse
  field},}\ }\href {\doibase 10.1103/PhysRevA.95.062317} {\bibfield  {journal}
  {\bibinfo  {journal} {Phys. Rev. A}\ }\textbf {\bibinfo {volume} {95}},\
  \bibinfo {pages} {062317} (\bibinfo {year} {2017})}\BibitemShut {NoStop}%
\bibitem [{\citenamefont {Borisov}(2000)}]{Borisov00a}%
  \BibitemOpen
  \bibfield  {author} {\bibinfo {author} {\bibfnamefont {V.~F.}\ \bibnamefont
  {Borisov}},\ }\bibfield  {title} {\enquote {\bibinfo {title} {Fuller's
  phenomenon: Review},}\ }\href {\doibase 10.1007/s10958-000-0001-9} {\bibfield
   {journal} {\bibinfo  {journal} {Journal of Mathematical Sciences,}\ }\textbf
  {\bibinfo {volume} {100}},\ \bibinfo {pages} {2311--2354} (\bibinfo {year}
  {2000})}\BibitemShut {NoStop}%
\bibitem [{\citenamefont {fuller}(1960)}]{Fuller60a}%
  \BibitemOpen
  \bibfield  {author} {\bibinfo {author} {\bibfnamefont {A.~T.}\ \bibnamefont
  {fuller}},\ }\bibfield  {title} {\enquote {\bibinfo {title} {Relay control
  systems optimized for various performance criteria},}\ }in\ \href@noop {}
  {\emph {\bibinfo {booktitle} {Proc. First World Congress IFAC}}}\ (\bibinfo
  {year} {1960})\ pp.\ \bibinfo {pages} {510--519}\BibitemShut {NoStop}%
\bibitem [{\citenamefont {Muthukrishnan}\ \emph {et~al.}(2015)\citenamefont
  {Muthukrishnan}, \citenamefont {Albash},\ and\ \citenamefont
  {Lidar}}]{muthukrishnan15a}%
  \BibitemOpen
  \bibfield  {author} {\bibinfo {author} {\bibfnamefont {Siddharth}\
  \bibnamefont {Muthukrishnan}}, \bibinfo {author} {\bibfnamefont {Tameem}\
  \bibnamefont {Albash}}, \ and\ \bibinfo {author} {\bibfnamefont {Daniel~A}\
  \bibnamefont {Lidar}},\ }\href@noop {} {\enquote {\bibinfo {title} {When
  diabatic trumps adiabatic in quantum optimization},}\ } (\bibinfo {year}
  {2015}),\ \Eprint {http://arxiv.org/abs/arXiv:1505.01249} {arXiv:1505.01249}
  \BibitemShut {NoStop}%
\bibitem [{\citenamefont {Muthukrishnan}\ \emph {et~al.}(2016)\citenamefont
  {Muthukrishnan}, \citenamefont {Albash},\ and\ \citenamefont
  {Lidar}}]{Muthukrishnan2016}%
  \BibitemOpen
  \bibfield  {author} {\bibinfo {author} {\bibfnamefont {Siddharth}\
  \bibnamefont {Muthukrishnan}}, \bibinfo {author} {\bibfnamefont {Tameem}\
  \bibnamefont {Albash}}, \ and\ \bibinfo {author} {\bibfnamefont {Daniel~A.}\
  \bibnamefont {Lidar}},\ }\bibfield  {title} {\enquote {\bibinfo {title}
  {Tunneling and speedup in quantum optimization for permutation-symmetric
  problems},}\ }\href {\doibase 10.1103/PhysRevX.6.031010} {\bibfield
  {journal} {\bibinfo  {journal} {Phys. Rev. X}\ }\textbf {\bibinfo {volume}
  {6}},\ \bibinfo {pages} {031010} (\bibinfo {year} {2016})}\BibitemShut
  {NoStop}%
\bibitem [{\citenamefont {Callison}\ \emph {et~al.}(2018)\citenamefont
  {Callison}, \citenamefont {Rennison-Jones}, \citenamefont {Chancellor},\ and\
  \citenamefont {Kendon}}]{callison18a}%
  \BibitemOpen
  \bibfield  {author} {\bibinfo {author} {\bibfnamefont {Adam}\ \bibnamefont
  {Callison}}, \bibinfo {author} {\bibfnamefont {Christian}\ \bibnamefont
  {Rennison-Jones}}, \bibinfo {author} {\bibfnamefont {Nicholas}\ \bibnamefont
  {Chancellor}}, \ and\ \bibinfo {author} {\bibfnamefont {Viv}\ \bibnamefont
  {Kendon}},\ }\href@noop {} {\enquote {\bibinfo {title} {Solving spin-glasses
  with quantum walks},}\ } (\bibinfo {year} {2018}),\ \bibinfo {note} {in
  preparation}\BibitemShut {NoStop}%
\bibitem [{pyt(2016)}]{python}%
  \BibitemOpen
  \href {https://www.python.org/} {\enquote {\bibinfo {title} {Python 2.7 \&
  {P}ython 3.5},}\ } (\bibinfo {year} {2016}),\ \bibinfo {note}
  {https://www.python.org/, accessed August 10th, 2016}\BibitemShut {NoStop}%
\bibitem [{num(2016)}]{numpy}%
  \BibitemOpen
  \href {http://www.numpy.org/} {\enquote {\bibinfo {title} {Numpy 1.11.1},}\ }
  (\bibinfo {year} {2016}),\ \bibinfo {note} {http://www.numpy.org/, accessed
  August 10th, 2016}\BibitemShut {NoStop}%
\bibitem [{sci(2016)}]{scipy}%
  \BibitemOpen
  \href {https://www.scipy.org/} {\enquote {\bibinfo {title} {Scipy 0.17.1},}\
  } (\bibinfo {year} {2016}),\ \bibinfo {note} {https://www.scipy.org/,
  accessed August 10th, 2016}\BibitemShut {NoStop}%
\bibitem [{mat(2016)}]{matplotlib}%
  \BibitemOpen
  \href {http://matplotlib.org/} {\enquote {\bibinfo {title} {Matplotlib
  1.5.1},}\ } (\bibinfo {year} {2016}),\ \bibinfo {note}
  {http://matplotlib.org/, accessed August 10th, 2016}\BibitemShut {NoStop}%
\end{thebibliography}%
%%%%%%%%%%%%%%%%%%%%%%%%%%%%%%%%%%%%%%%%%%%%%%%%%%%%%%%%%%%%%%%%%%%%%%%%%%%%%%%
%
\end{document}